# A Frequency-Space Terahertz Transceiver Chip for Multi-Agent Communications and Spatial Awareness

Xiaoyue Xia[1†], Zhicheng Lin[1†*], Hao Guo[1], Siran Wang[1], Jingyuan Zhang[1,2], Xinyu Fang[1], Ka Fai Chan[1], Kam Man Shum[1], Geng-Bo Wu[1,2*], and Chi Hou Chan[1,2*]

[1]State Key Laboratory of THz and Millimeter Waves, City University of Hong Kong, Hong Kong, 999077, China

[2]Department of Electrical Engineering, City University of Hong Kong, Hong Kong, 999077, China

[†]These authors contributed equally to this work

[*]Corresponding author. Email: zhichelin2@cityu.edu.hk, bogwu2@cityu.edu.hk, eechic@cityu.edu.hk

## ABSTRACT

Future indoor embodied-intelligence systems require scalable hardware platforms that support high-capacity multi-agent connectivity and mutual spatial awareness. The terahertz (THz) spectrum offers abundant bandwidth and spatial selectivity for integrated sensing and communication (ISAC); however, conventional phased arrays and programmable metasurfaces rely on dense beamforming networks, element-level control, or external THz illumination, making scalable multibeam operation challenging. Here, we report a fully integrated 208-258 GHz 65-nm complementary metal oxide semiconductor (CMOS) THz transceiver chip that monolithically integrates broadband front ends with heterogeneous leaky-wave metasurface (HLM) apertures within a 1.5 × 4.9 $mm^2$ area. The HLM generates strongly dispersive leaky modes, enabling frequency-controlled multibeam scanning across 75° with only four meta-atoms. The architecture cascades frequency-domain and spatial-domain mixing to establish spectrally clean, bidirectional frequency-space mapping between low-frequency signals and frequency-addressed free-space THz beams, enabling spatial-frequency division multiple access (SFDMA) communication. The THz chip demonstrates simultaneous multi-agent transmission and reception, two-dimensional localization, and sensing-enhanced communication, providing a scalable hardware platform for future THz embodied-intelligence networks.

The emergence of embodied artificial intelligence (AI) is transforming factories, warehouses, and other indoor spaces into densely connected environments populated by autonomous mobile robots, robotic manipulators, quadruped robots, inspection drones, and service machines[1-3]. Fig. 1a illustrates an example of the future lights-out smart factory. A highly integrated radio platform that combines concurrent directional connectivity with mutual spatial awareness is becoming a fundamental infrastructure requirement for scalable embodied-AI networks[4-6].

**a**

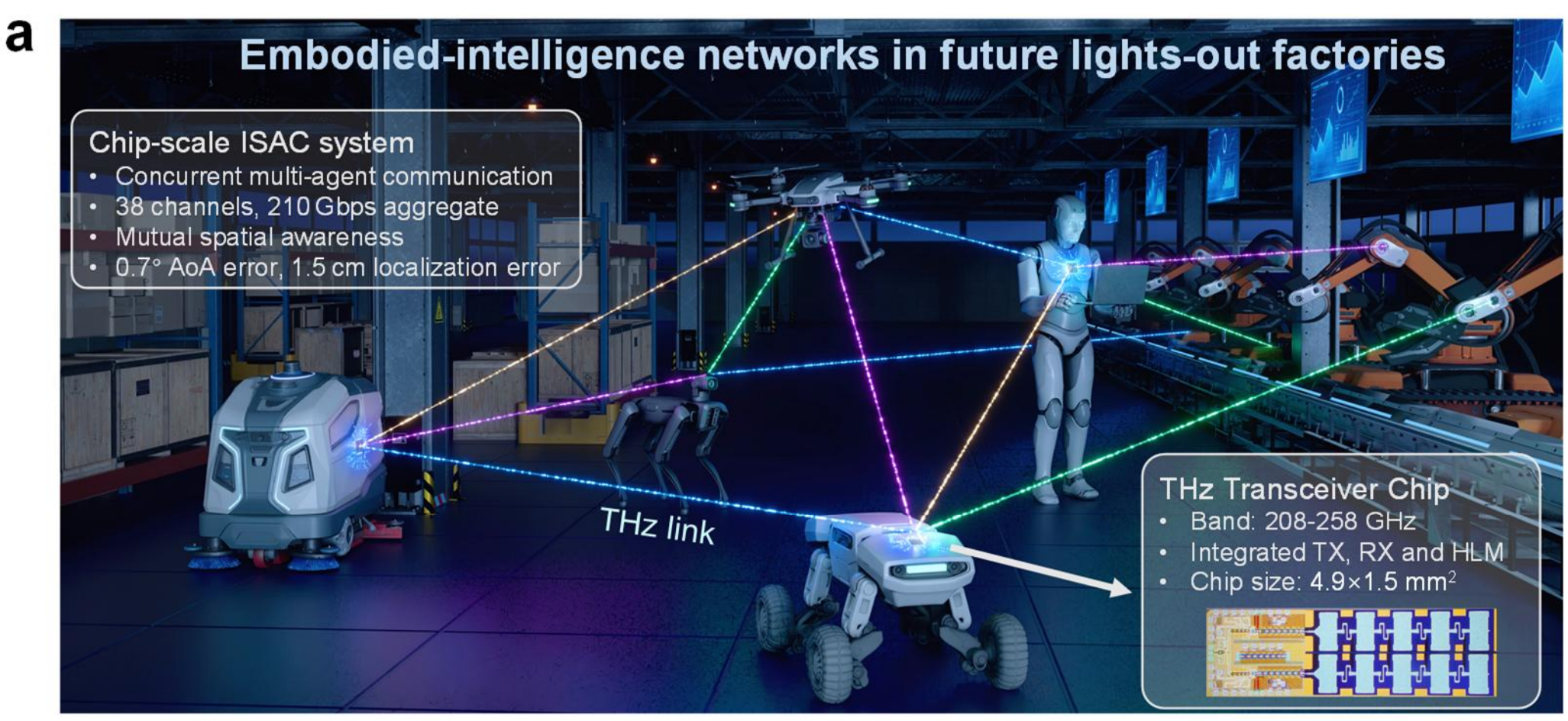

**b**

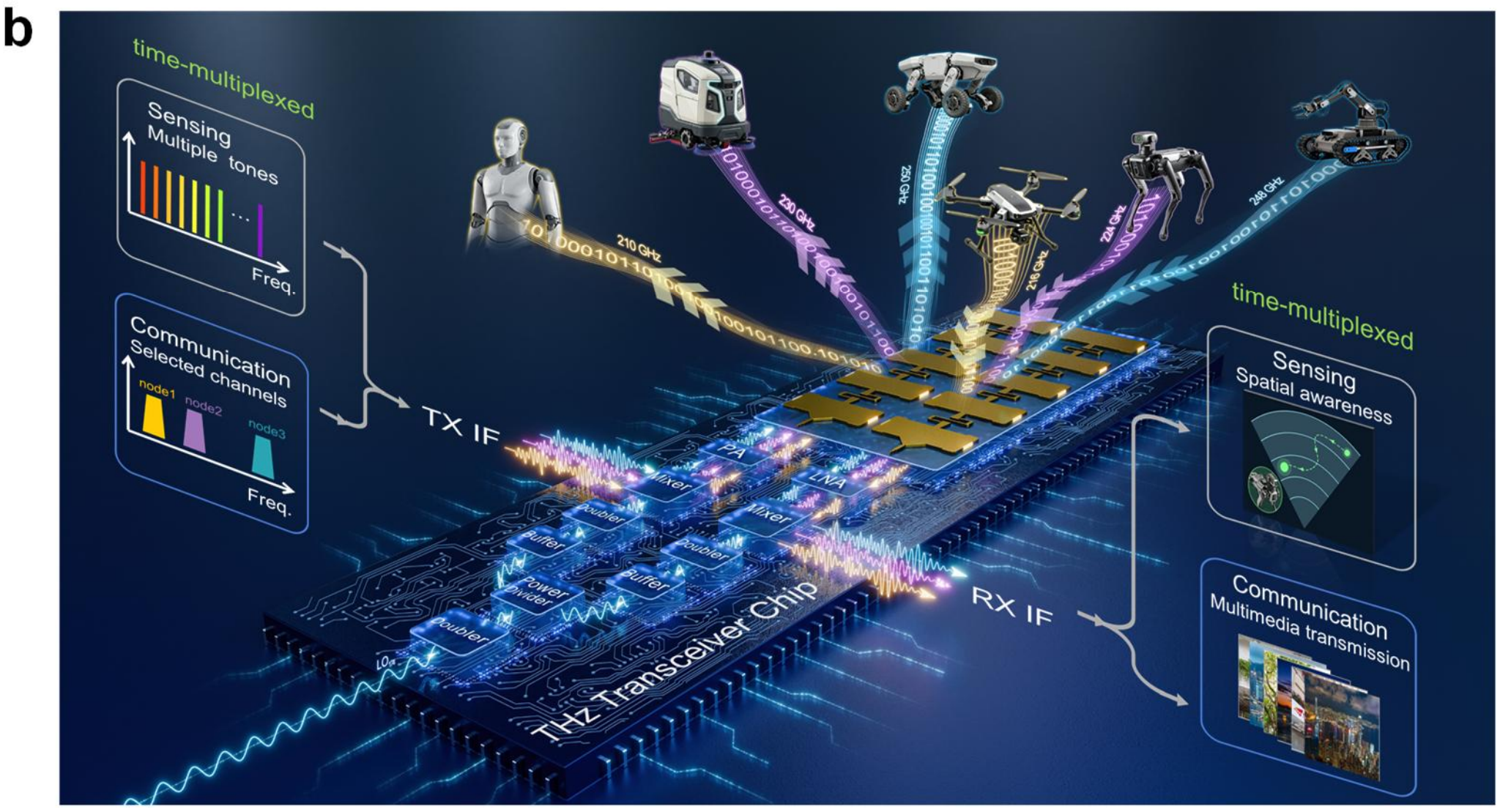

**Fig. 1 | Chip-level THz integrated sensing and communication for embodied-intelligence networks.** **a,** A lights-out smart factory with heterogeneous embodied-AI agents enabled by the proposed THz chip for multi-agent communication and mutual spatial sensing. Colored rays indicate THz links at different frequencies. **b,** Frequency-addressed transceiver concept with a time-multiplexed sensing-communication protocol, where multiple tones support spatial awareness and selected carriers enable concurrent frequency-spatial communication channels.

Recent years have witnessed substantial progress in integrated sensing and communication (ISAC), yet most demonstrations remain centred on point-to-point (P2P) communication and sensing of passive targets or external nodes[7-13], as summarised in Table 1. Embodied-AI collectives demand ISAC systems beyond this conventional paradigm, requiring scalable concurrent directional communication among dense mobile agents, together with mutual spatial awareness acquired directly between communicating peers[14,15]. In addition, dense embodied-AI networks can demand aggregate data rates beyond 100 Gbps, especially when multiple agents exchange high-resolution sensory streams, maps, or model states[16]. Such requirements are difficult to satisfy within millimeter-wave bandwidths, motivating a move towards the terahertz (THz) regime[17-28]. However, THz ISAC demonstrations remain scarce and insufficient for embodied-AI collectives (Table 1): a 122-168-GHz complementary metal oxide semiconductor (CMOS) transceiver (TRX) chip provides 12 Gbps P2P communication with frequency-modulated continuous-wave (FMCW) ranging of passive targets[12]. A portable 220-GHz phased-array platform achieves 64 Gbps communication and THz imaging but remains limited to a single user and a bulky $20 \times 20 \times 10$ cm$^3$ form factor[13].

**Table 1 | Comparison of representative integrated ISAC systems from microwave to THz frequencies.**

| Reference/ Frequency | Chip-level | TX-RX integration | Aperture integration | Sensing among TRX nodes | Concurrent multi-link operation | Communication channels | Aggregate data rate |
|---|---|---|---|---|---|---|---|
| [7]: 10.3 GHz | × | × | ✓ | × | × | 1 | 0.5 Mbps |
| [8]: 28 GHz | ✓ | × | × | × | × | 1 | N.R. |
| [9]: 30.5 GHz | ✓ | × | ✓ | × | × | 1 | 3.2 Gbps |
| [10]: 28.7-36.2 GHz | ✓ | ✓ | × | × | × | 1 | 2.4 Gbps |
| [11]: 60 GHz | ✓ | ✓ | × | × | × | 1 | >7 Gbps |
| [12]: 122-168 GHz | ✓ | ✓ | × | × | × | 1 | 12 Gbps |
| [13]: 220 GHz | × | ✓ | × | × | × | 1 | 64 Gbps |
| **This work: 208-258 GHz** | ✓ | ✓ | ✓ | ✓ | ✓ | **38** | **210 Gbps** |

Note: The table is restricted to systems that experimentally demonstrate both sensing and communication. For this work, 38 denotes frequency–spatial channels spanning 209.6–257.7 GHz, each with a 1-GHz bandwidth. The 210-Gbps value is obtained by summing the demonstrated per-channel data rates.

Another critical requirement for practical THz platforms is the ability to overcome the severe free-space path loss inherent to THz propagation. State-of-the-art THz beamforming chips address this challenge by generating and scanning highly directional beams. These approaches, including phased arrays[29-37] and reflect/transmitarrays[38-41], rely on dense phase-amplitude networks, element-level control, or external THz illumination, resulting in substantial hardware and control complexity. Their reported system demonstrations are limited to sequential beam steering or single-user operation, rather than concurrent multiuser access. Leaky-wave metasurfaces or antennas offer an alternative by integrating wave propagation and beam formation within a distributed interface[42-44]. Their intrinsic dispersion maps carrier frequency to radiation direction, enabling frequency-addressed multibeam links without beamforming networks. Previous implementations have demonstrated THz beam scanning[45], wireless connectivity[46], radar[47], localization[48], multiplexing[49], and link discovery[50], but each addresses only single function. Moreover, most rely on passive structures driven by external sources, probes, or multiplier chains, limiting system integration and link performance. Monolithic integration of dispersive apertures with active front ends, particularly THz amplifiers, is therefore essential for compact, high-performance multifunctional systems.

Here, we propose a fully integrated 65-nm CMOS THz ISAC TRX that monolithically integrates broadband transmitter (TX) and receiver (RX) front ends with heterogeneous leaky-wave metasurface (HLM) apertures on a 1.5 × 4.9 $mm^2$ chip. To the best of our knowledge, this work is the first CMOS implementation operating above 200 GHz to combine on-chip beamforming, single-chip TX/RX integration, and THz power amplifier (PA)/low-noise amplifier (LNA) front ends into a tiny chip platform. Periodic engineering of the phase and leakage responses in heterogeneous waveguide structures creates a strongly dispersive leaky mode, enabling rapid wide-angle frequency scanning with only four meta-atoms. By combining frequency-domain mixing in the CMOS TRX with spatial-domain mixing in the periodic HLM, the co-designed architecture selects and amplifies the desired in-band frequency-wavevector modes while suppressing image-frequency components and undesired spatial harmonics. Together with aperture-level self-interference cancellation, this circuit-to-space architecture generates spectrally and spatially clean THz frequency-space mapping for simultaneous transmit-receive operation. Experimentally, the chip achieves a 75° frequency-controlled beam scan from 208 to 258 GHz and supports 38 frequency-spatial channels with an aggregate capacity of 210 Gbps, obtained by summing the individually demonstrated per-channel rates. We further demonstrate three concurrent communication links, including two uplinks and one downlink, as well as angle of arrival (AoA) estimation with a mean error of 0.7° and two-dimensional trajectory reconstruction with a mean localization error of 1.49 cm. As compared in Table 1, our work uniquely enables concurrent multi-link communication together with direct spatial sensing of communicating peer TRX nodes. These results and advantages establish a compact hardware-native spatial-frequency division multiple access (SFDMA) platform for multi-agent connectivity and spatial awareness in future embodied-AI networks.

## Results

### Chip-level ISAC transceiver concept and architecture

Fig. 1b shows the concept of the TRX chip, and Fig. 2a details the architecture and signal flow. For TX, the frequency-multiplexed intermediate frequency (IF) input is upconverted by a common local oscillator (LO), amplified by a broadband seven-stage PA, and coupled to the TX HLM, whose frequency-to-space mapping directs different carrier frequencies toward different directions to generate multiple frequency-addressed THz beams. Reciprocally, the RX HLM captures THz signals from different directions and maps them to the corresponding frequency channels, which are then amplified by the LNA and downconverted to produce a frequency-multiplexed IF output. The proposed THz frequency-space mapping is further demonstrated through multi-agent communication, trajectory reconstruction, and representative wireless multimedia transmission. A micrograph of the chip is included as an inset in Fig. 1a.

The proposed ISAC chip operates with a time-multiplexed protocol. During sensing, one chip radiates multiple tones, while surrounding chips receive the resulting spectral signatures to estimate the AoA of the signal from the transmitting chip. During communication, carrier frequencies are selected based on the estimated AoAs, enabling agents at different locations to communicate concurrently over distinct frequency-spatial channels. Embodied-AI agents equipped with the proposed TRX chip could form large-scale networks for multi-node communication and spatial sensing, as exemplified by a lights-out smart factory in Fig. 1a. Different colors indicate THz links operating at distinct frequencies, highlighting the proposed architecture's frequency-to-space mapping capability.

### HLM-enabled frequency-to-space mapping

Frequency-space mapping of the HLM apertures provides the physical basis for both SFDMA communication and AoA sensing. The measured radiation patterns of TX and RX (Fig. 2b) demonstrate continuous frequency-controlled beam scanning over a 75° angular range. Remarkably, frequency-addressed multibeam generation or reception is realized using only four meta-atoms in a CMOS process, owing to the periodic alternation of three guided-wave structures within the leaky-wave metasurface: the chip-integrated-waveguide (CIW), parallel-plate-waveguide (PPW), and microstrip line (MSL). We therefore refer to the resulting architecture as an HLM, with details provided in Supplementary Section 1.1. As shown in Fig. 2c, each rotationally symmetric meta-atom sequentially comprises CIW1, PPW1, an S-shaped MSL, PPW2, and CIW2. The transition apertures and open-ended MSL stubs are modeled as six equivalent magnetic-current sources, $\boldsymbol{M}_{s1}$-$\boldsymbol{M}_{s6}$, which define the distributed leakage sites within each meta-atom. The HLM consists of four complete periods, with an input MSL-to-CIW transition for direct impedance matching to the front-end PA/LNA. Joint control of the longitudinal phase accumulation and spatial leakage distribution constitutes the phase-amplitude co-engineering of the HLM, enabling wide-angle, high-gain frequency scanning within a compact aperture.

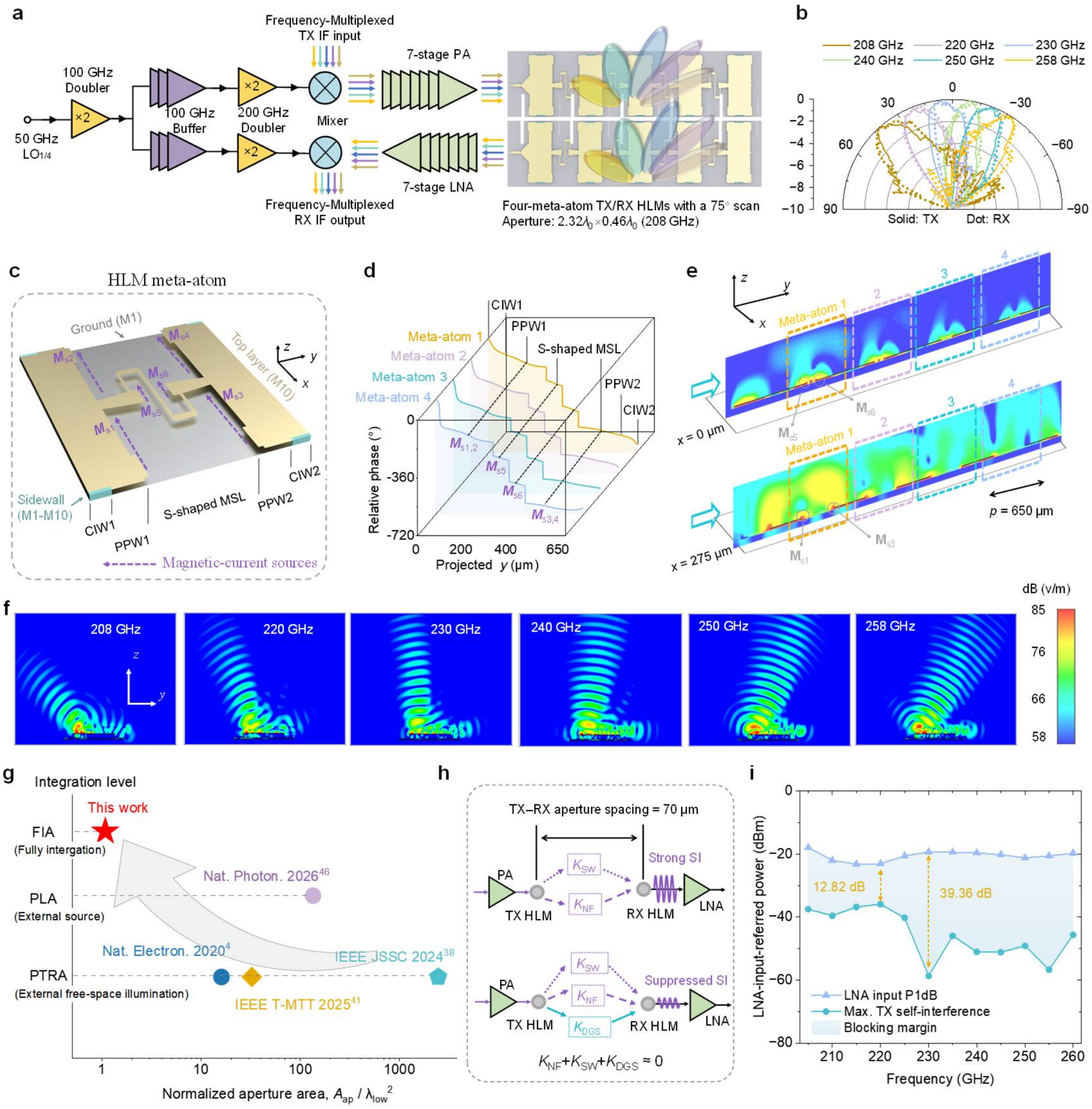

**Fig. 2 | Terahertz frequency-scanned ISAC TRX enabled by on-chip HLMs. a,** Architecture of the proposed frequency-space-multiplexed THz TRX. **b,** Measured frequency-scanned patterns of the TX and RX at different frequencies. **c,** Three-dimensional layout of HLM meta-atom implemented in 65-nm CMOS process, consisting of two CIWs, two PPWs, and an S-shaped MSL. **d,** Phase evolution of the proposed periodic structure at 235 GHz. The engineered S-shape MSL and CIW/PPW transitions provide controlled phase accumulation across the meta-atoms, enabling the desired spatial phase distribution. **e,**

Full-wave simulated E-field distributions in $x$=0 μm and $x$=275 μm planes at 235 GHz, showing the controlled leakage and radiation generated by the periodically distributed magnetic currents. **f,** Full-wave simulated E-field distributions of the HLM at frequencies from 208 to 258 GHz. **g,** Comparison of experimentally demonstrated on-chip THz wavefront-manipulation apertures in terms of integration level and aperture area. FIA: Fully integrated aperture; PLA: Passive leaky-wave aperture; PTRA: Programmable transmitarray/reflectarray aperture. **h,** Self-interference coupling topology between the TX and RX. The proposed DGS introduces an additional coupling path that destructively interferes with the intrinsic near-field and surface-wave coupling, leading to suppressed TX-to-RX leakage. **i,** Calculated LNA input-referred TX self-interference power with the DGS-based decoupling structure, compared with the LNA input 1-dB compression point.

Fig. 2d shows the phase evolution within the four meta-atoms at 235 GHz, plotted along the projected periodic $y$-axis. The CIW, PPW, and S-shaped MSL sections exhibit distinct phase-evolution rates because of their different modal propagation constants, while their cascaded transitions modify the local phase gradient. The folded MSL further introduces additional electrical length within a compact longitudinal footprint. The annotated source positions relate the local guided-wave phase to the effective leakage sites, thereby establishing the phase relationships required for coherent far-field synthesis. The similar phase-evolution trends across the four meta-atoms are consistent with the periodic spatial phase modulation of the HLM.

Fig. 2e shows the near-field electric-field (E-field) magnitude distributions in two cross-sections of the HLM at 235 GHz, with the $z$-axis exaggerated for clarity. In the $x$ = 0 μm plane, the fields emanate from the open-ended MSL stubs and couple into free space, thereby identifying $\boldsymbol{M}_{s5}$ and $\boldsymbol{M}_{s6}$ as radiating magnetic current sources. In the $x$ = 275 μm plane, localized fields emerge from the PPW-MSL transition apertures corresponding to $\boldsymbol{M}_{s1}$ and $\boldsymbol{M}_{s3}$, with $\boldsymbol{M}_{s2}$ and $\boldsymbol{M}_{s4}$ formed symmetrically on the opposite side. These field distributions repeat with a period of 650 μm, forming a distributed leaky aperture, and providing field-level evidence for the engineered amplitude modulation, whereby guided-wave energy is selectively coupled into free space along the HLM aperture.

Supplementary Section 1.2 presents full-wave simulation results demonstrating the impedance matching, conversion efficiency, and realized gain of the HLMs over the operating bandwidth. The simulated E-field distributions across 208-258 GHz (Fig. 2f) illustrate the guided-to-space-wave conversion and the resulting frequency-to-space mapping. Fig. 2g compares on-chip THz wavefront-manipulation apertures that enable negative-to-positive-angle scanning through aperture-level beamforming and have been experimentally validated at the system level. Apertures that provide only single-sided angular scanning are excluded to ensure a consistent comparison criterion[39,48]. State-of-the-art reflectarray[38] and transmitarrays[40,41] have demonstrated impressive angular scanning performance using electrically large apertures and sophisticated wavefront-control schemes. However, they often rely on complex element-level implementations and external THz free-space illumination. A recent leaky-wave metasurface[46] has demonstrated wide spatial coverage and wireless links, yet still relied on external THz coupling. In contrast, the proposed four-meta-atom HLM achieves a 75° continuous scan with the smallest normalized aperture and reduced control complexity. More importantly, the proposed architecture monolithically integrates the THz TX, RX, and HLM apertures on a single chip, providing a compact and scalable solution for low-cost, large-scale deployment of embodied-AI ISAC systems.

Closely spaced TX and RX apertures, separated by only 70 μm, introduce strong TX leakage that may compress the LNA, driving it to its input 1-dB compression point ($P_{1dB}$), during simultaneous operation. To suppress this coupling, the defected-ground structure (DGS) creates a controlled path that generates equal-amplitude, opposite-phase coupling to cancel the intrinsic surface-wave and near-field coupling between the two HLMs (Fig. 2h), without requiring an active cancellation loop. Supplementary Section 1.3 details the design and the analysis. As shown in Supplementary Fig. S1-8, the DGS provides 38–61 dB TX-to-RX isolation across 208–258 GHz, with a peak improvement of 26 dB over the no-DGS structure. Fig. 2i shows that the resulting TX self-interference remains below the LNA input $P_{1dB}$, with a minimum blocking margin of 12.82 dB and a margin of more than 39 dB at 230 GHz. Simultaneous TX/RX experiments under strong adjacent-channel TX excitation further confirm a stable demodulated signal-to-noise ratio (SNR) (Supplementary Fig. S1-10), validating adjacent-channel simultaneous transmission and reception in the fully integrated THz TRX.

## Bridging CMOS circuits and electromagnetic apertures

Fig. 3 presents the dual-domain circuit-aperture co-design of the fully integrated THz TRX. Frequency-domain mixing and spatial-domain mixing exhibit a formal frequency-wavevector duality: temporal mixing translates low-frequency electrical signals into THz spectral components, whereas spatial periodicity translates guided-wave momenta into Floquet harmonics. Cascading these two processes establishes a bidirectional frequency-to-space mapping between low-frequency signals and frequency-addressed free-space THz beams.

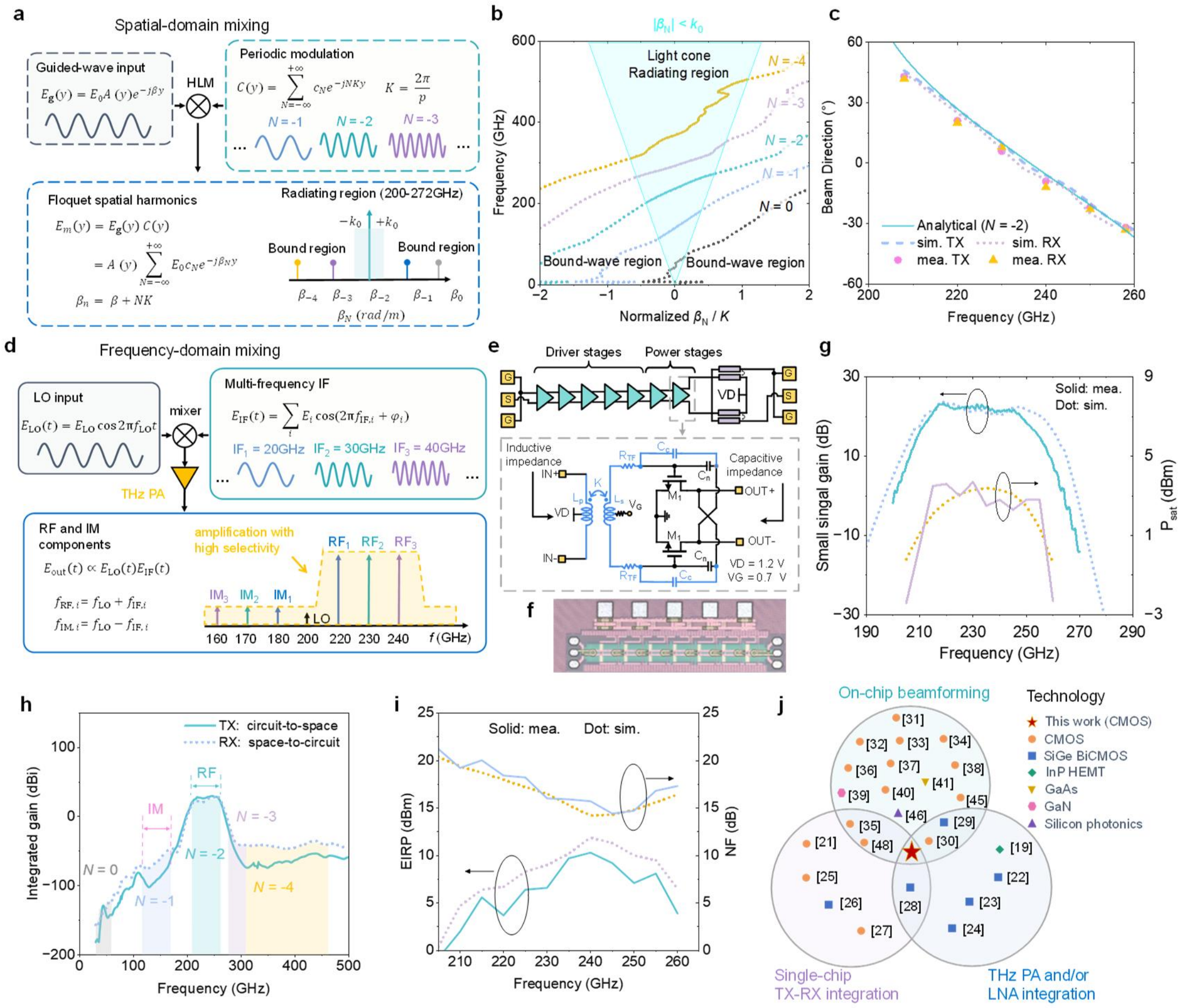

**Fig. 3 | Bridging CMOS front ends and HLM apertures through dual-domain mixing. a,** Spatial wavevector mixing based on the proposed HLM, enabling radiation through engineered periodic perturbations. **b,** Normalized phase constants $\beta_N/K$ of Floquet orders versus frequencies, where harmonics satisfying $|\beta_N| < k_0$ fall inside the light cone and radiate into free space. **c,** Beam direction of the $N$ = -2 harmonic from analytical calculation, full-wave simulation, and TX/RX measurements, confirming consistent frequency-to-space mapping in the operating band. **d,** Frequency-domain mixing scheme for THz signal generation and IM suppression enabled by the frequency-selective PA. **e,** Schematic of the proposed PA architecture and the detailed power-stage implementation. **f,** Die micrograph of the PA. **g,** Measured and simulated small-signal gain and saturated output power of the THz PA across the operating frequency range. **h,** Calculated integrated conversion gain of the TX and RX, combining the PA or LNA gain with the HLM circuit-to-space or space-to-circuit conversion response. **i,** Measured and simulated TX EIRP and RX noise figure of the complete HLM-integrated TRX. **j,** Technology landscape of representative integrated THz platforms categorized according to on-chip beamforming, single-chip TX-RX integration, and THz PA and/or LNA integration. The proposed 65-nm CMOS TRX chip is located at the intersection of all three categories.

We first describe the spatial-domain mixing mechanism underlying the frequency-to-space mapping of the periodic HLM. As illustrated in Fig. 3a, the guided wave incident on the HLM can be expressed as:

$$E_{\mathrm{g}}(y) = E_0 A(y) \exp(-j\beta y) \quad (1)$$

where $E_0$ is a constant reference amplitude, $A(y)$ is the normalized slowly varying envelope, and $\beta$ is the propagation constant of the unmodulated guided mode. The periodic HLM provides a spatial coupling function decomposed into Floquet components,

$$C(y) = \sum_{N=-\infty}^{+\infty} c_N \exp(-jNKy), \quad K = \frac{2\pi}{p} \quad (2)$$

where $p$ is the meta-atom periodicity, $K$ is the fundamental lattice wavevector, and $c_N$ is the complex coupling coefficient of the $N$-th spatial order. Multiplication of the guided field by the periodic coupling function generates a modulated aperture field:

$$E_{\mathrm{m}}(y) = E_{\mathrm{g}}(y) C(y) = E_0 A(y) \sum_{N=-\infty}^{+\infty} c_N \exp(-j\beta_N y) \quad (3)$$

$$\beta_N = \beta + NK, \quad N = 0, \pm 1, \pm 2, \dots \quad (4)$$

Thus, the periodic HLM spatially mixes the guided-wave momentum $\beta$ with lattice wavevectors $NK$, producing Floquet harmonics $\beta_N$. Using full-wave-simulated S-parameters of the HLM meta-atom, we extract the effective Bloch phase constant and construct the Floquet dispersion. As shown in Fig. 3b, radiation occurs when the harmonic enters the free-space light cone,

$$|\beta_N| < k_0 \quad (5)$$

Within the operating band of 208-258 GHz, the $N$ = -2 branch satisfies this condition and forms the desired radiating harmonic, while the other branches remain bound. Under the adopted angular convention, the radiation angle follows

$$\theta_N = \sin^{-1}\left(\frac{\beta_N}{k_0}\right) \quad (6)$$

As shown in Fig. 3c, the beam directions predicted from the analytical ($N$ = -2) Floquet branch agree closely with the full-wave simulations and the measured TX and RX radiation patterns. This agreement shows that the measured continuous 75° scan from 208 to 258 GHz originates from the engineered ($N$ = -2) dispersion branch and confirms the matched frequency-to-space mappings of the transmit and receive HLMs.

We next consider frequency-domain mixing. As illustrated in Fig. 3d, multiple low-frequency IF signals are mixed with a common 200-GHz LO. The mixing process can

be expressed as:

$$E_{\mathrm{mix}}(t) \propto E_{\mathrm{LO}}(t)E_{\mathrm{IF}}(t) = \frac{E_{\mathrm{LO}}}{2}\sum_i E_i\left[\cos\left(2\pi f_{\mathrm{RF},i}t + \varphi_i\right) + \cos\left(2\pi f_{\mathrm{IM},i}t - \varphi_i\right)\right] \quad (7)$$

where $f_{\mathrm{RF},i} = f_{\mathrm{LO}} + f_{\mathrm{IF},i}$ denotes the desired sum-frequency component, and $f_{\mathrm{IM},i} = f_{\mathrm{LO}} - f_{\mathrm{IF},i}$ denotes the corresponding difference-frequency image (IM). In the transmit path, the frequency-selective PA amplifies the desired 208-258-GHz radio frequency (RF) band while attenuating the image-frequency products. The LNA provides a corresponding frequency-selective response in the receive path. The TX and RX use bidirectional double-balanced fundamental mixers, with circuit-level details provided in Supplementary Section 2.5.

A key feature of the proposed architecture is the monolithic integration of an amplifier-last TX and an amplifier-first RX with broadband THz PA and LNA stages. Compared with many THz mixer-last TX[35] and mixer-first RX[31,35,37], placing the active amplifiers directly at the HLM interfaces improves the transmitted power and RX sensitivity, while their band-selective responses provide RF preselection against image-band products and noise (Supplementary Fig. S2-1). Enabled by THz-aware transistor layout and broadband transformer-embedded over-neutralization (TEON) gain stages, the measured seven-stage PA and simulated seven-stage LNA each achieve a peak gain exceeding 20 dB near the maximum oscillation frequency ($f_{\mathrm{max}}$) limit of the 65-nm CMOS process (Supplementary Sections 2.2 and 2.3).

A standalone PA chip validates the TEON design (Figs. 3e-g), achieving a measured peak small-signal gain of 23 dB, more than 11 dB gain from 208 to 258 GHz, and a maximum saturated output power of 3.7 dBm. Among the reported 65-nm CMOS H-band amplifiers in the open literature, the PA achieves the largest gain-bandwidth product of 497 GHz, the smallest core area of 0.04 mm$^2$, and the highest saturated-output-power density of 58.6 mW mm$^{-2}$ (Supplementary Table S2-1). The LNA also employs the TEON technique, with transistor sizing, matching networks, and bias conditions specifically optimized for low-noise operation. Together with the noise-optimized source impedance presented by the HLM, these design choices enable a minimum simulated noise figure (NF) of 9.3 dB (Supplementary Fig. S2-4).

Fig. 3h combines the simulated PA/LNA gain with the HLM conversion gain to evaluate the integrated circuit-to-space and space-to-circuit responses. The corresponding frequency windows and integrated gain ranges are summarized in Supplementary Fig. S2-9. For TX, the desired 208-258-GHz band is dominated by the engineered $N$ = -2 Floquet harmonic and provides 20.3 to 30.0 dBi integrated conversion gain, while the IM band and the $N$ = 0, $N$ = -1, $N$ = -3, and $N$ = -4 harmonic windows are suppressed by at least 34.6, 148.3, 79.6, 26.4, and 73.4 dB, respectively. This joint response selects the desired frequency-wavevector mode, yielding spectrally clean and spatially selective THz frequency-to-space mapping.

At the system level, the complete TRX achieves a measured peak effective isotropic radiated power (EIRP) of 10.3 dBm and a minimum RX NF of 14.4 dB (Fig. 3i). The measurement methods and the corresponding TX/RX conversion gains are provided in Supplementary Section 2.6. Compared with the reported THz phased-array TRXs, this work achieves the widest operating bandwidth, together with the lowest reported RX NF. It also delivers a per-channel saturation output power of 3.6 dBm while consuming only 0.43 W, representing the lowest power consumption among the compared designs. More importantly, the proposed HLM-enabled dispersion-engineering architecture is the only design in the comparison that supports multibeam operation, while its wide steering range from −33° to +42° is also highly competitive (Supplementary Table S2-2).

Fig. 3j summarises the technology landscape of representative integrated THz platforms across three capabilities: on-chip beamforming, single-chip TX-RX integration, and integrated THz PA/LNA front ends. Among the systems compared, the proposed 65-nm CMOS TRX chip is the only one that combines all three, providing a system-level architecture that jointly addresses integration density, spatial coverage, and front-end performance.

**SFDMA-enabled multi-agent THz communication**

The dispersive HLM provides an inherent frequency-to-space mapping for SFDMA, assigning different frequency-spatial channels to users in different directions. With matched TX and RX frequency-angle responses, the carrier frequency corresponding to a given link direction automatically aligns both beams when the apertures are parallel. Channel bandwidth can be flexibly allocated according to link requirements: wider channels provide higher data rates for nearby agents, whereas narrower channels improve link robustness and extend communication range for distant agents.

As a representative implementation, the available THz operating bandwidth is divided into 38 frequency-spatial channels with 1-GHz bandwidth and 0.3-GHz guard spacing (Fig. 4a). With the first channel centered at 209.6 GHz and adjacent centers spaced by 1.3 GHz, the center frequency of the $i$-th channel for communication is given as

$$f_i^{\mathrm{comm}} = 209.6 + 1.3\,(i-1)\ \mathrm{GHz},\ \ i = 1, 2, \ldots, 38 \tag{8}$$

The signal power received at the RX IF output at the $i$-th channel can be expressed as

$$P^{\mathrm{sig}}_{RX_IF,i} = P^{\mathrm{comm}}_{TX_IF,i}\, G_{u,i}\, G_{PA,i}\, G_{TXap,i}(\theta)\left(\frac{c}{4\pi\, f_i^{\mathrm{comm}}\, R}\right)^2 G_{RXap,i}(\theta) G_{LNA,i}\, G_{d,i} \tag{9}$$

Here, $P^{\mathrm{comm}}_{TX_IF}$ denotes average effective input power of the modulated IF signal at the TX mixer, $G_u$ and $G_d$ are the conversion gains of the TX upconversion and RX downconversion, $G_{PA}$ and $G_{LNA}$ are the gains of the PA and LNA, $G_{TXap}(\theta)$ and $G_{RXap}(\theta)$ are the frequency- and angle-dependent responses of the TX and RX leaky-metasurface apertures, $c$ is the light speed, and $R$ is the distance between TX and RX.

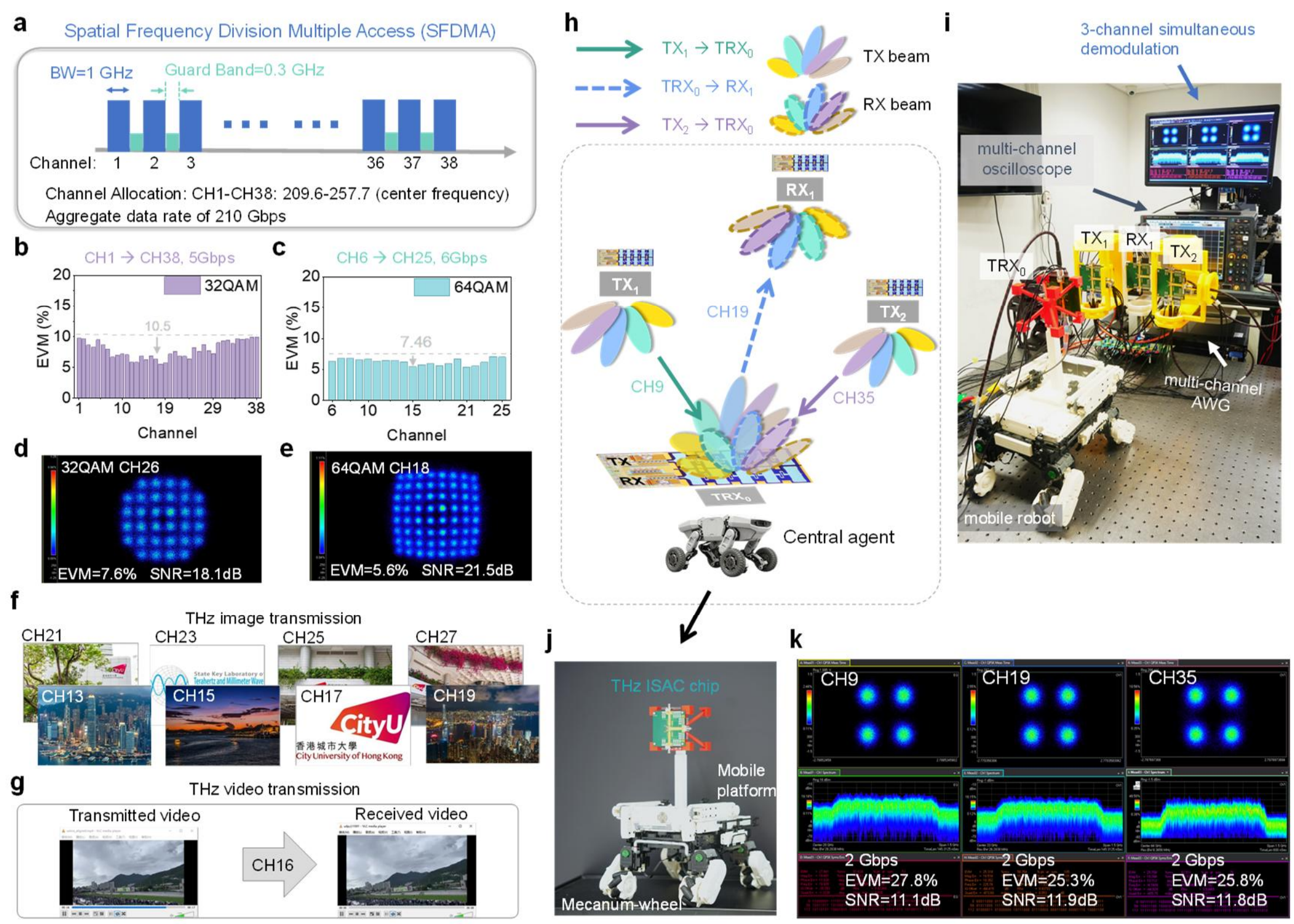

**Fig. 4 | SFDMA communication performance.** **a**, Channel allocation from CH1 to CH38. **b, c,** Measured EVM across the allocated channels for 32-QAM at 5 Gbps and 64-QAM at 6 Gbps, respectively. **d, e,** Demodulated constellations of 32-QAM at CH26 and 64-QAM at CH18, respectively. **f,** Over-the-air THz image transmission to a mobile robot through CH13, CH15, CH17, CH19, CH21, CH23, CH25, and CH27. **g,** Video transmission over the THz link at CH16. **h, i,** Conceptual illustration and experimental setup of one-to-three simultaneous TX/RX SFDMA communication. **j,** Photograph of the THz ISAC chip mounted on a mobile robotic platform. **k,** Measured QPSK constellations and received spectra for CH9, CH19, and CH35 at 2 Gbps per link, demonstrating simultaneous uplink and downlink communication over three spatial-frequency channels with an aggregate concurrent rate of 6 Gbps.

The 38 spatial-frequency channels are characterized through over-the-air modulation experiments using a mobile RX positioned over a 75° angular range around a fixed TX (Supplementary Fig. S3-1). Root-raised-cosine (RRC) pulse shaping and matched filtering are applied with a roll-off factor of $\alpha = 0.1$. For a bit-error-rate (BER) below $10^{-3}$, the error vector magnitude (EVM) limits are 10.5% and 7.46% for 32- and 64-quadrature amplitude modulation (QAM), respectively. As shown in Figs. 4b and c, all 38 channels support 32-QAM, while CH6-CH25 also support 64-QAM. The individually measured per-channel data rates yield an aggregate of 210 Gbps. Representative constellations of 32-QAM (CH26) and 64-QAM (CH18) are respectively shown in Figs. 4d and e.

With the 1-GHz channel allocation used for SFDMA, a maximum link distance of 82 cm is demonstrated using 1-Gbps binary phase-shift keying (BPSK) (Supplementary Fig. S3-2) at CH20. With a 5-GHz channel bandwidth, the proposed system achieves a maximum data rate of 25 Gbps using 5-Gbaud 32-QAM modulation (Supplementary Fig. S3-3). The link budgets for both cases are given in Supplementary Section 3.2.

Practical multimedia transmission is demonstrated using a pair of the TRX chips, with a detailed setup provided in Supplementary Section 3.3. One chip is fixed as the TX, while the other chip operates as the RX on a mobile platform and moves along predefined positions. Image transmission is demonstrated over eight representative SFDMA channels, with channel switching selecting the corresponding spatial directions and enabling successful image recovery at different positions over 0.5-1.0 m links (Fig. 4f). Real-time video transmission is further demonstrated through CH16 over a 1-m link, with continuous frame reconstruction and stable link operation (Fig. 4g). The video is provided as Supplementary Video 1.

Finally, simultaneous multi-agent SFDMA operation is demonstrated in a one-to-three communication experiment using four chips (Figs. 4h, i and j). $TRX_0$ is mounted on a mobile robot, while the remote nodes ($TX_1$, $TX_2$, and $RX_1$) are positioned 42-45 cm from $TRX_0$ in different angular directions. The omnidirectional translation capability of the Mecanum-wheel platform enables motion without rotation, keeping the $TRX_0$ aperture parallel to those of the remote nodes. In the demonstrated static three-link configuration, $TRX_0$ concurrently receives data from $TX_1$ and $TX_2$ through CH9 and CH35, respectively, and transmits data to RX1 through CH19. Given the link distances, quadrature phase-shift keying (QPSK) modulation is used to ensure reliable communication, with each link operating at 2 Gbps, yielding an aggregate concurrent rate of 6 Gbps. As illustrated in Fig. 4k, successful demodulation of all three data streams provides a proof-of-concept demonstration of concurrent SFDMA operation, including simultaneous uplink and downlink communication over independently configured frequency-spatial channels. Owing to limitations in the available test instrumentation, concurrent operation of all 38 channels was not experimentally evaluated.

**Neural-network-assisted AoA estimation for sensing-assisted communication and 2D trajectory reconstruction**

Real-time mutual awareness of spatial orientation is essential for coordinated interaction among embodied intelligent agents. In the proposed SFDMA communication scheme, the carrier frequency must also be selected according to the relative TX-RX direction. The physical basis of the proposed AoA estimation lies in the frequency-to-space mapping of the HLM apertures. Following the link model in Eq. (9), we can see that the angle information is encoded in the frequency-dependent received spectrum. Therefore, we develop a neural-network-assisted AoA estimator to learn these angle-dependent spectral signatures. Fig. 5a shows the sample-acquisition experiment, with details provided in Supplementary Section 4.1. The TX radiates 50 tones from 207.8 to 256.8 GHz, spaced 1 GHz apart. After over-the-air propagation, the THz tones are captured by the RX, downconverted to IF, and recorded by a spectrum analyzer. During sample acquisition, the TX-RX link angle $\theta$ is varied from -45° to +45°, yielding 95 measured spectral samples. The valid AoA estimation range is defined as -40° to +40°, within which 85 spectra are acquired, while the remaining 10 spectra, acquired over the boundary ranges from -45° to -41° and from +41° to +45°, are used as boundary-support samples. At each angular position, the known $\theta$ is assigned as the ground-truth AoA label, and the received powers of the 50 IF tones are extracted for subsequent feature construction. Positive AoA is defined from broadside towards the -$x$ direction (Fig. 5a).

Fig. 5b shows the feature extraction, training, and inference workflow of the prototype-similarity neural regressor. Each measured 50-tone spectrum is converted to linear power and normalized by the total received power to form a probability density function (PDF), thereby suppressing range-dependent power variations while preserving the angle-dependent relative spectral distribution, as derived in Supplementary Eqs. (S4-1)-(S4-10). Its cumulative distribution function (CDF) is then computed, and the PDF and CDF are concatenated into a 100-dimensional feature vector that captures both local spectral energy distribution and cumulative frequency-order information. Details and an example are provided in Supplementary Section 4.2. Owing to the range-normalized feature extraction, this short-range proof-of-concept demonstration could be extended to long-range ISAC with higher TX power and a larger HLM aperture.

For an unknown feature vector $\boldsymbol{x}$, the estimator predicts its AoA through similarity-weighted interpolation among $N_p$ known-angle spectral prototypes:

$$\hat{\theta} = \sum_{k=1}^{N_p} \alpha_k \, \theta_k \tag{10}$$

Here, $\theta_k$ is the known AoA of the $k$-th spectral prototype, and $\alpha_k$ is the normalized weight assigned to that prototype according to its learned spectral similarity to the unknown sample. The construction of $\alpha_k$, feature weighting, cosine-similarity evaluation, and softmax normalization, is provided in Supplementary Section 4.3.

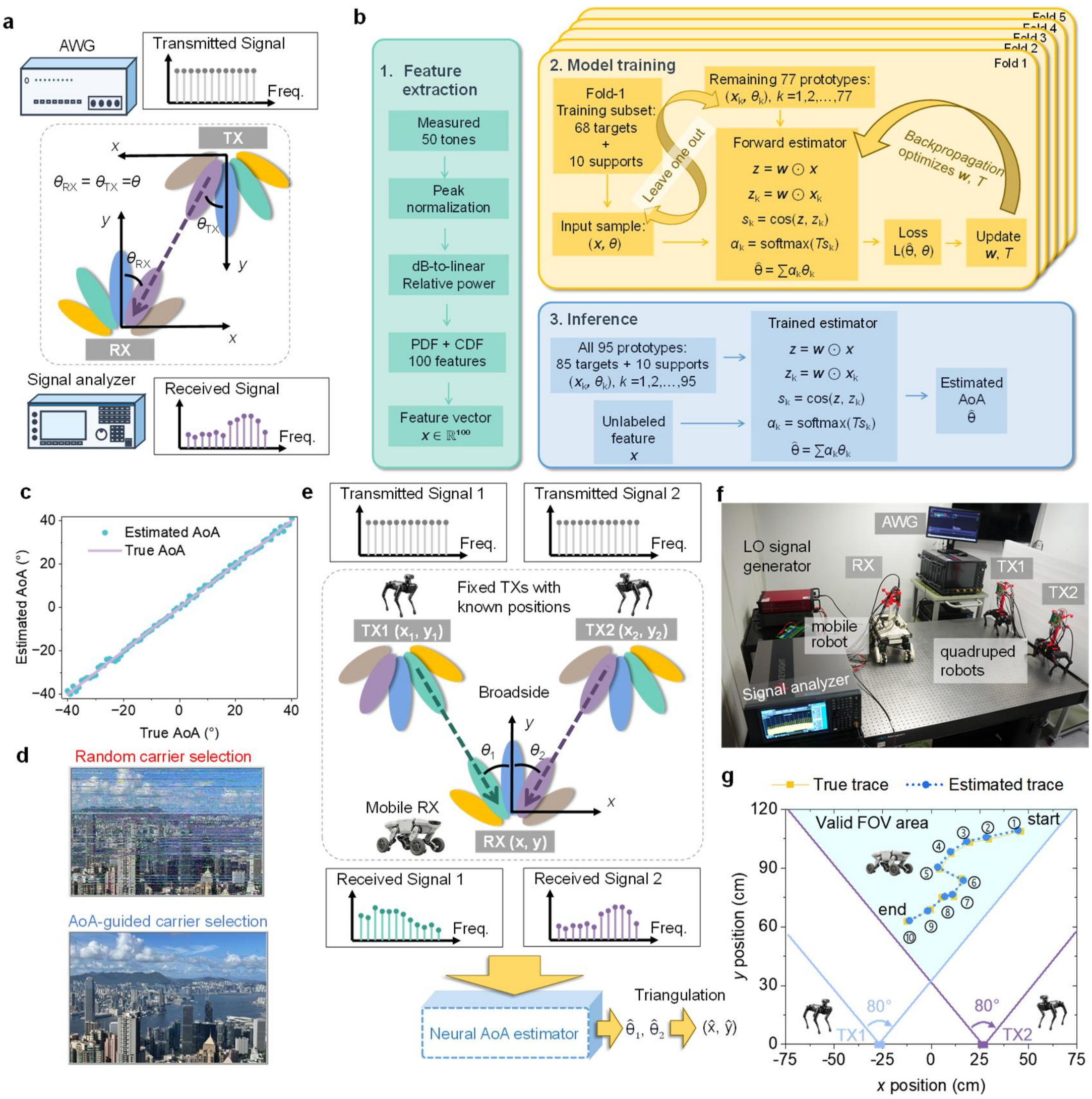

**Fig. 5 | Neural-network-assisted AoA estimation, sensing-enhanced communication, and 2D trajectory reconstruction. a,** AoA sample acquisition experiment: at a specified TX-RX angle $\theta$, with $\theta_{TX} = \theta_{RX} = \theta$, the TX emits equal-amplitude tones, and the frequency-space mapping of the TX/RX apertures produces an angle-dependent received spectrum, which is recorded as the labeled dataset sample. **b,** Workflow of the neural-network-assisted AoA estimator. **c,** Estimated AoA versus true AoA obtained from the five-fold cross-validation results. **d,** Comparison of image transmission using random and sensing-assisted carrier selection. AoA estimation identifies the target direction and selects the corresponding carrier frequency, improving beam alignment and communication quality. **e,** 2D

localization experiment: two fixed TXs transmit multiple tones to a mobile RX. The neural AoA estimator processes the two received spectra to obtain $\theta_1$ and $\theta_2$, which are then used to reconstruct the RX position $(x,y)$ through geometric triangulation. **f,** Photograph of the 2D localization measurement setup. **g,** Reconstructed 2D RX trajectory obtained by AoA-based triangulation compared with true positions.

The AoA estimation accuracy is evaluated using five-fold cross-validation. The out-of-fold estimates closely agree with the ground-truth angles, achieving a mean absolute error of 0.672° (Fig. 5c). The detailed model training procedure and cross-validation are provided in Supplementary Section 4.4. Importantly, Supplementary Fig. S4-5 confirms that the PDF-CDF feature-extraction method outperforms the PDF-only and CDF-only inputs.

Furthermore, the AoA estimate can guide carrier selection for SFDMA communication. As illustrated in Fig. 5d, AoA-guided carrier selection aligns the TX and RX beams through frequency-to-space mapping, substantially improving image transmission compared with random carrier selection.

As illustrated in Figs. 5e and f, the trained AoA estimator is further used for two-dimensional localization through triangulation. In our experiment, two TXs, TX1 and TX2, are mounted on quadruped robots at fixed, known positions separated by 56 cm, while the RX is mounted on a Mecanum-wheel mobile platform and moves along a predefined two-dimensional trajectory. At each trajectory point, two AoA measurements are performed separately using TX1 and TX2. The two estimated AoAs are then converted into the RX position using geometric triangulation, as detailed in Supplementary Section 4.5.

Fig. 5g compares the reconstructed and ground-truth trajectories at 10 measured positions that form an S-shaped path. The overlapping -40° to +40° angular coverage of TX1 and TX2 defines the detectable region, within which the estimated trajectory closely follows the ground truth. The farthest measured position is (46.1 cm, 108.7 cm), and the resulting mean and maximum 2D position errors are 1.49 cm and 1.81 cm, respectively, demonstrating a centimetre-level localization and trajectory reconstruction over a metre-scale range using the proposed THz chip. Supplementary Fig. S4-6 provides the detailed error analysis.

## Conclusion

We have demonstrated a fully integrated CMOS THz ISAC TRX that monolithically combines broadband TX and RX front ends with dedicated HLM apertures. Dispersion engineering of the HLM establishes a frequency-to-space mapping over a 75° angular range, enabling frequency-addressed directional communication and AoA sensing without element-level phase control. The chip supports 38 frequency-spatial channels with an aggregate capacity of 210 Gbps, and demonstrates three concurrent communication links, sub-degree AoA estimation, and a centimetre-scale two-dimensional trajectory reconstruction.

The present implementation constitutes a proof-of-concept demonstration, and its communication and sensing ranges could be further extended by increasing the transmitted power through PA power combining and enhancing the aperture gain using a larger HLM or a package-integrated lens. Furthermore, system-level 360° coverage can be obtained by arranging multiple ISAC chips with complementary orientations around an agent. Future work will extend the HLM from one-dimensional frequency scanning to two-dimensional frequency-to-space mapping, enabling a single integrated aperture to resolve and address nodes over both azimuth and elevation.

## Methods

### CMOS chip fabrication and assembly

The TRX chip was fabricated in a standard bulk 65-nm CMOS process. The die monolithically integrates the TX and RX front ends, the shared LO-generation chain, biasing networks, and two HLM apertures. The CMOS process provides ten metal layers. M1 forms the HLM ground plane and M10 forms its upper conducting surface, while the intermediate metal layers and vias implement the required interconnections and vertical sidewalls.

Each TX or RX aperture comprises four cascaded meta-atoms with a period of 650 μm. The two apertures are arranged in parallel with an edge-to-edge separation of 70 μm. A defected-ground structure is incorporated between them to suppress TX-to-RX coupling. A 500-μm-thick quartz superstrate was aligned with and bonded to the HLM apertures using a UV-curable adhesive. The fabricated die was mounted on a Rogers 5880 printed circuit board and wire-bonded to grounded coplanar-waveguide traces. The complete HLM geometry, CMOS metal stack and assembly procedure are provided in Supplementary Section 1.1.

### Electromagnetic and circuit simulations

The HLM apertures were simulated using CST Studio Suite 2023. The full-wave model included the CMOS metal and dielectric stack, conductor loss, silicon substrate and quartz superstrate. The HLM modelling and simulated characteristics are detailed in Supplementary Sections 1.1 and 1.2.

The transformers, matching networks, transmission lines and circuit interconnects were simulated using Ansys HFSS 2022. Their multiport electromagnetic responses were imported into Keysight ADS 2024 for electromagnetic–circuit co-simulation. Transistor-level and post-layout simulations were performed using Keysight ADS 2024 and Cadence Spectre 6.1.8 with the foundry-supplied device models. The electromagnetically simulated passive networks and extracted interconnect parasitics were included in the final simulations.

Small-signal, noise and large-signal analyses were performed for the PA, LNA, mixers and 4× LO chain. The PA stages were biased at $V_D = 1.2$ V and $V_G = 0.7$ V, while the LNA stages used $V_D = 1.2$ V and $V_G = 0.6$ V. The TX- and RX-mode mixer gate voltages were 0.3 V and 0.5 V, respectively. Detailed circuit implementations and simulation procedures are provided in Supplementary Sections 2.1–2.5.

### Standalone PA characterization

The standalone seven-stage PA was characterized on wafer. Small-signal S-parameters were measured using a Keysight N5227B network analyzer, an OML WR3.4 VNA extender and GGB WR3.4 probes with a 50-μm pitch. The measurement reference planes were established at the probe tips, and the losses of the external cables and waveguide transitions were removed.

Large-signal measurements were performed using a VDI WR3.4 up-conversion module as the input source and an Erickson PM5B power meter to measure the PA output power. The input- and output-path losses were independently calibrated. The output 1-dB compression point was determined relative to the extrapolated small-signal response, and the saturated output power was obtained from the maximum measured output level. The stability quantities $K_f$ and $|\Delta|$ were calculated from the measured complex S-parameters. The PA architecture, stability definitions and measured results are provided in Supplementary Section 2.2.

**Over-the-air TRX characterization**

The over-the-air measurements were performed at room temperature. Before characterization, the losses of the IF and reference-LO cables, end-launch connectors and printed-circuit-board transmission lines were measured and de-embedded. The standard horn antenna had a nominal gain of 22 dBi, which was verified using a two-antenna calibration.

A VDI WR3.4 up-conversion module was used as the calibrated THz source for the RX measurements, while a VDI WR3.4 down-conversion module was used to receive and downconvert the radiated TX signal. An Erickson PM5B power meter established the absolute THz power reference. A Keysight M8199A arbitrary waveform generator (AWG), installed in an M9505A AXIe chassis and controlled using M8070B system software, generated the IF signals. A Keysight AP5021A analogue signal generator supplied the reference-LO signal.

For TX characterization, the radiated signal was collected by the 22-dBi horn and downconverted by the VDI WR3.4 down-conversion module. For RX characterization, the calibrated VDI WR3.4 up-conversion module and horn illuminated the RX HLM from the far field. Radiation-pattern measurements were performed by changing the angular position of the WR3.4 module while keeping the chip fixed.

The TX EIRP, TX conversion gain and RX conversion gain were extracted from the calibrated power measurements, while the free-space path loss was calculated from the link distance and carrier frequency. Their definitions, calibration reference planes and calculation equations are provided in Supplementary Section 2.6.

The RX noise figure was measured using the output-noise method. With the RF source disabled, an AT Microwave AT-BB-0050-3820 broadband IF amplifier was inserted before the PXA N9030B to raise the measured noise above the analyzer noise floor. The displayed noise power was normalized to a 1-Hz bandwidth, and the independently characterized frequency-dependent gain of the external IF amplifier was de-embedded. Because the receiver output noise was substantially higher than the input-referred added noise of the external amplifier, the latter contribution was negligible. The complete noise-figure extraction method is provided in Supplementary Section 2.6.

**One-to-one communication measurements and per-channel data-rate characterization across 38 spatial–frequency channels**

The modulated IF waveforms were generated using the M8199A AWG. Pulse shaping was applied with a roll-off factor of 0.1. One chip operated as the fixed TX, while a second chip operated as the RX and was mounted on a mobile robotic platform. The received IF waveform was captured using a Keysight UXR0404AP real-time oscilloscope.

Waveform generation was controlled using M8070B system software. Synchronization, equalization, demodulation and error-vector-magnitude evaluation were performed using the software supplied with the Keysight measurement instruments. An equalizer was enabled before constellation and error-vector-magnitude evaluation. No additional third-party waveform-generation or demodulation software was used.

A 1-GBaud BPSK link and a 5-GBaud 32-QAM link were used to evaluate long-distance and high-data-rate operation, respectively. The complete communication setup, carrier frequencies, link distances and representative modulation results are provided in Supplementary Section 3.1, while the adopted EVM and SNR criteria are described in Supplementary Section 3.2.

The link budgets were evaluated in linear power units using the measured circuit and HLM responses. Thermal noise was calculated at ($T_0$=290) K using noise-equivalent bandwidths corresponding to the transmitted symbol rates. The adopted SNR requirements were 5.16 dB for BPSK and 19.58 dB for 32-QAM. Coding and protocol overheads were excluded from the quoted raw data rates. The link-budget equations and stage-by-stage parameters are provided in Supplementary Section 3.2.

**One-to-one multimedia transmission**

A USRP-2944 generated the QPSK-modulated image and video streams. An external IF frequency-conversion stage and an AT Microwave AT-BB-0050-3820 amplifier translated and amplified the waveform before it was applied to the chip. The experimental arrangement and multimedia-transmission procedure are provided in Supplementary Section 3.3.

**Simultaneous TX and RX operation**

To evaluate simultaneous on-chip TX–RX operation, an external chip wirelessly transmitted the desired signal to the RX, while the TX on the same chip simultaneously operated in an adjacent frequency–spatial channel. The desired RX waveform was captured using the Keysight UXR0404AP and demodulated using the instrument-supplied software. Receiver degradation was quantified from the change in demodulated SNR as the TX IF amplitude was increased.

The DGS and the associated coupling paths were analyzed using CST Studio Suite 2023. The receiver blocking margin was calculated from the CST-simulated TX-to-RX coupling and the PA $P_{sat}$ and LNA $IP_{1dB}$ simulated using Cadence Spectre 6.1.8. The DGS geometry, coupling-path analysis, simulated isolation, experimental arrangement, selected frequency channels and TX-amplitude sweep are described in Supplementary Section 1.3.

**One-to-three communication measurements**

The one-to-three SFDMA experiment employed four TRX chips. The central node, $TRX_0$, was mounted on a Mecanum-wheel mobile robot, while two uplink transmitters ($TX_1$ and $TX_2$) and one downlink receiver ($RX_1$) were positioned at different angular locations, 42–45 cm from $TRX_0$. Three spatial–frequency channels were activated concurrently: $TX_1$ and $TX_2$ transmitted to $TRX_0$ through CH9 and CH35, respectively, while $TRX_0$ simultaneously transmitted to $RX_1$ through CH19. Each link carried a 2-Gbps QPSK waveform, corresponding to an aggregate concurrent data rate of 6 Gbps.

The three modulated IF waveforms were generated using separate output channels of the M8199A AWG. At the RX side, the composite IF output of $TRX_0$, containing the concurrently received CH9 and CH35 signals, was divided into two identical measurement paths using an AT Microwave AT-PD2-0465 two-way power divider. The two outputs of the power divider were connected to separate input channels of a Keysight UXR0404AP real-time oscilloscope, allowing the received CH9 and CH35 waveforms to be independently synchronized, equalized and demodulated using the instrument-supplied software. The IF output of $RX_1$, corresponding to the CH19 downlink, was connected to a third oscilloscope channel and processed in the same manner. The measured constellations and spectra of the three concurrent links are presented in Fig. 4k.

**AoA spectral-data acquisition**

The AoA dataset was acquired using one TX chip and one RX chip mounted on two independently controlled rotation stages, with the TX–RX separation fixed at 45 cm. The sensing-tone waveform was generated using the M8199A AWG. Two Keysight AP5021A analogue signal generators supplied the quarter-rate reference LO signals to the TX and RX chips, and the downconverted IF tones were recorded using a Keysight PXA N9030B signal analyser.

At each angular position, the experiment produced a 50-dimensional received-power spectrum comprising 50 tone-power values across the operating band. The complete tone allocation, two-segment frequency-acquisition procedure, angle definition, rotation method and dataset composition are provided in Supplementary Section 4.1.

**Spectral-feature construction**

The AoA data processing was implemented using Python 3.14. The measured tone powers were first converted from decibel units to linear units and normalized by their total power. The cumulative distribution of the normalized spectrum was then calculated, and the normalized and cumulative spectral distributions were concatenated to form a 100-dimensional feature vector.

This representation suppresses the common power scaling caused primarily by propagation distance while retaining the frequency-dependent spectral signature produced by the HLM. The mathematical derivation and complete feature definitions are provided in Supplementary Section 4.2.

**Prototype-similarity AoA estimation**

The AoA estimator used measured spectra with known angles as prototypes. A positive learnable feature-weight vector was applied before calculating the cosine similarities between an unknown spectrum and the prototype spectra. A positive learnable similarity scale was then used in a softmax operation. The AoA was estimated by interpolating the prototype angles using the resulting normalized similarity weights. The complete model equations are provided in Supplementary Section 4.3.

The AoA estimator was implemented in CPython 3.14.0 using NumPy 2.5.1. Model optimization used a deterministic, full-batch NumPy implementation of Adam with decoupled weight decay, equivalent to an AdamW update. The learning rate was 0.02, with beta1 = 0.9, beta2 = 0.999, epsilon = $1 \times 10^{-8}$, and a weight-decay coefficient of $1 \times 10^{-4}$. The feature-weight parameters q were initialized to zero and converted into positive weights as w = softplus(q) + $1 \times 10^{-4}$. The positive similarity-scale parameter T was initialized to 35 and constrained to the range from 1 to 250. Training minimized a mean Smooth-L1 loss with beta = 0.5 together with a regularization term of $1 \times 10^{-4}$ times the mean squared value of q. Each model was optimized for 2,500 epochs, and the checkpoint with the minimum recorded leave-one-out training objective was retained. The seed argument was set to 42, although the implementation contained no stochastic training operations.

Five-fold cross-validation was applied to the target-range dataset. Boundary-support samples were included only in model training and were excluded from the validation statistics. PDF-only and CDF-only models were trained using the same protocol for the feature-ablation analysis. The dataset division, training procedure and validation metrics are detailed in Supplementary Section 4.4.

**Two-dimensional localization**

Two fixed TXs separated by 56 cm were used to estimate the position of a mobile receiver. At each trajectory point, the two TXs were activated separately, and the trained model estimated the corresponding bearing angles from the independently measured spectra. The RX coordinates were then obtained by intersecting the two angle-defined lines of bearing. The M8199A AWG generated the sensing-tone waveforms for the two TXs, and the downconverted IF tones of the RX were recorded using a Keysight PXA N9030B signal analyzer.

All spectra used in the trajectory experiment were acquired independently and were excluded from the training, cross-validation and prototype datasets. The triangulation equations, ground-truth trajectory and localization-error definition are provided in Supplementary Section 4.5.

**Data Availability Statement**

The data that support the findings of this study are available from the corresponding authors upon reasonable request.

**Code availability**

The code used to assist in implementing the AoA neural network is available from the corresponding authors upon reasonable request.

**Acknowledgments**

This research was supported in part by the Research Grants Council of Hong Kong under the Collaborative Research Fund (grant C1009-22G) and the General Research Fund (grants CityU 11214123 and CityU 11203225).

**Author Contributions**

C.H.C. initiated the plan, supervised the entire study, and led the project. X.X. and Z.L. conceived the idea of this work. Z.L. proposed the circuit scheme. X.X., Z.L., and H.G. designed the THz chip and performed the theoretical analysis. Z.L., X.X., and G.B.W. designed the experiments. K.F.C. and K.M.S. assisted in preparing the experimental setup. X.X. and Z.L. conducted the experiments and performed the data analysis. H.G., S.W., J.Z., and X.F. assisted with the experiments. X.X. and Z.L. wrote the manuscript. G.B.W. and C.H.C. provided critical feedback and revised the manuscript. All authors discussed the results and reviewed the manuscript.

**Conflict of Interest**

The authors declare no conflict of interest.

**Correspondence and requests for materials should be addressed to**

Zhicheng Lin, Geng-Bo Wu, or Chi Hou Chan

Supporting Information for

# A Frequency–Space Terahertz Transceiver Chip for Multi-Agent Communications and Spatial Awareness

# Catalogue

# 1. CMOS heterogeneous leaky-wave metasurfaces

## 1.1 Design of the HLM

The proposed terahertz (THz) integrated sensing and communication (ISAC) chip is implemented in a bulk 65-nm Complementary metal-oxide-semiconductor (CMOS) process. As shown in Supplementary Fig. S1-1, the process provides ten metal layers, from M1 to M10, interconnected by vias. For the heterogeneous leaky-wave metasurface (HLM) implementation, M1 serves as the ground plane, and M10 forms the upper conducting surface, while the intermediate metal layers and vias provide the required interconnections and vertical sidewalls.

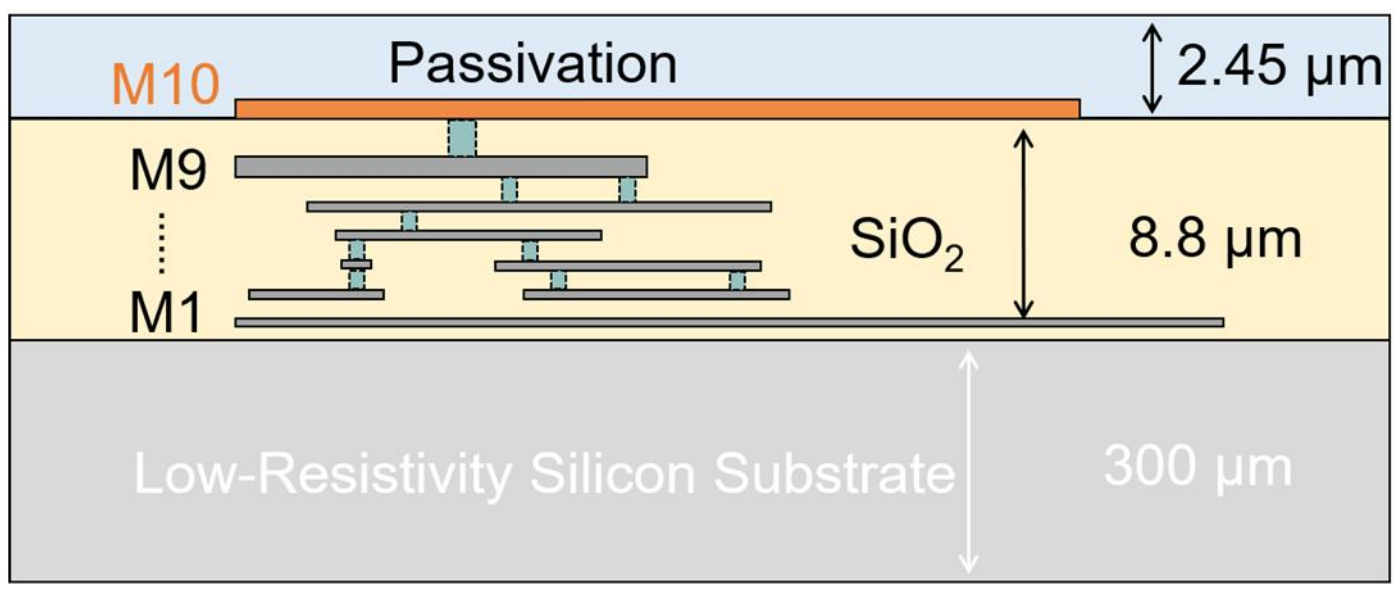

**Supplementary Fig. S1-1 | Cross-sectional stack of the bulk 65-nm CMOS process.**

The detailed geometry and dimensions of the paired transmitter (TX) and receiver (RX) HLMs are shown in Supplementary Fig. S1-2. Each transmit or receive aperture comprises four cascaded meta-atoms along the periodic direction. The two apertures are placed in parallel with an edge-to-edge separation of only 70 μm, enabling compact integration while imposing a stringent requirement on TX–RX isolation, as detailed in Supplementary Section 1.3. Each rotationally symmetric meta-atom combines chip-integrated-waveguide (CIW), parallel-plate-waveguide (PPW), and S-shaped microstrip line (MSL) sections with a period of $p = 650$ μm. These heterogeneous wave-guided sections jointly engineer the phase accumulation and distributed coupling to free space.

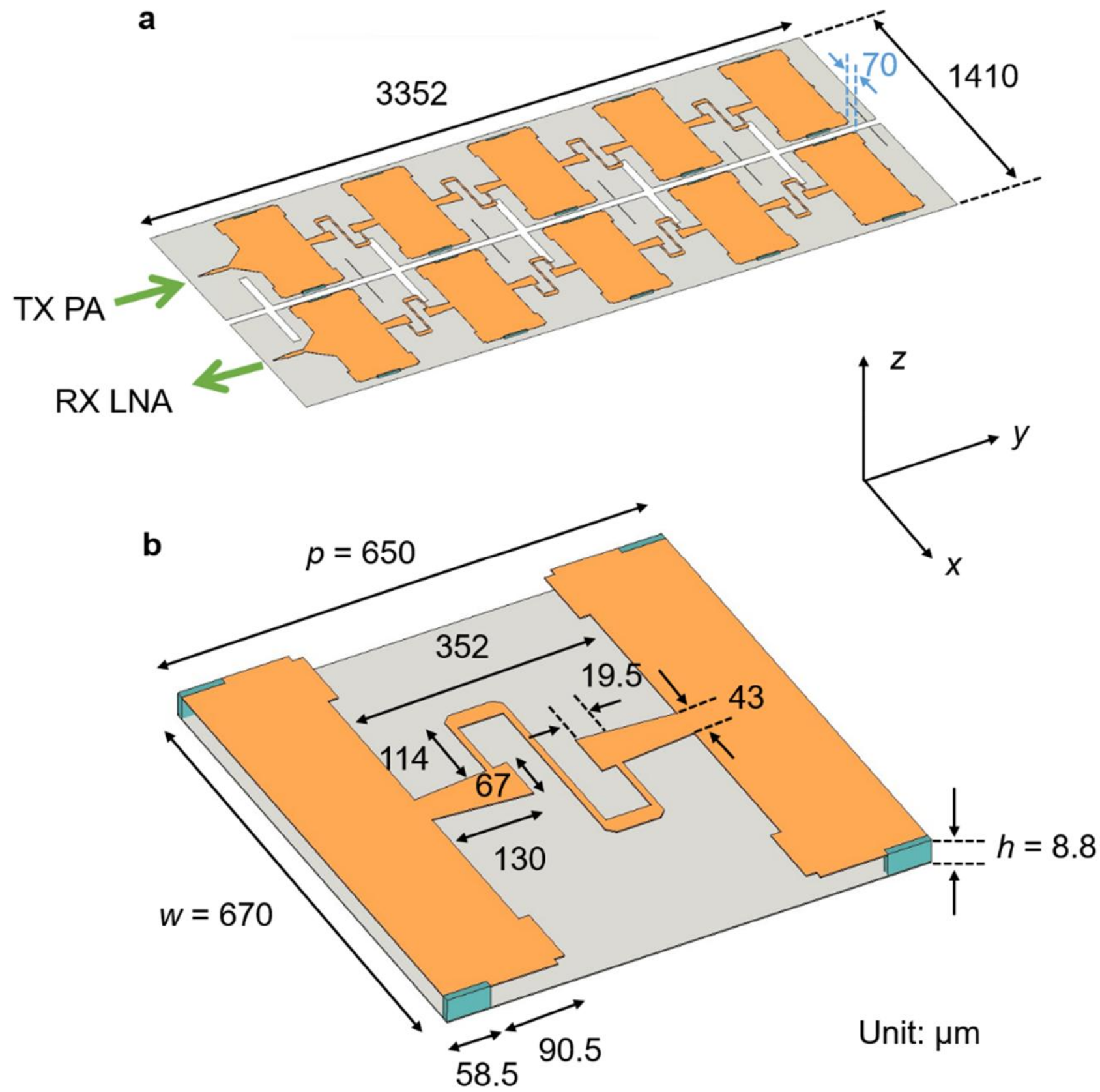

**Supplementary Fig. S1-2 | Geometry and dimensions of the HLM. a,** Parallel arrangement of the four-meta-atom TX and RX apertures. **b,** Enlarged view of the rotationally symmetric meta-atom.

To enhance the radiation efficiency and provide mechanical rigidity, a 500-µm-thick quartz superstrate was bonded over each TX and RX HLM aperture using a ultraviolet-curable adhesive [S1], as shown in Supplementary Fig. S1-3.

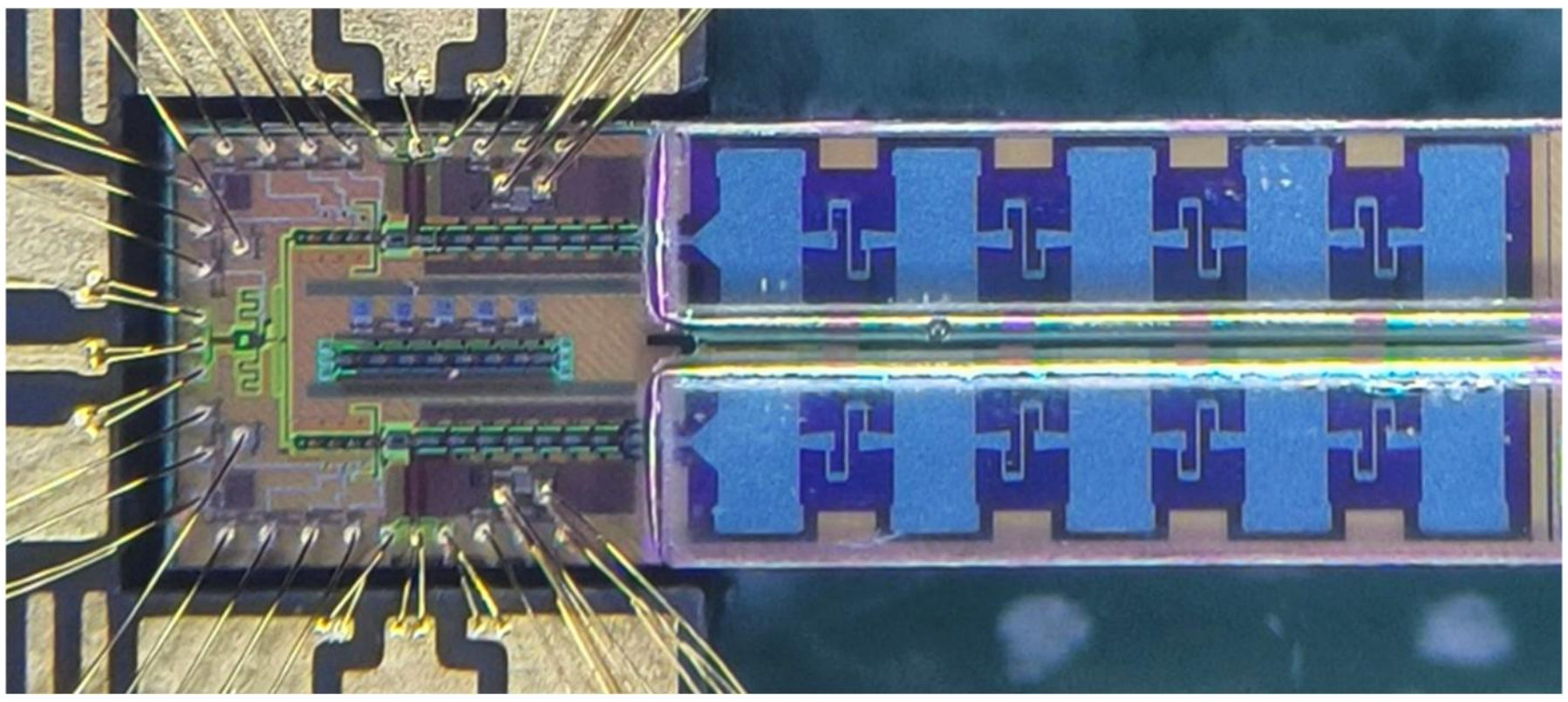

**Supplementary Fig. S1-3 | Micrograph of the chip, with quartz placed upon the HLM apertures and the chip wire-bonded to the printed circuit board (PCB).**

### 1.2 Characteristics of the HLM

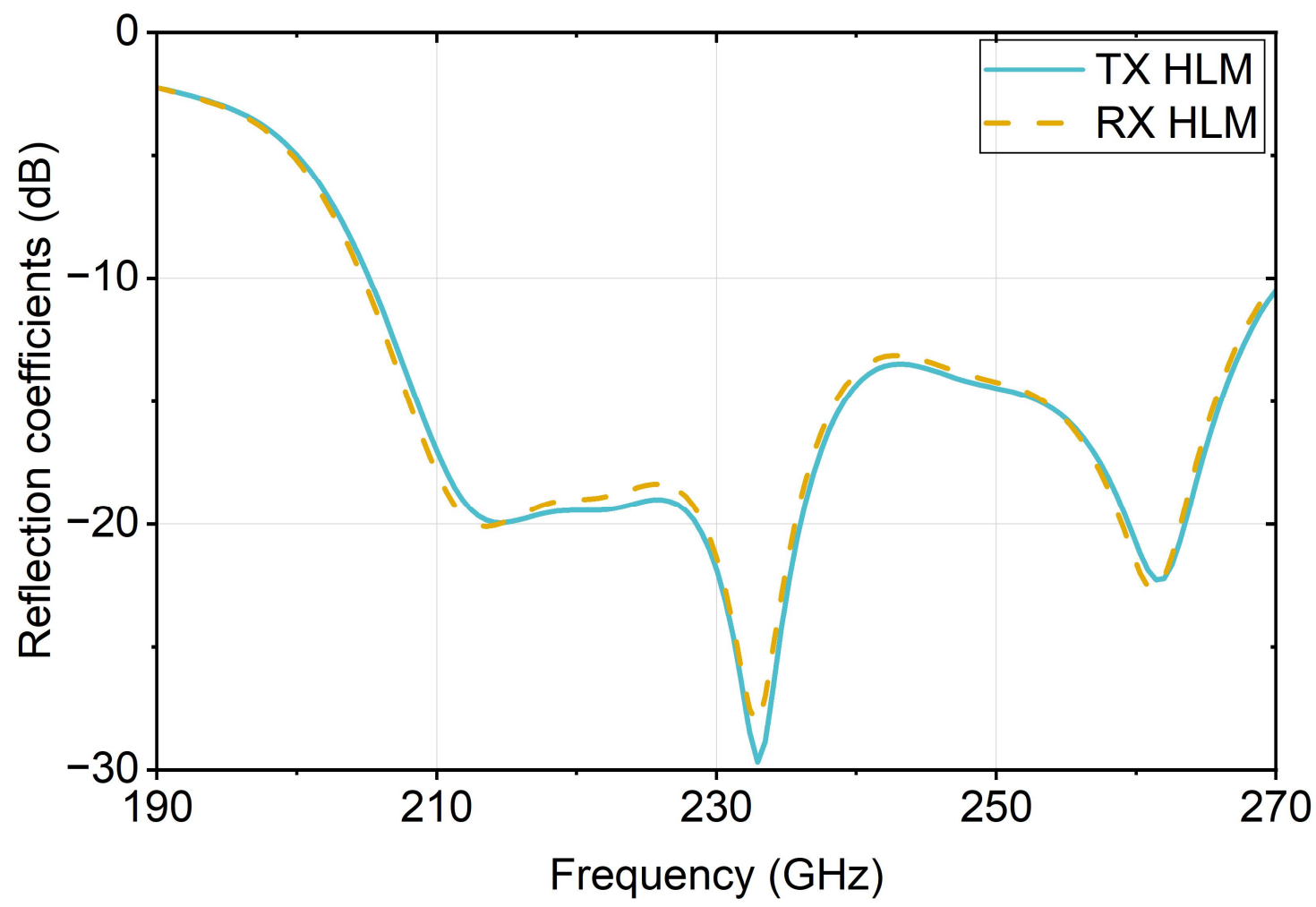

**Supplementary Fig. S1-4 | Simulated reflection coefficients of the TX HLM and the RX HLM.**

Full-wave simulations of the on-chip HLMs are conducted to evaluate their electromagnetic performance. The simulated reflection coefficients of the TX and RX HLMs are shown in Supplementary Fig. S1-4. Both apertures exhibit similar broadband impedance characteristics, with reflection coefficients remaining below -10 dB from 205 to 270 GHz. Within the chip's operating bands from 208 to 258 GHz, the corresponding reflection coefficients are below -13 dB, indicating good broadband matching between the power amplifier (PA)/low-noise amplifier (LNA) and the HLM apertures.

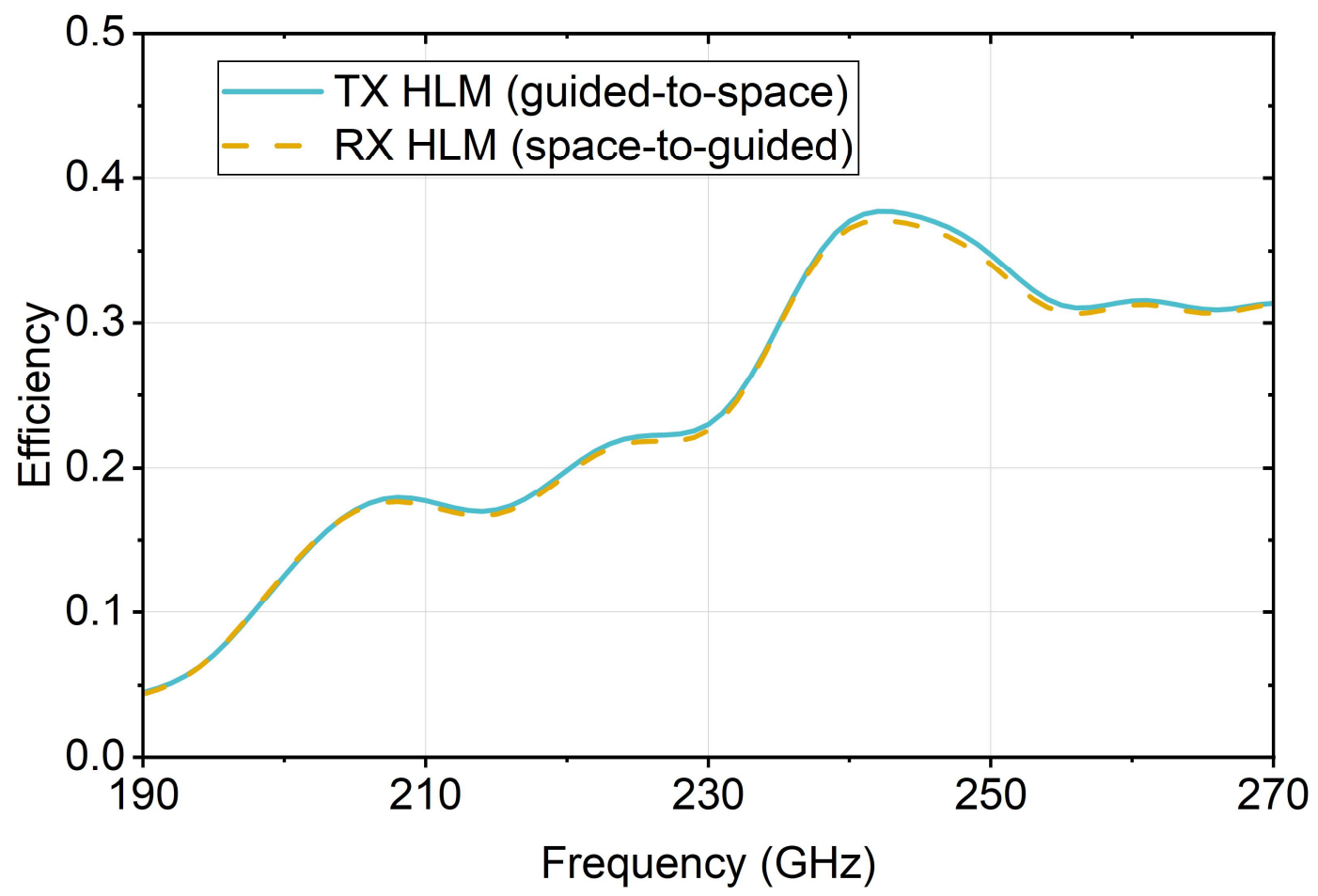

**Supplementary Fig. S1-5 | Simulated guided-to-space efficiency of the TX HLM and the space-to-guided efficiency of the RX HLM.**

Supplementary Fig. S1-5 further presents the guided-to-space and space-to-guided conversion efficiencies of the TX and RX HLMs, respectively. Owing to the reciprocal aperture design, both responses exhibit nearly identical frequency dependence. Across the 208–258-GHz operating band, the conversion efficiency ranges from 18% to 37%, demonstrating efficient coupling between guided waves and free-space radiation. Furthermore, the corresponding realized gains are shown in Supplementary Fig. S1-6. Both the TX and RX HLMs achieve broadband gain responses, with the peak realized gain reaching approximately 8.2 dBi around 240 GHz.

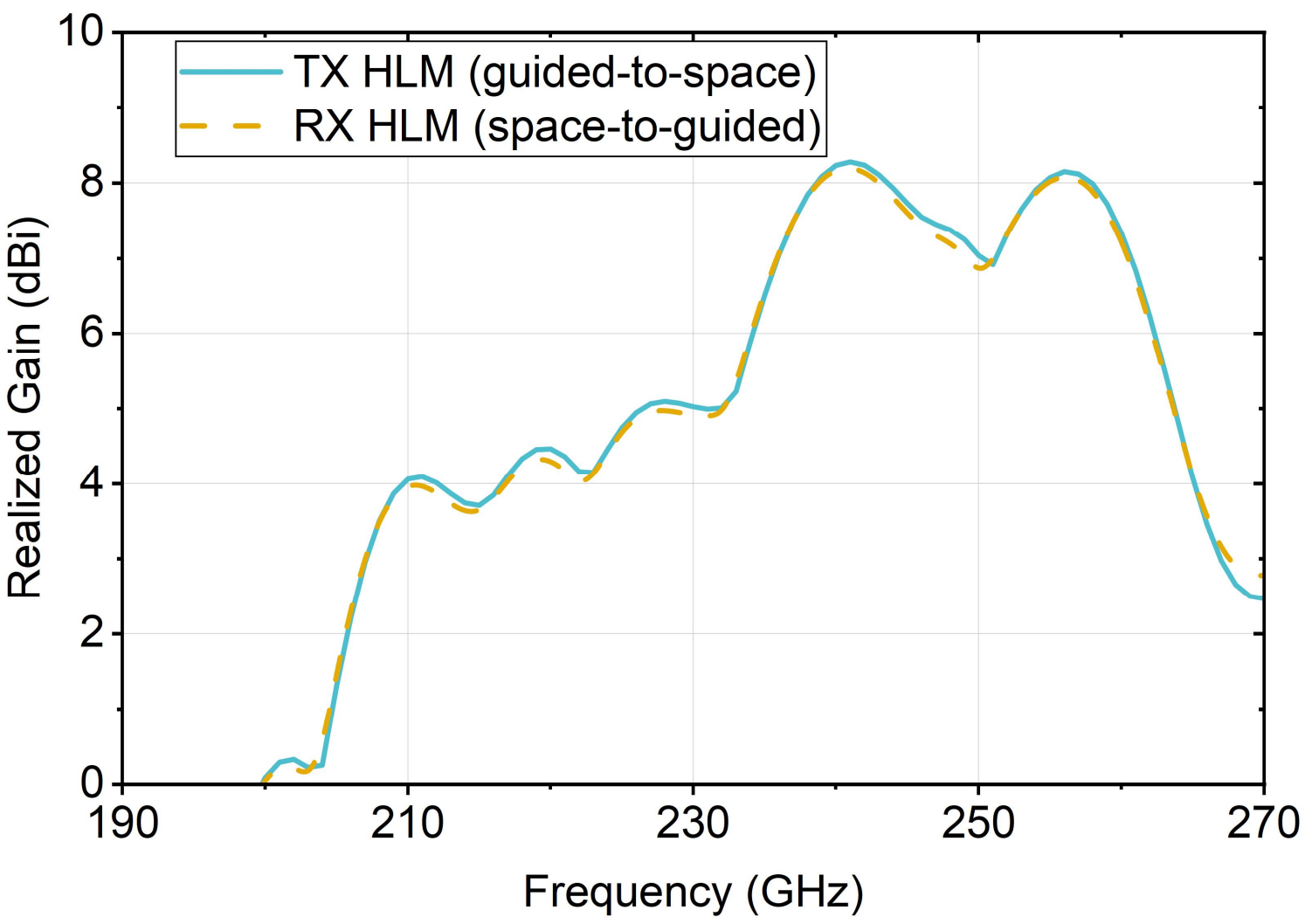

**Supplementary Fig. S1-6 | Simulated guided-to-space gain of the TX HLM and the space-to-guided gain of the RX HLM.**

### 1.3 TX-to-RX decoupling and simultaneous TX/RX operation experiments

To enable simultaneous transmit-and-receive operation over different spatial-frequency channels, high isolation between the transmitting and receiving metasurface apertures is required. Although the proposed system does not operate as an in-band full-duplex transceiver (TRX), TX leakage can still be problematic because the TX and RX HLM apertures are placed in close proximity on the same chip. In this configuration, intrinsic coupling can occur through multiple paths, including surface-wave coupling $K_{SW}(f)$ and near-field coupling $K_{NF}(f)$. These coupling terms are treated as complex-valued frequency responses, which contain both amplitude and phase information. When the TX delivers high output power, part of the transmitted energy may couple into the receiving metasurface aperture and be delivered to the LNA, potentially causing gain compression, desensitization, or receiver blocking.

To suppress this self-interference, a defected-ground-structure (DGS)-assisted decoupling technique is employed. As shown in Supplementary Fig. S1-7, the DGS consists of a continuous longitudinal ground slot and periodically loaded transverse slots. The longitudinal slot also behaves as a slotline-like channel along the aperture direction, enabling a controlled artificial coupling path between selected TX and RX aperture units. The periodically loaded transverse slots provide local current perturbation, phase tuning, and controllable coupling between the HLM units and the longitudinal slot.

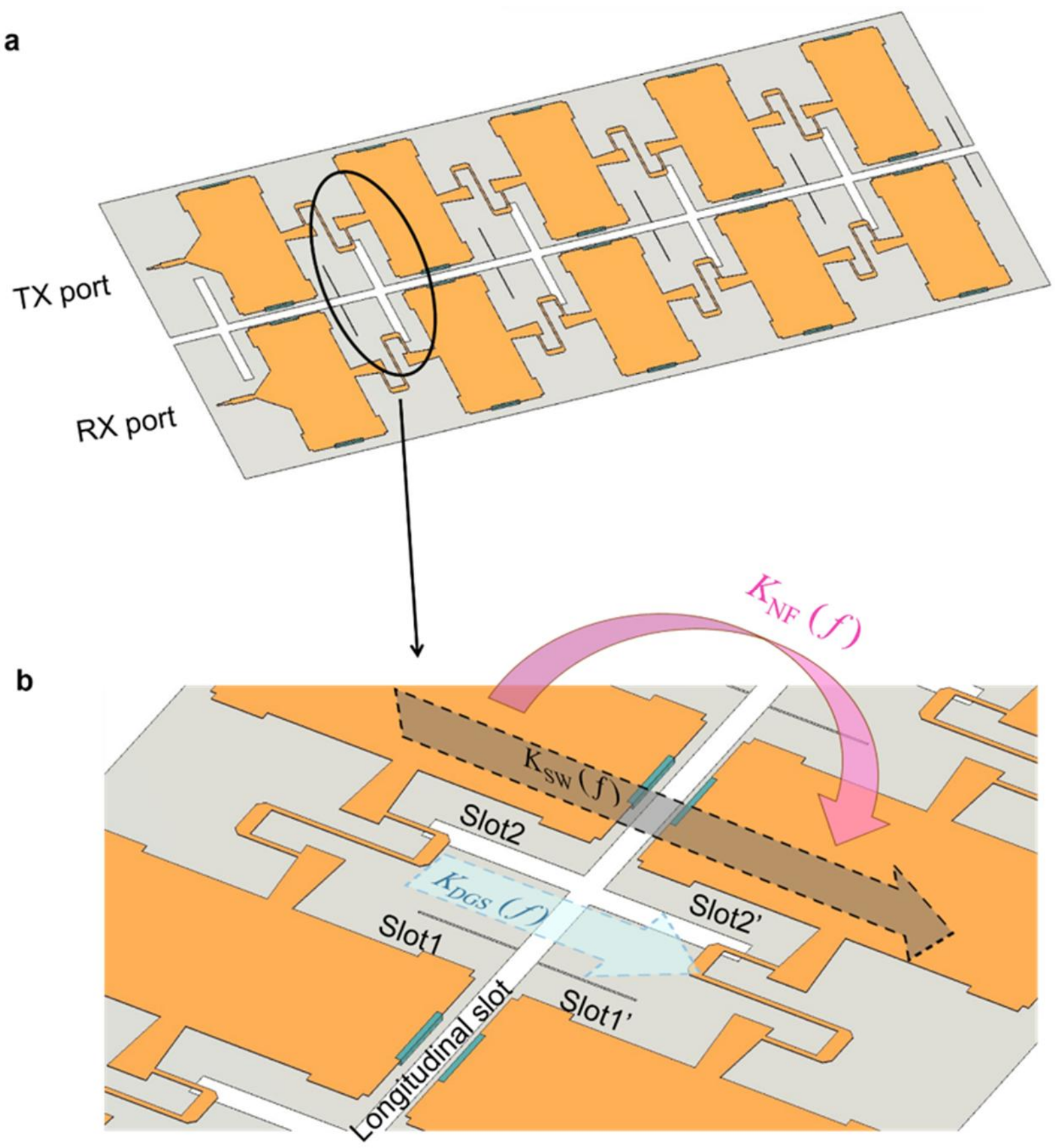

**Supplementary Fig. S1-7 | DGS-based decoupling structure between the transmitting and receiving metasurface apertures. a,** Layout of the two closely spaced HLM apertures with the defected ground structure (DGS). **b,** Close-up view of the DGS, showing the longitudinal ground cut and periodic transverse slots used to form the engineered coupling path.

As shown in the close-up view in Supplementary Fig. S1-7b, Slot1 and Slot2 are placed on the TX side, while Slot1′ and Slot2′ are placed on the RX side. Slot2 and Slot2′ act as the primary artificial coupling ports. They extract a controlled portion of the guided energy near the bent meandered microstrip section of the TX HLM unit and couple it through the longitudinal slot into the corresponding RX HLM unit. In contrast, Slot1 and Slot1′ mainly serve as tuning slots, which reshape the local ground-current return path and fine-adjust the phase response of the artificial coupling path.

With the DGS introduced, the equivalent TX-to-RX coupling becomes the coherent sum of the residual intrinsic coupling and the DGS-assisted artificial coupling path:

$$K_{\mathrm{eff}}(f) = K_{\mathrm{SW}}(f) + K_{\mathrm{NF}}(f) + K_{\mathrm{DGS}}(f) \tag{S1-1}$$

where $K_{\mathrm{DGS}}(f)$ denotes the artificial slot-mediated coupling introduced by the DGS. By engineering the aperture spacing and offset, as well as the length, width, and position of the transverse slots, the amplitude and phase of $K_{\mathrm{DGS}}(f)$ can be tuned to interfere with the intrinsic surface-wave and near-field coupling paths destructively. Ideally, the cancellation condition is

$$|K_{\mathrm{DGS}}(f)| \approx |K_{\mathrm{SW}}(f) + K_{\mathrm{NF}}(f)|, \tag{S1-2}$$

$$\angle K_{\mathrm{DGS}}(f) \approx \angle[K_{\mathrm{SW}}(f) + K_{\mathrm{NF}}(f)] + 180^{\circ} \tag{S1-3}$$

Under this condition, the two coupling contributions cancel at the RX aperture, and the magnitude of the total coupling approaches a minimum:

$$|K_{\mathrm{eff}}(f)| \rightarrow \text{minimum} \tag{S1-4}$$

Supplementary Fig. S1-8 compares the simulated isolation with and without the DGS. Without the DGS, the isolation strongly fluctuates over the operating band, with several pronounced resonant features. Although high isolation is observed at a few narrow frequencies, the isolation drops to around 27.5-35 dB over a large portion of the band, indicating significant intrinsic coupling between the TX and RX apertures. After introducing the DGS, the isolation is substantially improved and becomes more stable across the operating band. In particular, the isolation remains above approximately 35.5 dB over 205-265 GHz and exceeds 49 dB over a broad frequency range from around 227 to 258 GHz. This improvement indicates that the DGS effectively suppresses the intrinsic TX-to-RX coupling.

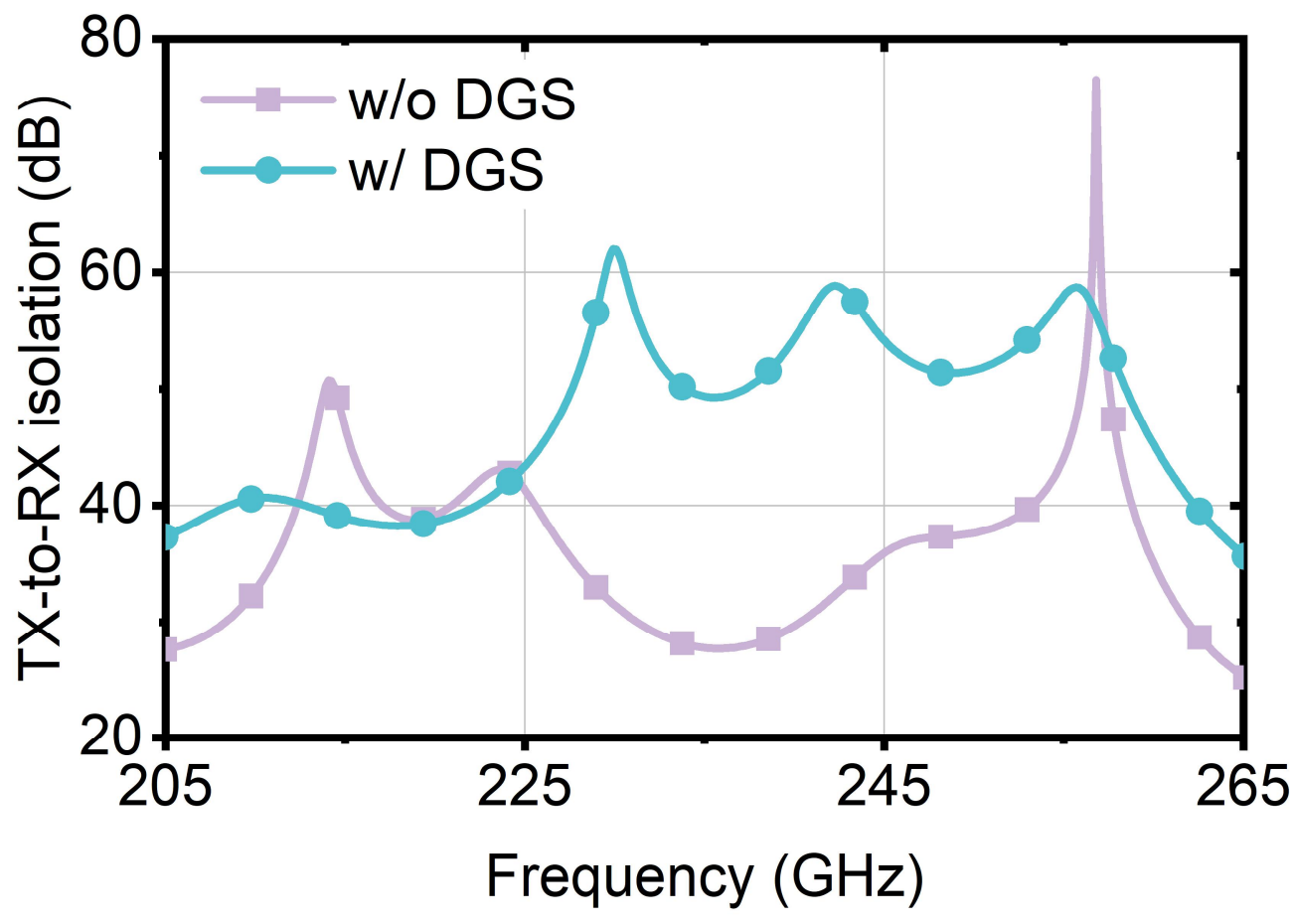

**Supplementary Fig. S1-8 | Simulated TX-to-RX isolation with and without DGS.**

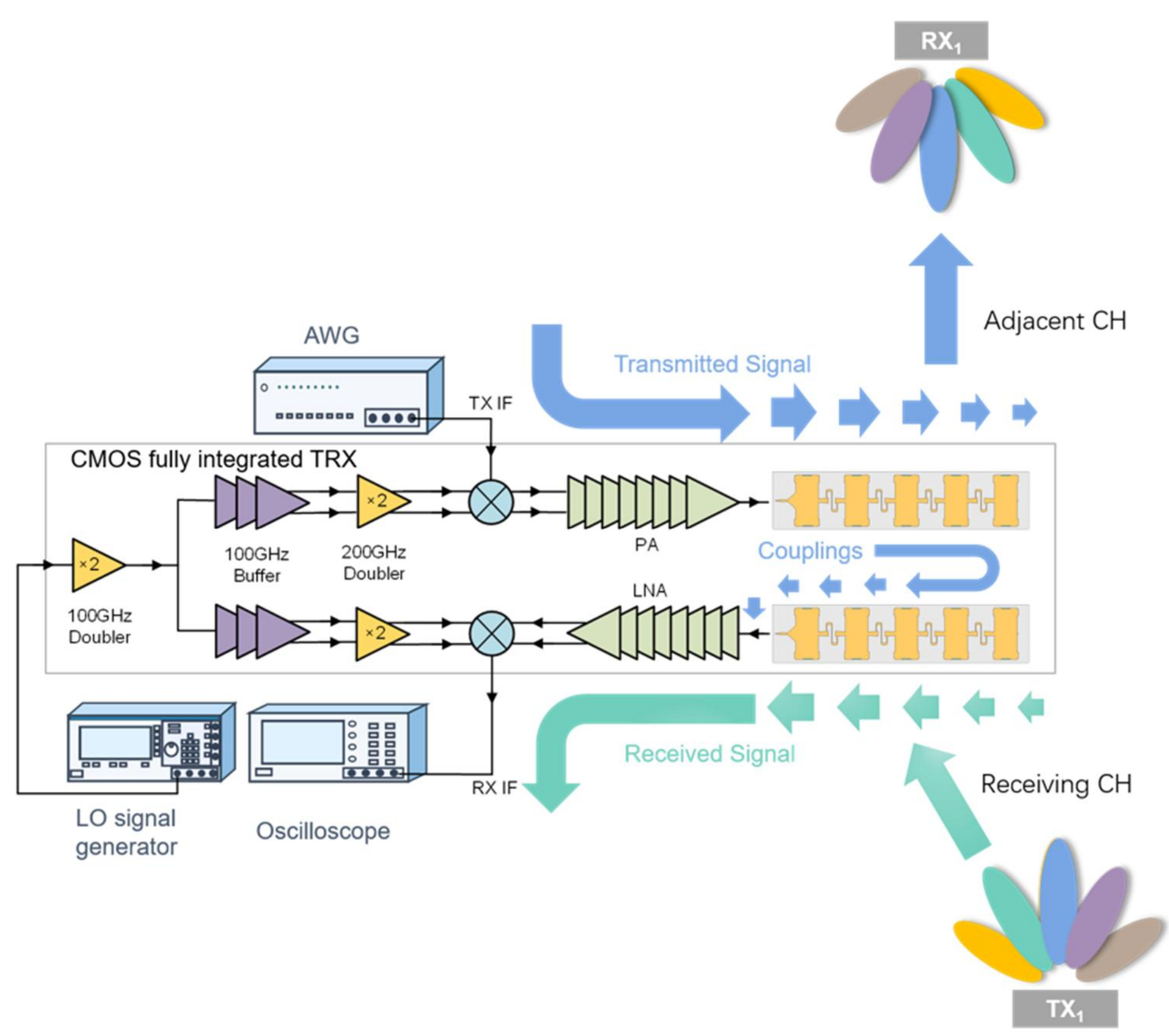

**Supplementary Fig. S1-9 | Measurement setup for TX self-interference evaluation during simultaneous TX/RX operation.**

Finally, since the two HLM apertures are monolithically integrated on the same chip, their mutual coupling cannot be directly measured at the aperture ports. Therefore, the self-interference tolerance is evaluated at the system level by monitoring the demodulated signal-to-noise ratio (SNR) of a desired receiving channel while the on-chip TX chain is simultaneously excited at adjacent frequency channels.

Supplementary Fig. S1-9 illustrates the measurement setup for the simultaneous TX/RX evaluation. The desired receiving signal is first generated by an external transmitter TX1, and radiated toward the receiving HLM aperture over a communication distance of 25 cm. After being collected by the receiving aperture, the signal is amplified by the LNA, downconverted to the RX intermediate frequency (IF), captured by the oscilloscope, and digitally demodulated to extract the SNR of the target receiving channel.

Meanwhile, the on-chip TX chain is driven by an independent IF signal at frequency channels adjacent to the desired receiving channel. This TX signal is upconverted using the shared local oscillator (LO), amplified by the PA, and delivered to the transmitting HLM aperture, where it is radiated toward RX1. The TX excitation is deliberately placed in neighboring channels rather than in the same channel as the desired received signal. Therefore, this experiment evaluates not only the receiver tolerance to TX-to-RX self-interference through the on-chip HLM coupling path, but also the effectiveness of frequency-domain isolation between adjacent channels. Any residual adjacent-channel interference, arising from TX spectral leakage or finite RX channel selectivity, would be directly folded into the target receiving channel and would severely degrade the demodulated SNR.

Two representative receiving channels are selected, and the measured SNR of the demodulated signal is illustrated in Supplementary Fig. S1-10. First, CH19 is used as the receiving channel, while TX self-interference is separately applied at the adjacent channels CH18 and CH20. As the TX IF input amplitude is swept from 0 to 800 mV, approaching the PA input P1dB region near 700 mV, the measured SNR varies by only 0.9 dB and 0.85 dB for CH18 and CH20 self-interference, respectively. Second, CH9 is selected as a more stringent case because it corresponds to the smallest blocking margin observed in Fig. 2i. In this case, TX self-interference is separately applied at the adjacent channels CH8 and CH10. Even under this worst-case condition, the measured SNR varies by only 0.8 dB and 1.2 dB, respectively. These results confirm the feasibility of simultaneous TX/RX operation and the effectiveness of inter-channel isolation in the proposed spatial-frequency division multiple access (SFDMA) scheme.

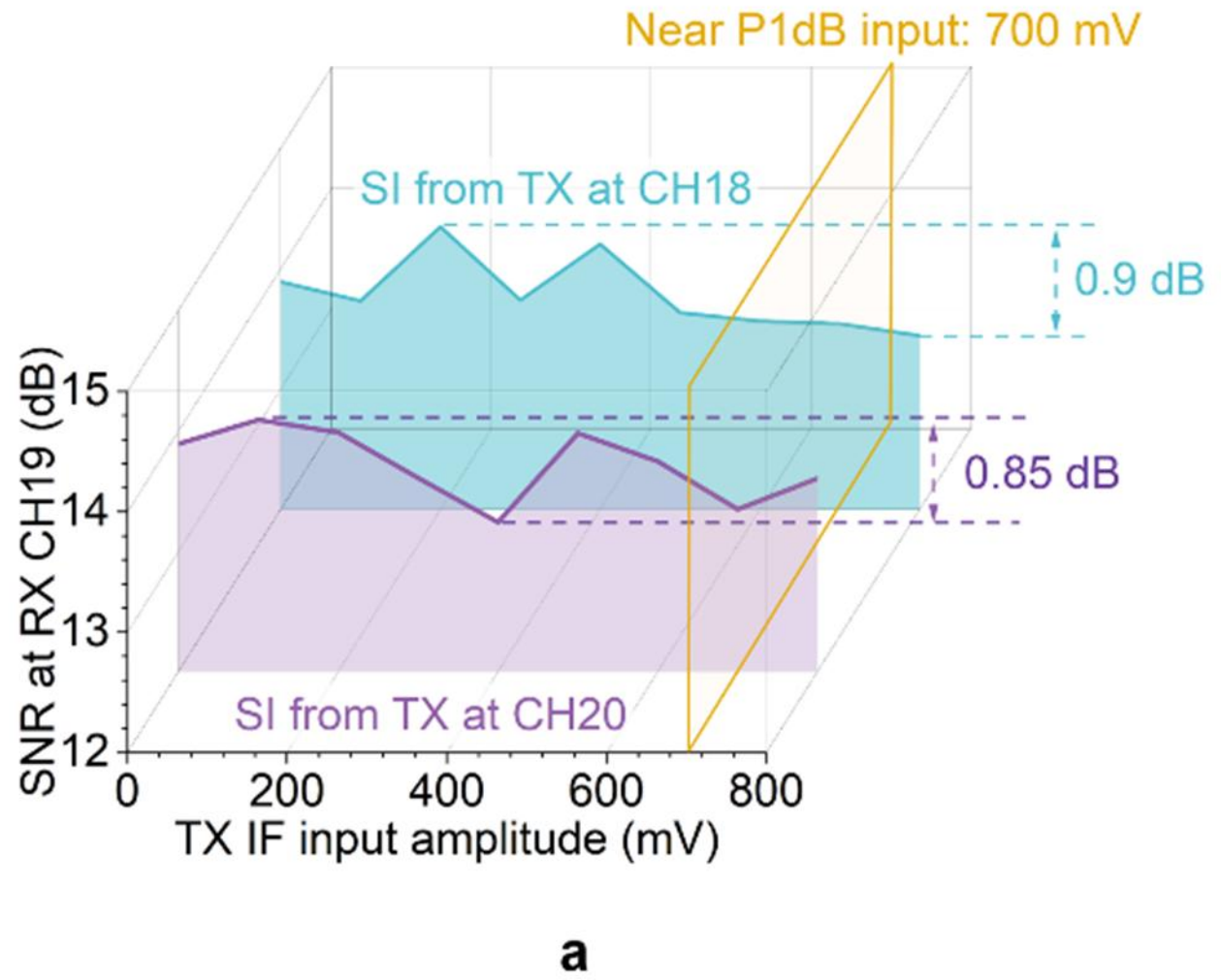

**a**

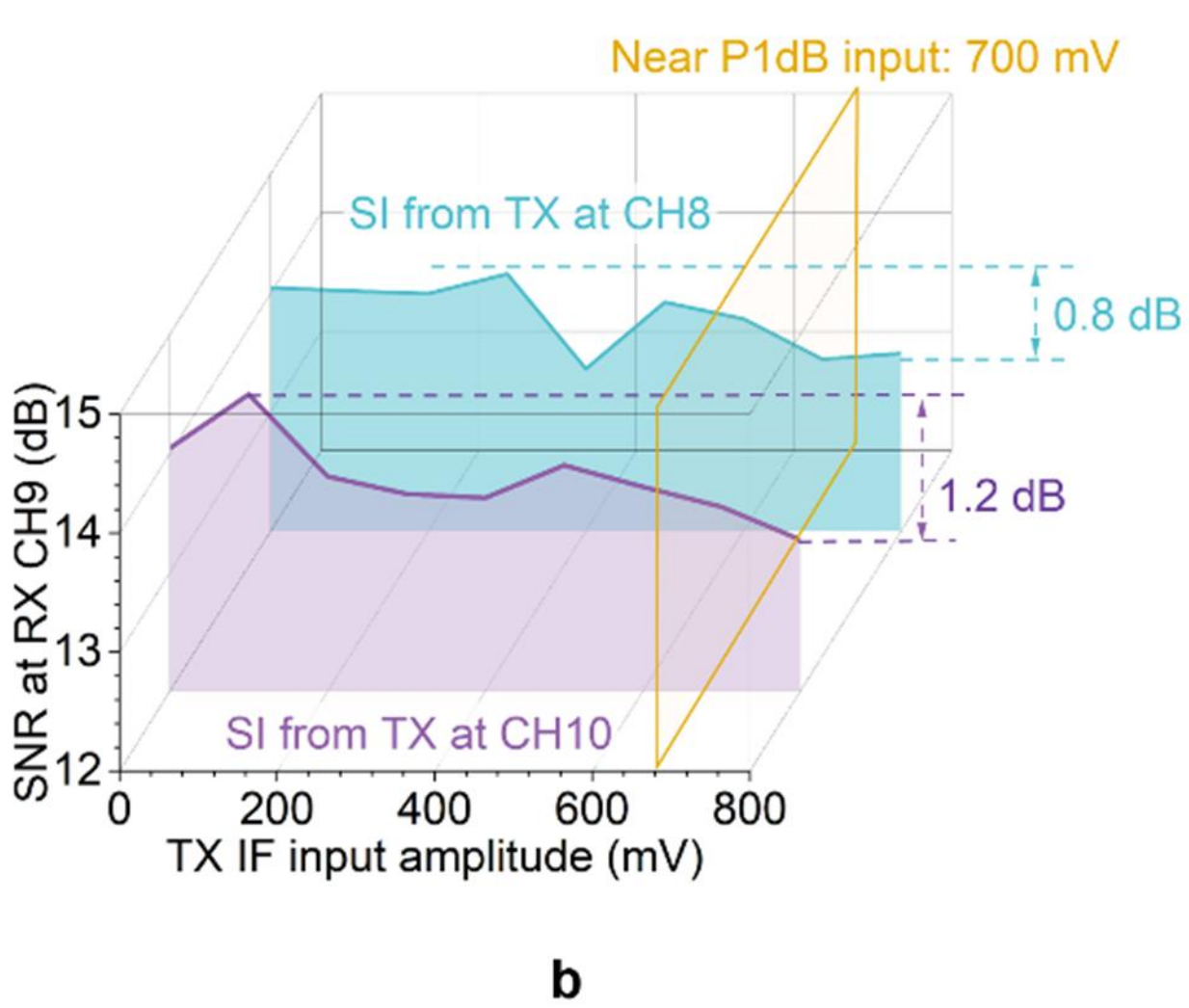

**b**

**Supplementary Fig. S1-10 | Measured SNR under TX self-interference versus TX IF input amplitude. a,** The desired receiving channel is fixed at CH19, while TX self-interference is applied at CH18 and CH20 in separate measurements. **b,** The desired receiving channel is fixed at CH9, while TX self-interference is applied at CH8 and CH10 in separate measurements.

# 2. THz Transceiver Architecture and Key Components

## 2.1 System architecture

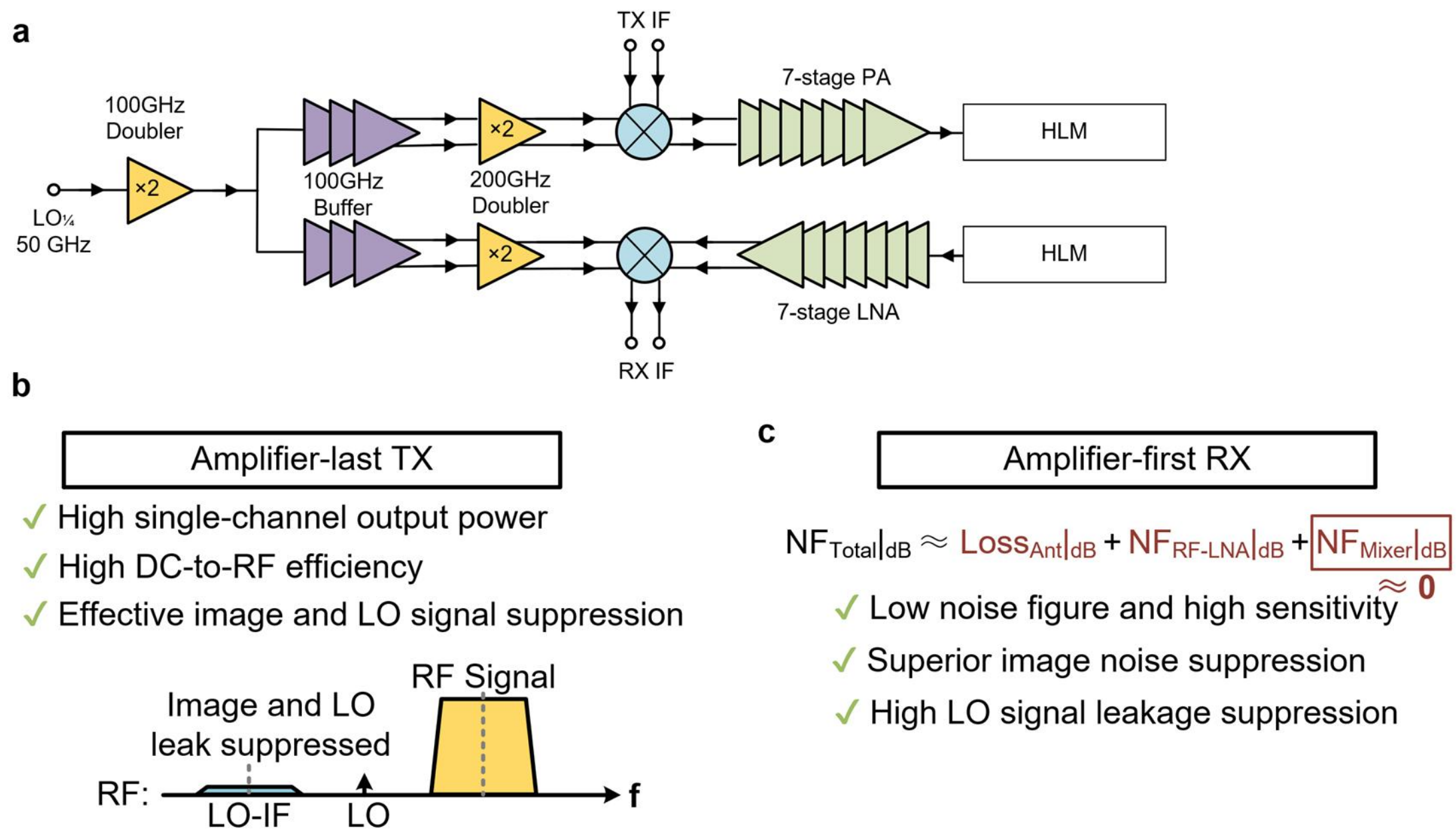

**Supplementary Fig. S2-1 | Architecture and operating principles of the proposed 208–258-GHz CMOS transceiver. a**, System architecture incorporating an amplifier-last TX, an amplifier-first RX, a shared ×4 LO-generation chain, and HLM apertures. **b**, TX image and LO-leakage suppression. **c**, RX noise-figure improvement, and image-noise and LO-leakage suppression.

Supplementary Fig. S2-1a presents the architecture of the proposed 208–258-GHz CMOS transceiver. The system combines an amplifier-last transmitter (TX) with an amplifier-first receiver (RX), each connected to a dedicated heterogeneous leaky-wave metasurface (HLM) aperture. The two signal paths share the first stage of the LO-generation chain. An on-chip frequency doubler converts the 50-GHz reference to 100 GHz before a two-way divider, generating coherent TX and RX branches. In each branch, a three-stage 100-GHz buffer amplifier compensates for the distribution loss and drives a second doubler. The resulting coherent 200-GHz signals are applied to the LO ports of the TX and RX double-balanced mixers for upconversion and downconversion, respectively.

In the TX path, the differential IF signal is mixed with the 200-GHz LO, and the upper sideband, $f_{RF} = f_{LO} + f_{IF}$, forms the desired 208–258-GHz output. As illustrated in Supplementary Fig. S2-1b, the seven-stage PA is placed after the mixer and directly drives the transmitting HLM. This amplifier-last arrangement minimizes RF loss after power amplification and preserves both the available radiated power and DC-to-RF efficiency. The PA passband is aligned with the desired upper sideband, whereas the lower-sideband image at $f_{LO}$ - $f_{IF}$ and the LO feedthrough near 200 GHz fall outside its high-gain region. The PA therefore functions simultaneously as the output-power stage and an on-chip spectral prefilter.

In the RX path, the signal collected by the receiving HLM is first amplified by the seven-stage LNA before entering the mixer. As summarized in Supplementary Fig. S2-1c, this amplifier-first arrangement places sufficient RF gain ahead of the lossy conversion stage, substantially reducing the mixer contribution to the cascaded receiver noise figure. The LNA also provides RF preselection: the desired 208–258-GHz signal lies within its passband, while image-band noise around $f_{LO}$ - $f_{IF}$ is attenuated before downconversion. Its out-of-band response and reverse isolation further suppress 200-GHz LO leakage from the mixer toward the receiving aperture. The resulting architecture combines high TX output power, low RX noise, image rejection, and coherent LO generation in a compact integrated transceiver.

### 2.2 H-band power amplifier design

Traditional CMOS H-band amplifiers have predominantly employed the $G_{max}$-core topology[S2-S10] to compensate for the limited intrinsic transistor gain near 300 GHz. In this approach, inductive-feedback and impedance-matching networks are embedded in the transistor at selected frequencies, so the gain enhancement is inherently linked to the embedding condition. A single embedding frequency generally produces a narrow boosted-gain response, whereas multiple embedding frequencies can extend the bandwidth with a reduction in peak gain[S11]. The two-frequency design achieves an approximately 30-GHz bandwidth[S5]. However, reported wideband $G_{max}$-core CMOS amplifiers still provide limited single-channel output power, including $P_{sat}$ = -3.3 dBm[S5] and -3.4 dBm[S9], and $OP_{1dB}$ = -13 dBm[S10]. Their inductive-feedback networks also increase the occupied area and constrain independent bias optimization. Although the DC-blocking capacitors separate the transistor bias domains and broaden the response, the resulting per-stage gain is only approximately 1–1.3 dB.

The proposed seven-stage PA addresses these limitations using the transformer-embedded over-neutralization (TEON) gain-boosting technique shown in Supplementary Fig. S2-2a. The first five stages provide broadband drive, and the last two stages are scaled as the power stage to increase the available output current and compression level. In each stage, the transformer is embedded directly in the neutralized transistor core rather than being used only as a separate interstage matching element. It therefore performs impedance transformation, couples adjacent stages, and supplies independent gate and drain bias through its centre taps within a compact signal path.

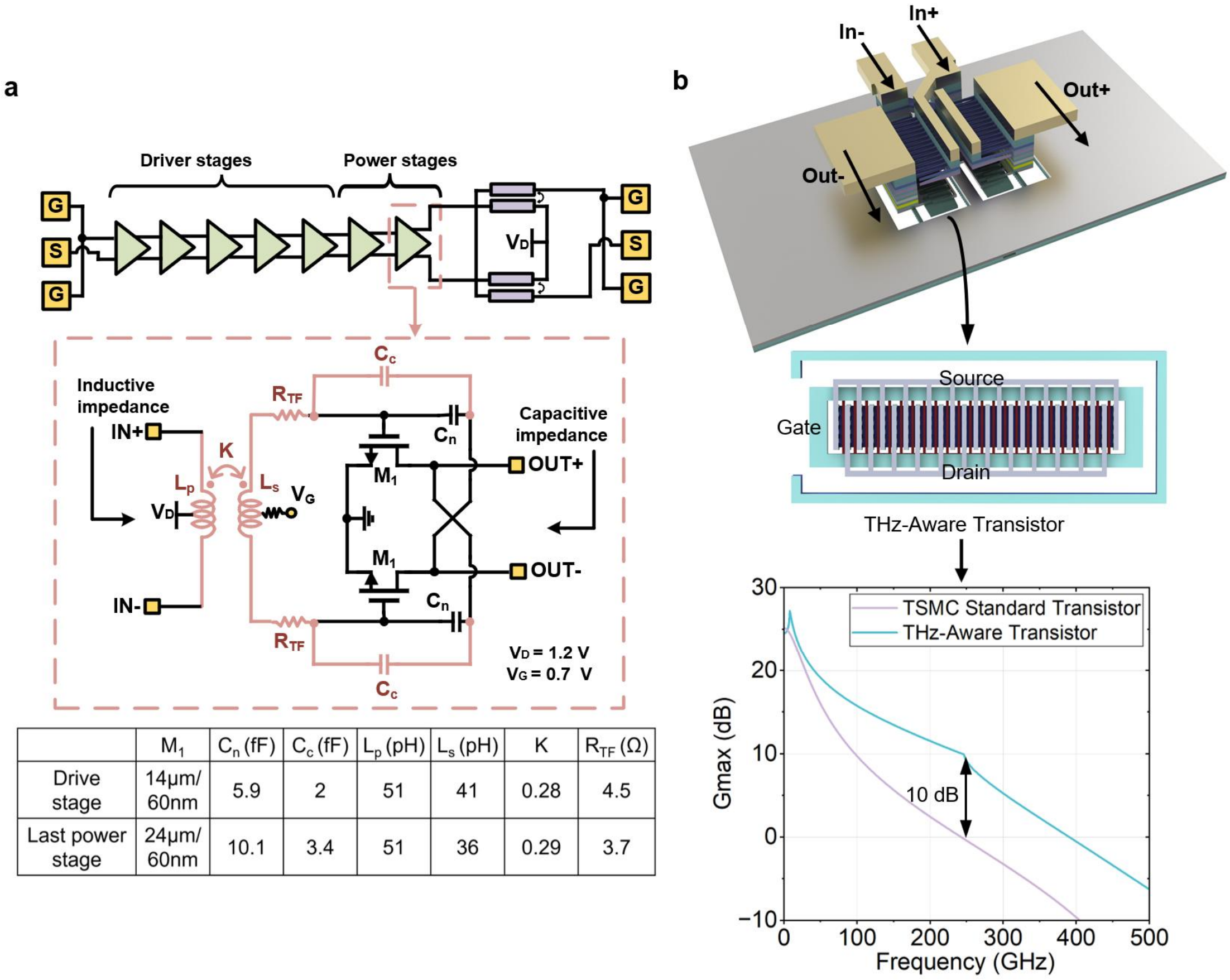

| | $M_1$ | $C_n$ (fF) | $C_c$ (fF) | $L_p$ (pH) | $L_s$ (pH) | K | $R_{TF}$ (Ω) |
|---|---|---|---|---|---|---|---|
| Drive stage | 14μm/ 60nm | 5.9 | 2 | 51 | 41 | 0.28 | 4.5 |
| Last power stage | 24μm/ 60nm | 10.1 | 3.4 | 51 | 36 | 0.29 | 3.7 |

**Supplementary Fig. S2-2 | Circuit and device implementation of the 208–258-GHz seven-stage TEON PA. a,** PA architecture, TEON gain-boosting-cell, and stage parameters. **b,** THz-aware transistor and PA interconnect layouts, with the simulated $G_{max}$ comparison against the standard transistor layout.

In a conventional over-neutralized pseudo-differential pair, the cross-coupled capacitor $C_n$ compensates the intrinsic gate–drain capacitance and increases the available gain, but excessive neutralization reduces the stability margin. In the TEON cell, the finite transformer loss, represented by $R_{TF}$, introduces controlled damping. This damping permits the auxiliary compensation capacitor $C_c$ to operate together with $C_n$, strengthening the effective neutralization while retaining a stable broadband response. The embedded transformer simultaneously transforms the capacitive output impedance of one stage to an impedance appropriate for driving the inductive input of the following stage, allowing the seven cells to be cascaded without an additional matching network between adjacent stages.

The driver and power-stage parameters are summarized in Supplementary Fig. S2-2a.

Each driver stage uses a 14 µm/60 nm transistor, $C_n$ = 5.9 fF, and $C_c$ = 2 fF. Its embedded transformer has $L_p$ = 51 pH, $L_s$ = 41 pH, K = 0.28, and $R_{TF}$ = 4.5 Ω. The final two power stages use larger 24 µm/60 nm transistors to increase output-current capability and improve large-signal compression. Because this scaling also increases the device parasitic capacitances and lowers the input impedance, the neutralization network and transformer secondary are re-optimized to $C_n$ = 10.1 fF, $C_c$ = 3.4 fF, and $L_s$ = 36 pH. The primary inductance remains 51 pH, while the coupling coefficient (K) = 0.29 and $R_{TF}$ = 3.7 Ω. All stages operate with $V_D$ = 1.2 V and $V_G$ = 0.7 V.

The THz-aware transistor and the PA interconnect implementation in Supplementary Fig. S2-2b further enable broadband operation above 240 GHz. The layout shortens the critical gate, drain, and source connections and reduces the parasitic resistance, capacitance, and electromagnetic coupling introduced by device routing and stage-to-stage interconnects. Relative to the standard transistor layout supplied by TSMC, the proposed implementation improves the simulated $G_{max}$ by approximately 10 dB at 250 GHz. The recovered device gain provides additional margin for broadband seven-stage cascading and reduces the gain burden placed on each passive TEON network.

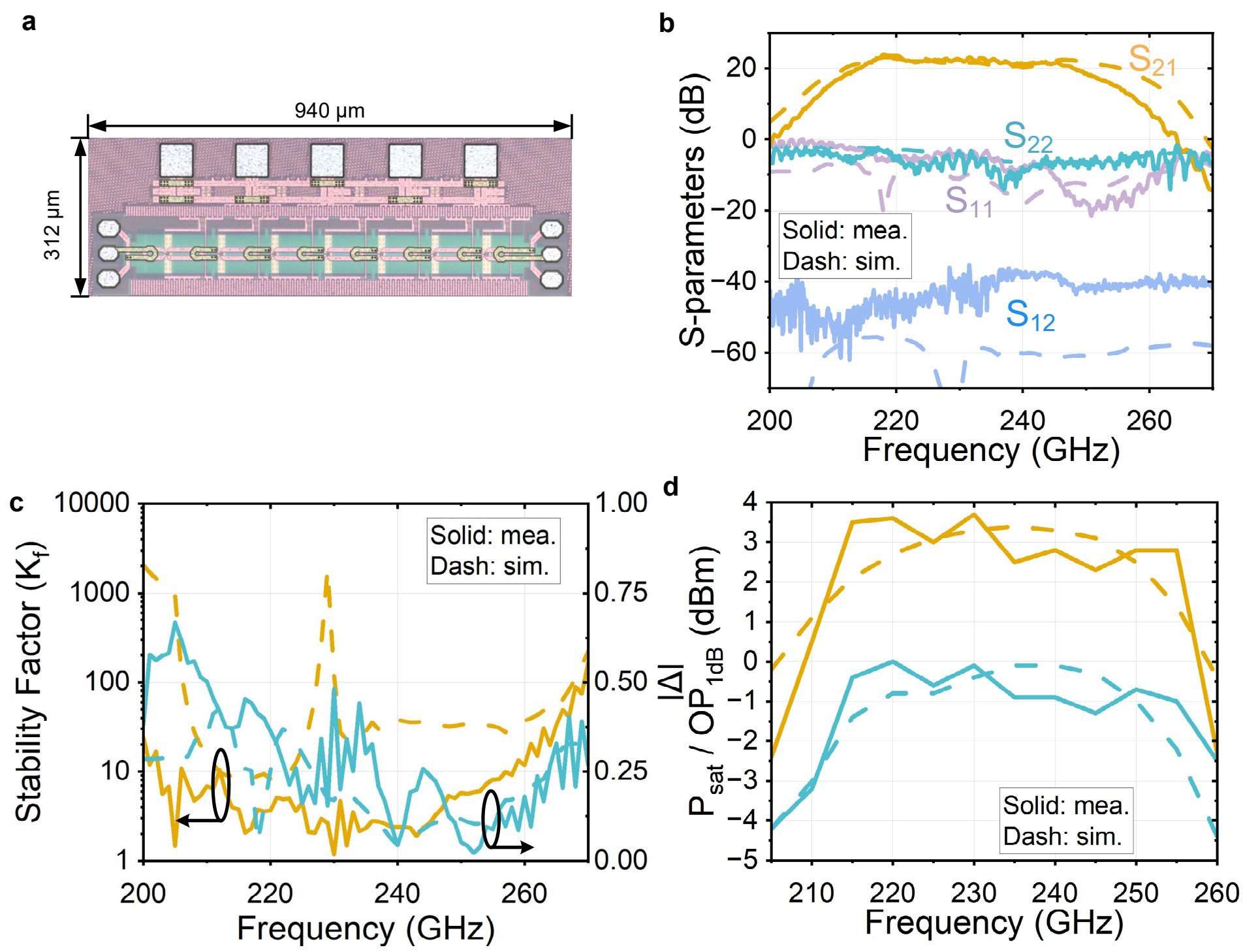

**Supplementary Fig. S2-3 | Fabricated PA and measured performance. a,** Chip micrograph. **b,** Simulated and measured S-parameters. **c,** Simulated and measured $K_f$ and |Δ|. **d,** Simulated and measured $P_{sat}$ and $OP_{1dB}$.

A micrograph of the fabricated PA is shown in Supplementary Fig. S2-3a; the circuit occupies 940 µm × 312 µm. Supplementary Fig. S2-3b compares the simulated and measured small-signal responses. The measured $|S_{21}|$ reaches 23.3 dB at 219 GHz, and the measured 3-dB gain bandwidth extends from 214.4 to 247.7 GHz. Across 208–258 GHz, $|S_{21}|$ remains above 11 dB, while $|S_{12}|$ remains below -35 dB, demonstrating broadband forward gain together with strong reverse isolation. The roll-off outside the intended passband also limits the amplification of mixer products outside the desired RF band.

The stability quantities derived from the simulated and measured S-parameters are presented in Supplementary Fig. S2-3c. The Rollet stability factor ($K_f$) remains greater than unity, and the determinant magnitude $|\Delta|$ remains below unity throughout the 208–258-GHz operating band. The two conditions are simultaneously satisfied despite the strong neutralization and seven-stage cascade, confirming unconditional small-signal stability over the intended band.

The large-signal measurements in Supplementary Fig. S2-3d verify the output capability of the enlarged final stage. The measured $P_{sat}$ remains positive from 210 to 255 GHz and reaches a peak value of 3.7 dBm at 230 GHz. The measured $OP_{1dB}$ approaches 0 dBm near 220 GHz and remains above approximately -1.4 dBm from 215 to 255 GHz. Together, the measured gain, stability, and compression results show that the TEON network, power-stage scaling, and THz-aware layout provide broadband amplification and positive saturated output power across the targeted H-band range.

**Supplementary Table S2-1 | Performance comparison with reported CMOS H-band amplifiers.**

| | This work | [S9] ISSCC'24 | [S5] JSSC'19 | [S6] JSSC'21 | | [S8] JSSC'23 | [S10] MWTL'19 | [S7] MWTL'21 | [S4] TMTT'25 | | [S2] JSSC'17 |
|---|---|---|---|---|---|---|---|---|---|---|---|
| Technology | 65-nm CMOS | | | | | | | | | | |
| Topology | TEON | $G_{max}$-core | | | | | | | | | |
| Supply (V) | 1.2 | 1 | 0.85 | 1 | 1 | 1 | 1 | 1.2 | 0.75 | 0.75 | 1 |
| Gain (dB) | 23.3 | 22* | 13.9 | 18 | 15 | 26 | 21 | 12.6 | 18.2 | 9.3 | 9.2 |
| Frequency (GHz) | 214-248 | 237-267 | 227.5-257.2 | 247 | 272 | 243 | 298 | 248.6 | 280.2 | 309.2 | 257 |
| 3-dB BW (GHz) | 34 | 30* | 29.7 | 2.9 | 3.7 | 7 | 2* | 5* | 3* | 3* | 12.2 |
| GBW (GHz) | 497 | 378 | 147 | 23 | 21 | 140 | 22 | 21 | 24 | 9 | 35 |
| Stage | 7 | 15 | 4 | 2 | 2 | 6 | 16 | 3 | 3 | 3 | 4 |
| $P_{sat}$ (dBm) | 3.7 | -3.4 | -3.3 | 0.09 | -2.36 | 10.5 (4-way) | N/A | 3.8 | -10.4 | -8.1 | -3.9 |
| $OP_{1dB}$ (dBm) | 0 | -11.2 | -5.1 | -9.52 | -10.18 | 7.7 | -13 | 2.3 | N/A | N/A | -8 |
| PAE (%) | 1.7 | N/A | 1.6 | 4.44 | 2.37 | 2.7 | N/A | 3.2 | 0.4 | 1.9 | 0.8 |
| $P_{dc}$ (mW) | 122 | 38 | 23.8 | 21.5 | 21.5 | 407 | 35.4 | 62.4 | 12.3 | 6.6 | 27.6 |
| Power Density ($mW/mm^2$) *** | 58.6 | 0.95 | 8.68 | 7.5 | 5.23 | 10.79 | N/A | 12.83 | 0.42 | 2.35 | 2.9 |
| Core Area ($mm^2$) | 0.04 | 0.48 | 0.053 | 0.136** | 0.111** | 1.04** | 0.39 | 0.187** | 0.212** | 0.066** | 0.14 |

*Estimated from the figures. **Core area estimated according to chip photos, excluding RF and DC pads. *** Power Density = $P_{sat}$ (mW)/Core Area ($mm^2$)

Supplementary Table S2-1 benchmarks the proposed seven-stage TEON PA against reported 65-nm CMOS H-band amplifiers, most of which employ the $G_{max}$-core topology. The proposed PA achieves a measured gain of 23.3 dB and a 34-GHz 3-dB bandwidth spanning 214-248 GHz, corresponding to a gain-bandwidth product of 497 GHz, the largest among the designs compared. Its measured $P_{sat}$ of 3.7 dBm and $OP_{1dB}$

of 0 dBm are competitive with prior broadband single-channel implementations. Despite its seven-stage architecture, the PA occupies a core area of only 0.04 mm$^2$, the smallest in the comparison, and achieves the highest saturated-output-power density of 58.6 mW/mm$^2$. These results show that TEON gain boosting relaxes the conventional trade-off among gain, bandwidth, output power, and silicon area, enabling broadband, high-gain H-band amplification in a compact CMOS implementation.

### 2.3 H-band low-noise amplifier design

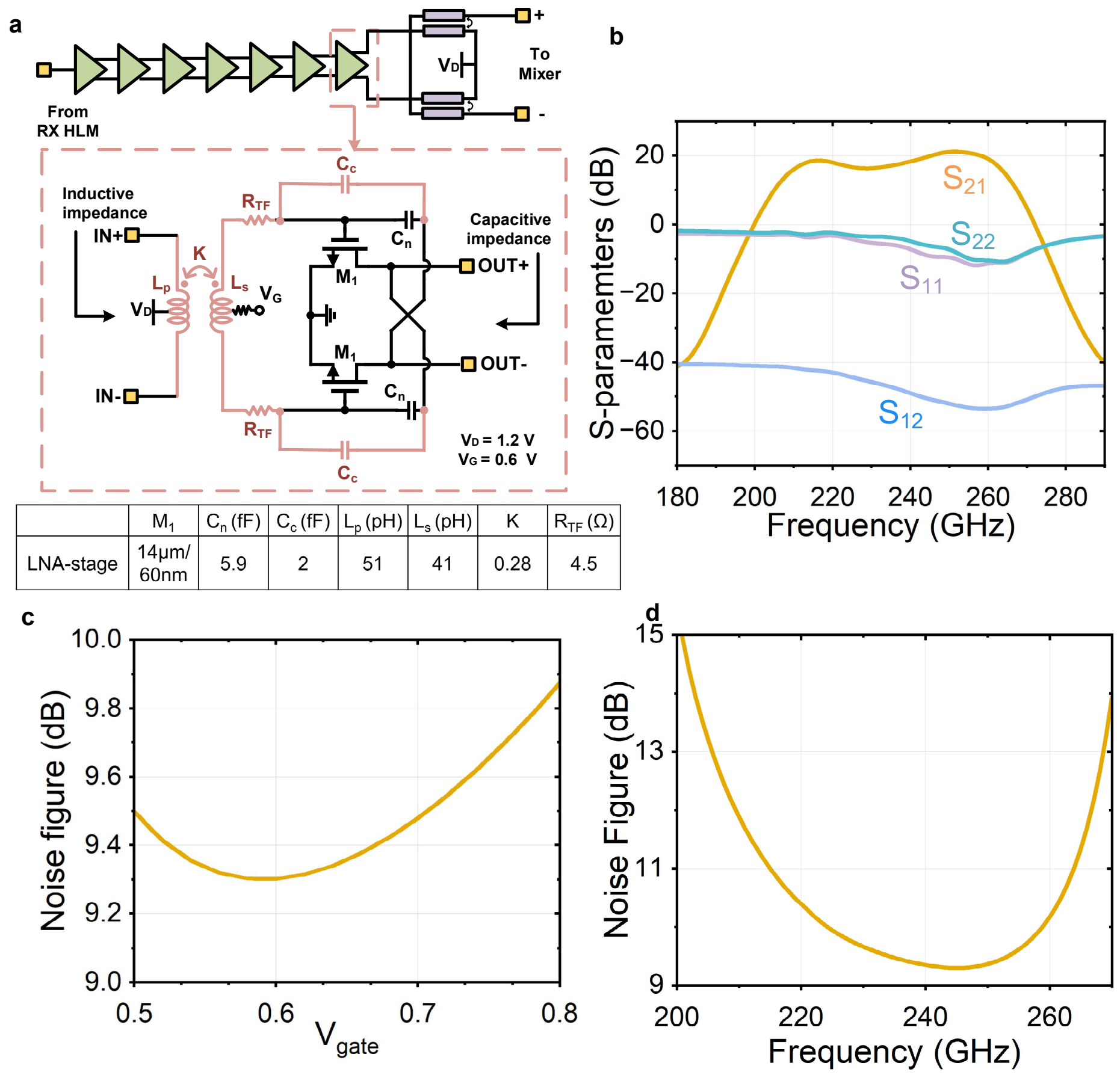

| | $M_1$ | $C_n$ (fF) | $C_c$ (fF) | $L_p$ (pH) | $L_s$ (pH) | K | $R_{TF}$ (Ω) |
|---|---|---|---|---|---|---|---|
| LNA-stage | 14μm/ 60nm | 5.9 | 2 | 51 | 41 | 0.28 | 4.5 |

**Supplementary Fig. S2-4 | Design and simulated performance of the 208–258-GHz seven-stage TEON LNA. a,** LNA architecture, TEON stage, and design parameters. **b,** Simulated S-parameters. **c,** Simulated noise figure versus gate-bias voltage. **d,** Simulated noise figure under the source impedance presented by the integrated HLM.

The proposed seven-stage LNA employs the transformer-embedded over-neutralization (TEON) gain-boosting technique introduced in Section 2.2. Although the PA and LNA share the same basic gain cell, they are optimized for different objectives. The enlarged final stage of the PA is designed for output swing and saturated output power, whereas the LNA stages are optimized for low noise and broadband gain over 208–258 GHz.

Because the LNA is directly connected between the on-chip HLM and the mixer, the source impedance presented by the HLM and the actual mixer input impedance are incorporated directly into the LNA design.

Supplementary Fig. S2-4a shows the LNA architecture, TEON stage, and design parameters. Each stage uses a 14 μm/60 nm transistor pair with $C_n = 5.9$ fF and $C_c = 2$ fF. The embedded transformer has $L_p = 51$ pH, $L_s = 41$ pH, $K = 0.28$, and an equivalent parasitic resistance $R_{TF} = 4.5\ \Omega$. The transformer simultaneously performs interstage impedance transformation and provides independent gate and drain bias access through its centre taps. All stages operate at $V_D = 1.2$ V and share the low-noise gate bias selected below.

The simulated S-parameters in Supplementary Fig. S2-4b show that the seven cascaded TEON stages provide a broadband gain response covering the targeted 208–258-GHz RF band while maintaining strong reverse isolation. The gain decreases rapidly outside the desired band, particularly below 200 GHz. Consequently, the LNA also acts as an RF preselection stage: the desired upper-sideband signal is amplified before downconversion, whereas image-band noise below the LO frequency receives substantially less gain before reaching the mixer.

The LNA noise performance is optimized first through its gate-bias condition. Supplementary Fig. S2-4c plots the simulated noise figure as a function of gate voltage. The noise figure reaches its minimum near $V_G = 0.6$ V, where the extracted transistor noise-current contribution is also minimized. This bias is therefore selected as the nominal operating point for all LNA stages.

The input interface is co-designed with the integrated HLM, while the actual mixer input impedance is retained as the LNA output load. Supplementary Fig. S2-4d presents the simulated LNA noise figure under the source impedance provided by the HLM. The resulting low-noise performance is further validated by the subsequent system-level receiver noise figure measurements.

### 2.4 4× multiplier chain design

To generate sufficient and coherent LO signals for the TX and RX mixers, the transceiver employs the shared 4× frequency multiplier and distribution chain shown in Supplementary Fig. S2-5a. A common 50-GHz reference is first converted to 100 GHz by an input-stage frequency doubler. The resulting signal is divided into two paths by a two-way 100-GHz Wilkinson power divider. Each path contains an identical three-stage fully differential 100-GHz PA followed by a push–push 100-to-200-GHz frequency doubler, providing separate LO signals for the TX and RX double-balanced mixers (DBMs). Deriving both LO signals from the same reference ensures intrinsic frequency coherence while reducing the area and power consumption compared with two independent multiplier chains.

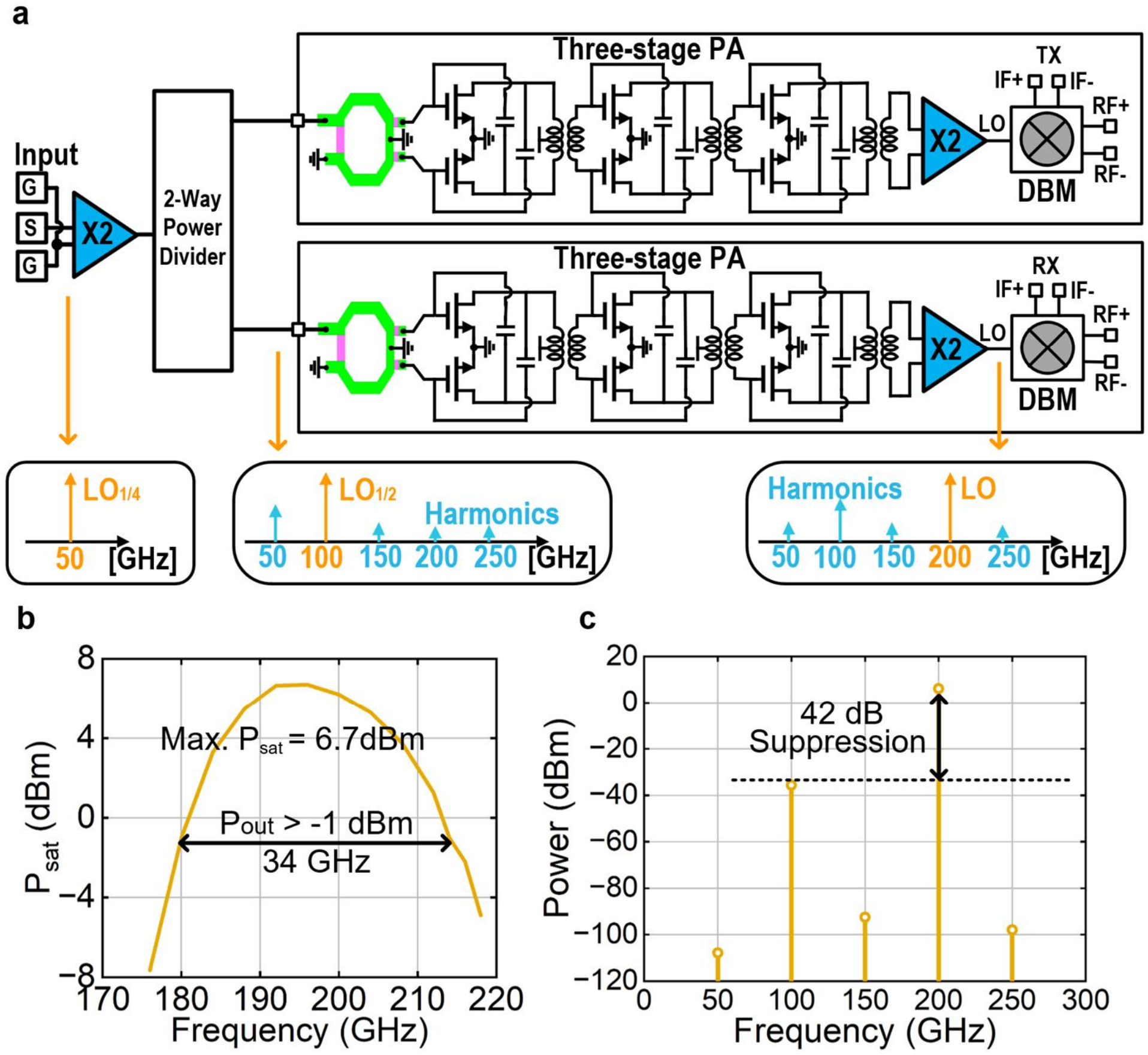

**Supplementary Fig. S2-5 | Architecture and simulated performance of the shared 4× LO multiplier chain. a,** Circuit architecture and representative spectra at the 50-, 100-, and 200-GHz nodes. **b,** Simulated saturated output power of the multiplier chain. **c,** Simulated output spectrum showing more than 42-dB harmonic suppression.

In each branch, the three-stage PA compensates for the divider and interconnect losses and provides sufficient input swing to the final doubler. The PA employs capacitively neutralized pseudo-differential stages to improve the available gain and stability at 100 GHz, while transformer-based interstage matching enables compact broadband cascading. The final doubler adopts a push–push topology, in which the fundamental components generated by the two transistor branches cancel at the differential output, whereas their second-harmonic currents combine constructively to produce the desired 200-GHz LO signal.

The gate-side second-harmonic termination of the push–push doubler is further optimized using a quarter-wavelength shorting structure. This termination suppresses the undesired second-harmonic component returning toward the input and prevents it from interacting destructively with the second-harmonic output current generated by the push–push core. Consequently, the doubler achieves a peak simulated conversion gain of -5 dB, including its input and output matching losses, and produces 6.2 dBm with a 12-dBm input.

As shown in Supplementary Fig. S2-5b, the complete multiplier chain achieves a peak simulated saturated output power of 6.7 dBm and maintains $P_{sat}$ > -1 dBm from 180 to 214 GHz. At the nominal 200-GHz LO frequency, approximately 6.7 dBm is available before the output transformer. After accounting for its 4.1-dB insertion loss, 2.6 dBm is delivered to the LO port of each DBM. The simulated output spectrum in Supplementary Fig. S2-5c exhibits more than 42-dB suppression of the undesired harmonics. The shared multiplier chain therefore provides sufficient LO drive, high spectral purity, and coherent TX–RX LO generation for broadband transceiver operation.

### 2.5 Terahertz double-balanced fundamental mixer design

Both the TX and RX front ends employ a bidirectional double-balanced fundamental mixer for frequency upconversion and downconversion, respectively. Although the double-balanced topology is well established at RF and millimetre-wave frequencies, its implementation above 200 GHz requires sufficient LO swing and accurately preserved differential symmetry among the RF, LO, and IF ports. Fundamental mixing is selected to avoid the conversion-gain degradation generally associated with harmonic or subharmonic conversion. This choice is enabled by the shared 4× LO chain described in Section 2.4, which delivers approximately 2.6 dBm of 200-GHz power to the LO port of each mixer after the approximately 4-dB loss of the output transformer.

At H-band frequencies, the compact arrangement of the three differential ports introduces routing imbalance and parasitic electromagnetic coupling[S12]. These nonidealities disturb the virtual-ground condition at the RF port of the switching core and increase direct LO-to-RF leakage. The proposed mixer, therefore, incorporates a layout-balance compensation technique that restores the effective symmetry of the signal paths without increasing the core area substantially.

The circuit schematic and three-dimensional layout are shown in Figs. S2-6a and S2-6b, respectively. The switching transistors use a gate width of 6 × 1 μm. Even with a nominally symmetric floorplan, the parasitic capacitance $C_p$ between the RF interconnects and the gates of the upper transistor pair cannot be eliminated completely.

The resulting capacitive imbalance introduces amplitude and phase errors into the LO excitation and prevents an ideal RF-port virtual ground, allowing part of the LO signal to couple directly to the RF path.

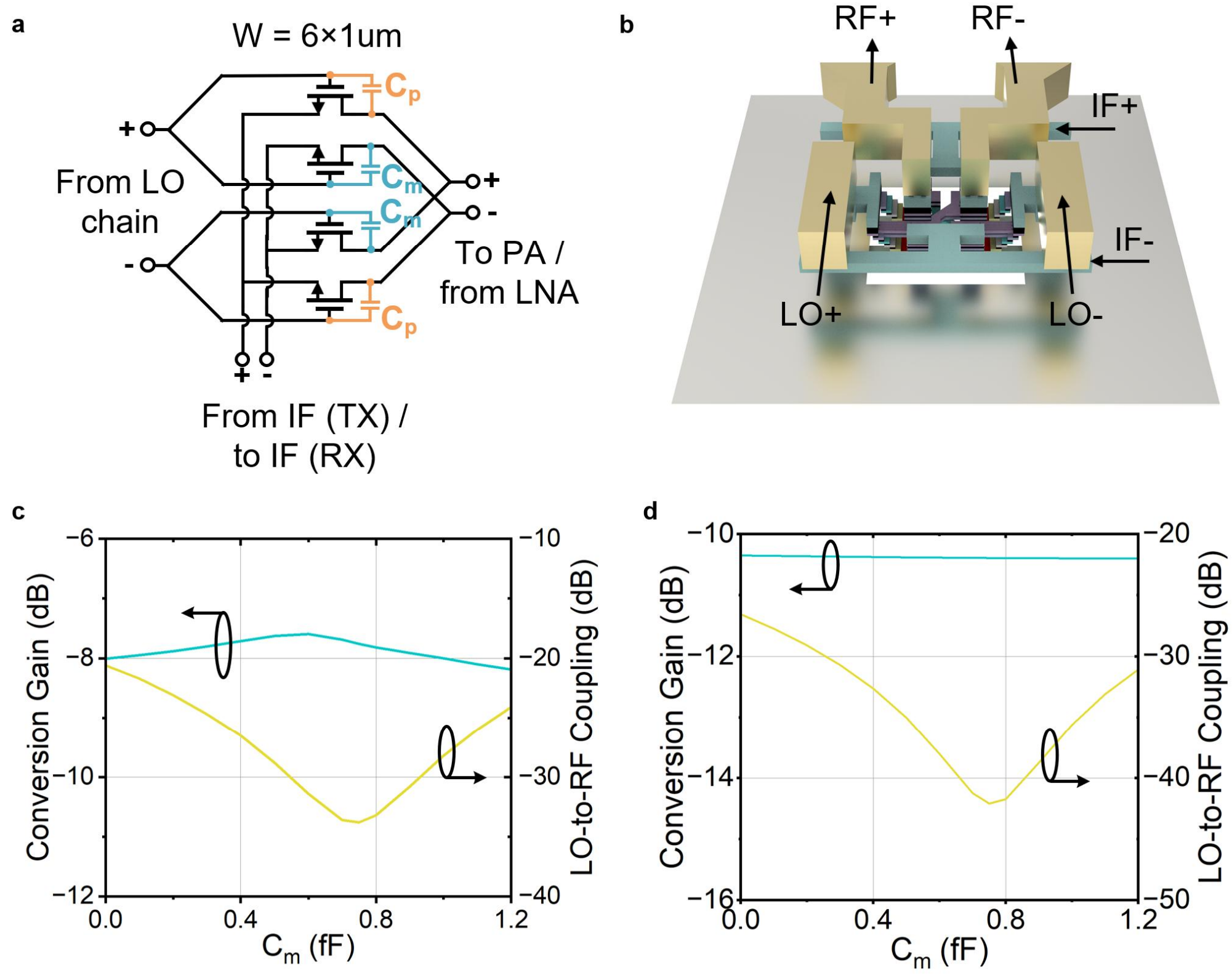

**Supplementary Fig. S2-6 | Bidirectional double-balanced fundamental mixer and layout-balance compensation. a,** Mixer schematic showing the parasitic and compensation capacitances. **b,** Three-dimensional mixer layout and port arrangement. **c,d,** Simulated conversion gain and LO-to-RF coupling versus $C_m$ in the TX and RX modes, respectively.

To counteract this parasitic path, a compensation capacitor $C_m$ is inserted between the gates and drains of the complementary lower transistor pair. $C_m$ is implemented using interdigitated metal fingers above the active devices, providing a controlled coupling path that counterbalances $C_p$. Figs. S2-6c and S2-6d show the simulated conversion gain and LO-to-RF coupling as functions of $C_m$ in the TX and RX modes, respectively. An optimized $C_m$ of 0.7 fF improves LO-to-RF suppression by more than 10 dB while introducing only a small conversion-gain variation. In the complete transceiver, the following TX PA and preceding RX LNA provide additional isolation between the mixer and the radiating interface.

The mixer is co-designed with the broadband PA, LNA, and shared 4× LO chain to support operation from 208 to 258 GHz. To balance conversion gain, linearity, and port matching in the two directions, the gate bias is set to 0.3 V in TX mode and 0.5 V in RX mode. The difference between the TX- and RX-mode conversion gains arises primarily from their different bias conditions and RF embedding impedances. In TX mode, the mixer drives the PA input through an RF load of approximately 20 + j47 Ω. In RX mode, the mixer RF input is embedded in approximately 22 - j35 Ω and transformed to serve as the LNA load. The differential IF ports are routed to the chip pads through 50-Ω differential transmission lines and wire-bonded to grounded coplanar-waveguide traces on the printed circuit board (PCB).

## 2.6 System performance

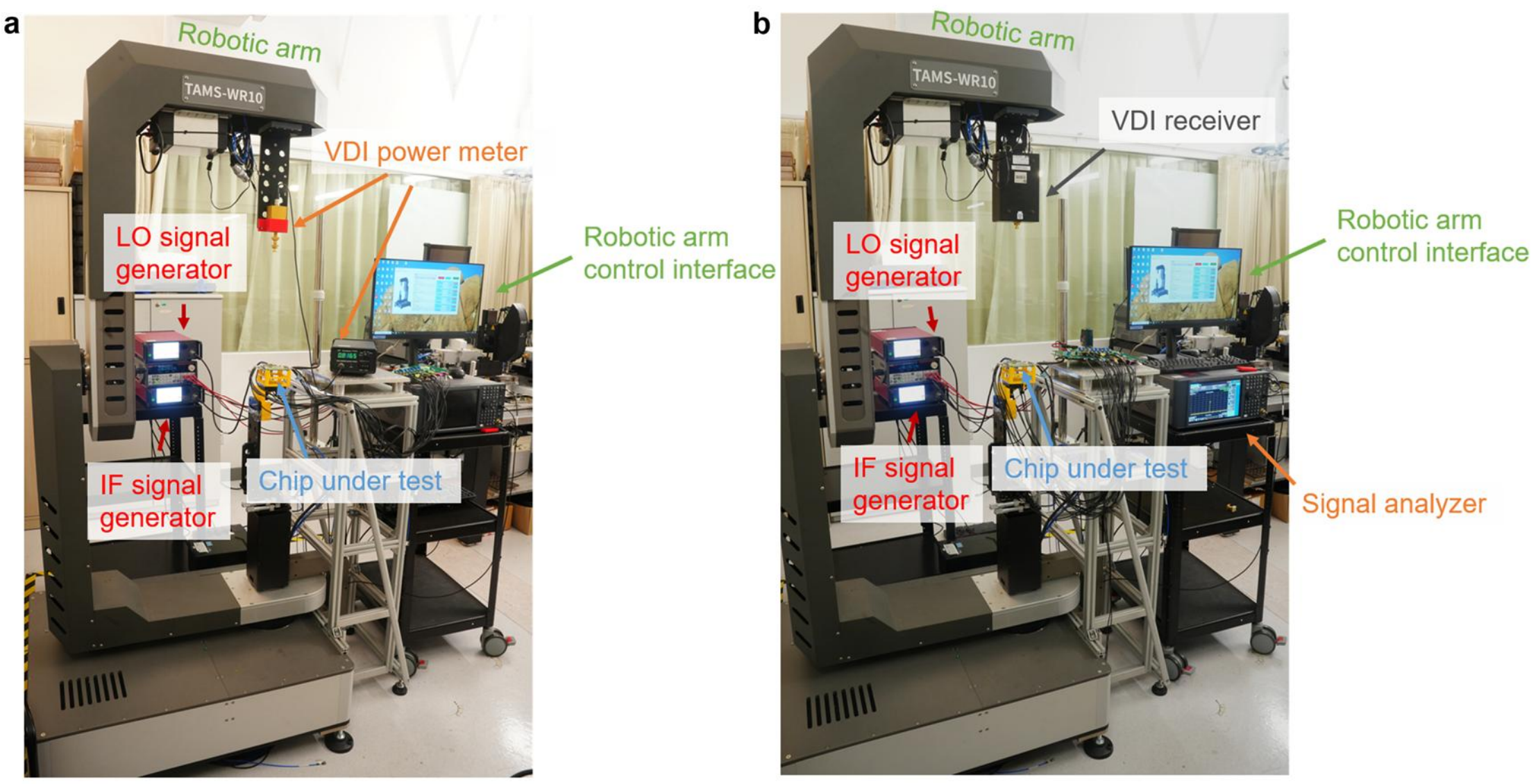

**Supplementary Fig. S2-7 | Over-the-air transceiver measurement setup. a,** Setup for TX EIRP and TX/RX conversion-gain characterization. **b,** Setup for HLM radiation-pattern and RX noise-figure measurements.

The over-the-air measurement arrangements in Supplementary Fig. S2-7 are used to characterize the fabricated transceiver. Supplementary Fig. S2-7a shows the setup for TX effective isotropic radiated power (EIRP) and conversion-gain measurements, while Supplementary Fig. S2-7b shows the configuration used for HLM radiation-pattern and RX noise-figure measurements. A continuous-wave signal generator drives the differential IF input, and a second generator supplies the 50-GHz reference for the integrated 4× LO chain. The device under test (DUT) and the WR3.4 extender are separated by a distance d that satisfies the far-field condition. During angular measurements, the extender is positioned by a robotic arm while the DUT remains fixed.

Before the transceiver metrics are extracted, the measured insertion losses of the IF and reference-LO cables, end-launch connectors, and PCB grounded coplanar-waveguide traces are de-embedded so that all reported powers are referred to the chip interfaces.

The gain of the standard horn antenna is established using a two-antenna calibration. A calibrated VDI WR3.4 source module and an Erickson PM5B thermal power meter establish the absolute sub-terahertz power reference for the transmit and receive measurement chains.

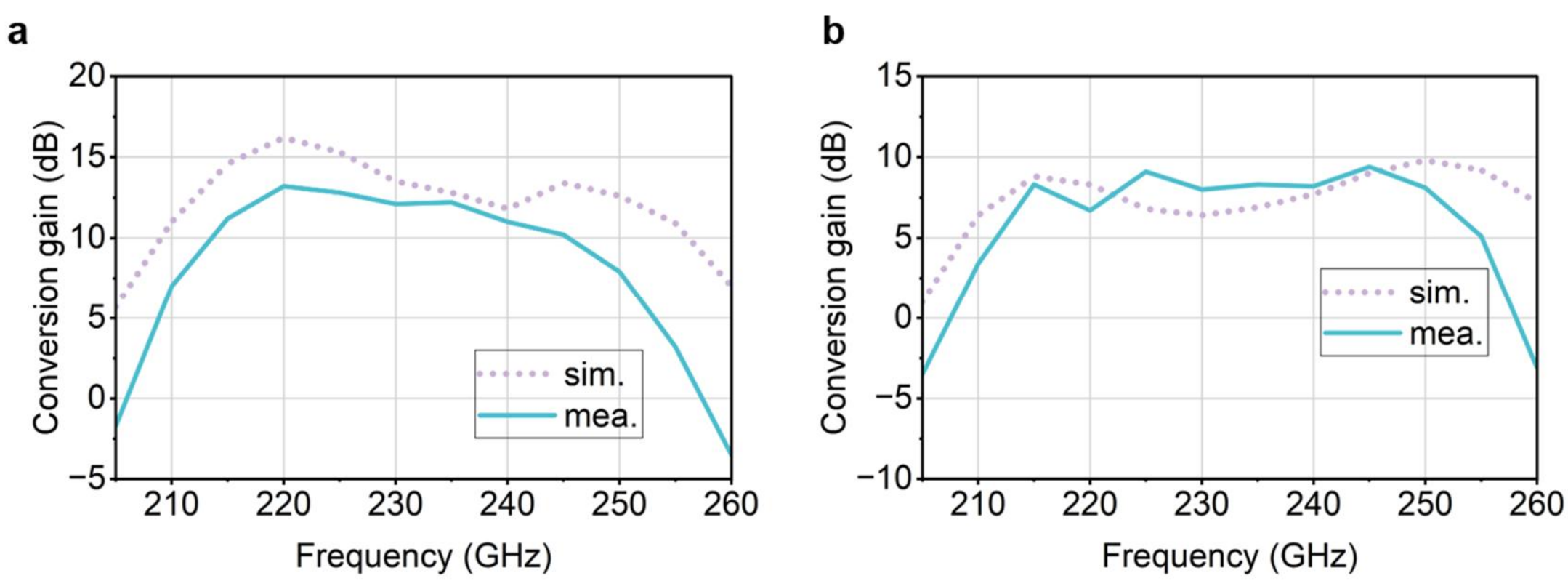

**Supplementary Fig. S2-8 | Simulated and measured system-level performance. a,** TX conversion gain and **b,** RX conversion gain.

For TX characterization, the radiated signal is collected by the calibrated horn, downconverted by the VDI WR3.4 receiver, and recorded by the signal analyzer. The RF frequency is swept across 205–260 GHz by varying the IF frequency while maintaining the on-chip LO near 200 GHz. From the calibrated received power $P_{RX}$, the TX EIRP is obtained as[S13]

$$EIRP_{TX}\,(dBm) = P_{RX}\,(dBm) - G_{RX,horn}\,(dBi) + FSPL(d,f)\,(dB) \tag{S2-1}$$

where

$$FSPL(d,f) = 20\,log_{10}(\frac{4\pi df}{c}) \tag{S2-2}$$

where $G_{RX,horn}$ is the calibrated receiving-horn gain and $FSPL(d,f)$ is the free-space path loss at separation d and frequency f. The TX HLM directivity $D_{TX}$ is extracted from the measured angular response. For conversion-gain characterization, the IF drive is backed off until the TX operates in its linear region. The small-signal TX conversion gain is then

$$CG_{TX}\,(dB) = EIRP_{TX}\,(dBm) - D_{TX}\,(dBi) - P_{IF,in}\,(dBm) \tag{S2-3}$$

where $P_{IF,in}$ is the available IF power at the chip reference plane. This definition includes the HLM conversion loss but excludes its directivity. The measured and simulated TX conversion gain are presented in Figs. S2-8a.

For RX characterization, the calibrated WR3.4 source module and transmitting horn illuminate the receiving HLM from the far field. The source power $P_{SRC}$ is referenced with the thermal power meter, and the downconverted IF signal is measured by the signal analyzer. After de-embedding the IF output path, the RX conversion gain is calculated as

$$CG_{RX}\,(dB) = P_{IF,out}(dBm) - [P_{SRC}\,(dBm) + G_{TX,horn}\,(dBi) - FSPL(d,f)\,(dB)] - D_{RX}\,(dBi) \tag{S2-4}$$

where $P_{IF,out}$ is the calibrated IF output power, $G_{TX,horn}$ is the transmitting-horn gain, and $D_{RX}$ is the receiving-HLM directivity obtained from its measured radiation pattern. As for the TX definition, the reported RX conversion gain includes the HLM loss but excludes its directional gain. The extracted RX conversion gain is shown in Supplementary Fig. S2-8b, compared with the simulation data.

The RX noise figure is determined using the output-noise method[S12]. With the RF source disabled, an external low-noise IF amplifier is inserted before the signal analyzer so that the measured noise remains above the analyzer floor. The displayed noise power is normalized to a 1-Hz bandwidth, and the independently characterized frequency-dependent gain of the external IF amplifier is de-embedded. Because the receiver output noise is substantially higher than the input-referred added noise of the external amplifier, the latter contribution is negligible. The RX noise figure is consequently obtained from

$$NF_{RX}\,(dB) \approx N_{SA}\,(dBm/Hz) - G_{ext}\,(dB) + 174 - CG_{RX}\,(dB) \tag{S2-5}$$

where $N_{SA}$ is the noise density measured by the signal analyzer, and $G_{ext}$ is the independently characterized gain of the external IF amplifier. The constant 174 represents the magnitude of the room-temperature thermal-noise density of -174 dBm/Hz. The extracted RX noise figure is shown in Fig. 3i.

The measured system-level responses in Fig. 3i and Supplementary Fig. S2-8 follow the simulated frequency trends over the operating band. The measured TX EIRP reaches 10.3 dBm near 240 GHz, while the small-signal TX conversion gain peaks at 13.2 dB near 220 GHz. The measured RX conversion gain reaches 9.1 dB at 225 GHz, and the minimum measured RX noise figure is 14.4 dB near 245 GHz. These over-the-air results jointly validate the amplifier-last TX, amplifier-first RX, and shared coherent-LO architecture at the system level.

**Supplementary Table S2-2 | System-level performance comparison with reported silicon-based H-band beamforming systems.**

| | [S14] | [S15] | [S16] | [S17] | [S18] | [S9] | [S19] | This Work |
|---|---|---|---|---|---|---|---|---|
| Process | 65nm CMOS | 40nm CMOS | 40nm CMOS | 40nm CMOS | 65nm CMOS | 65nm CMOS | 65nm CMOS | 65nm CMOS |
| RF Frequency (GHz) | 293-307 | 262-275 | 263-279 | 252-285 | 242-280 | 236-266 | 288-334 | 208-258 |
| Structure | RX: Mixer-first | RX: Mixer-first | TX: Mixer-last | RX: Mixer-first | TRX: Mixer-last/Mixer-first | TX: Amp-last | RX: Mixer-first | TRX: Amp-last/Amp-first |
| Beamforming Architecture | LO+IF Beamforming | LO+IF Beamforming | LO+IF Beamforming | LO Beamforming | LO Beamforming | LO Beamforming | IF Beamforming | HLM-enabled dispersion engineering |
| $P_{sat}$ Per Channel (dBm)* | N/A | N/A | N/A | N/A | N/A | -3.4 | N/A | 3.6 |
| TX EIRP (dBm) | N/A | N/A | -6.9 | N/A | N/A | 16.2 | N/A | 10.3 |
| RX SSB NF (dB) | 23.6 | N/A | N/A | 32** | 20# | N/A | 26.8 | 14.4 |
| $P_{DC}$(W) | 1.4 | 2.8 | 3.7 | 2.8 | 3+ | 13.44+ | 1.877 | 0.43 |
| Chip Area (mm²) | 2.4×2.5 | 6.23×4.73 | 6.23×4.73 | 3.52×1.37& 2.15×0.96 | 2.45×1.7 | 3.8×2.6 | 6.43×4.24 | 4.9×1.5 |
| Array Type | single-chip 2D array | single-chip 2D array | single-chip 2D array | multi-chip 1D array +movable lens | stacked-chip 1D array | stacked-chip 2D array | single-chip 2D array | single-chip HLM aperture |
| Array Size | 4×4 | 3×3* | 3×3* | 1×4 | 1×4 TRX | 4×16 | 4×4 | 1×TRX |
| Number of Chips | 1 | 1 | 1 | 5 | 4 | 16 | 1 | 1 |
| Ant. Integration | on-quartz ant. | PCB ant. | PCB ant. | horn+lens | PCB ant. | on-chip ant. | on-quartz ant. | on-chip ant. |
| Multi-Beam Operation | No | No | No | No | No | No | No | Yes |
| Steering Range (°) | ±24/±40 | ±20/±30v | ±30/±30v | ±14/±9 | ±18 | ±24/±28 | ±40/±44 | -33~+42 |

* For each element ** Estimated from figure v -6dB beam coverage +Calculated from 1-element result×element number

Supplementary Table S2-2 benchmarks the proposed transceiver against reported silicon-based H-band beamforming systems. Whereas previous implementations rely on LO- or IF-domain beamforming across multi-element arrays, the proposed architecture exploits HLM-enabled dispersion engineering to associate individual frequency channels with distinct radiation directions. It is the only system in the comparison that combines bidirectional transceiver operation, multibeam capability, and an on-chip aperture within a single 65-nm CMOS die. The 208-258-GHz operating range is also the broadest among the systems listed, while the frequency-scanned beam covers -33 degrees to +42 degrees without a multi-chip array, external lens, or PCB antenna.

The amplifier-last transmitter delivers a per-channel saturated output power of 3.6 dBm and an EIRP of 10.3 dBm without array combining. Its per-channel output power is 7 dB higher than that of the other entry reporting this metric. The amplifier-first receiver achieves a 14.4-dB single-sideband (SSB) noise figure, improving on the other reported RX and TRX systems by at least 5.6 dB. Despite integrating the TX, RX, LO chain, beam-steering aperture, and multibeam functionality within a 4.9 × 1.5 mm² die, the complete chip consumes only 0.43 W, less than one-third of the next-lowest reported value. These results demonstrate that the proposed circuit-aperture co-design simultaneously advances operating bandwidth, receiver sensitivity, per-channel output power, energy efficiency, and integration complexity, providing a compact system-level solution for broadband multibeam H-band communication.

## 2.7 Circuit-to-space and space-to-circuit conversion gain

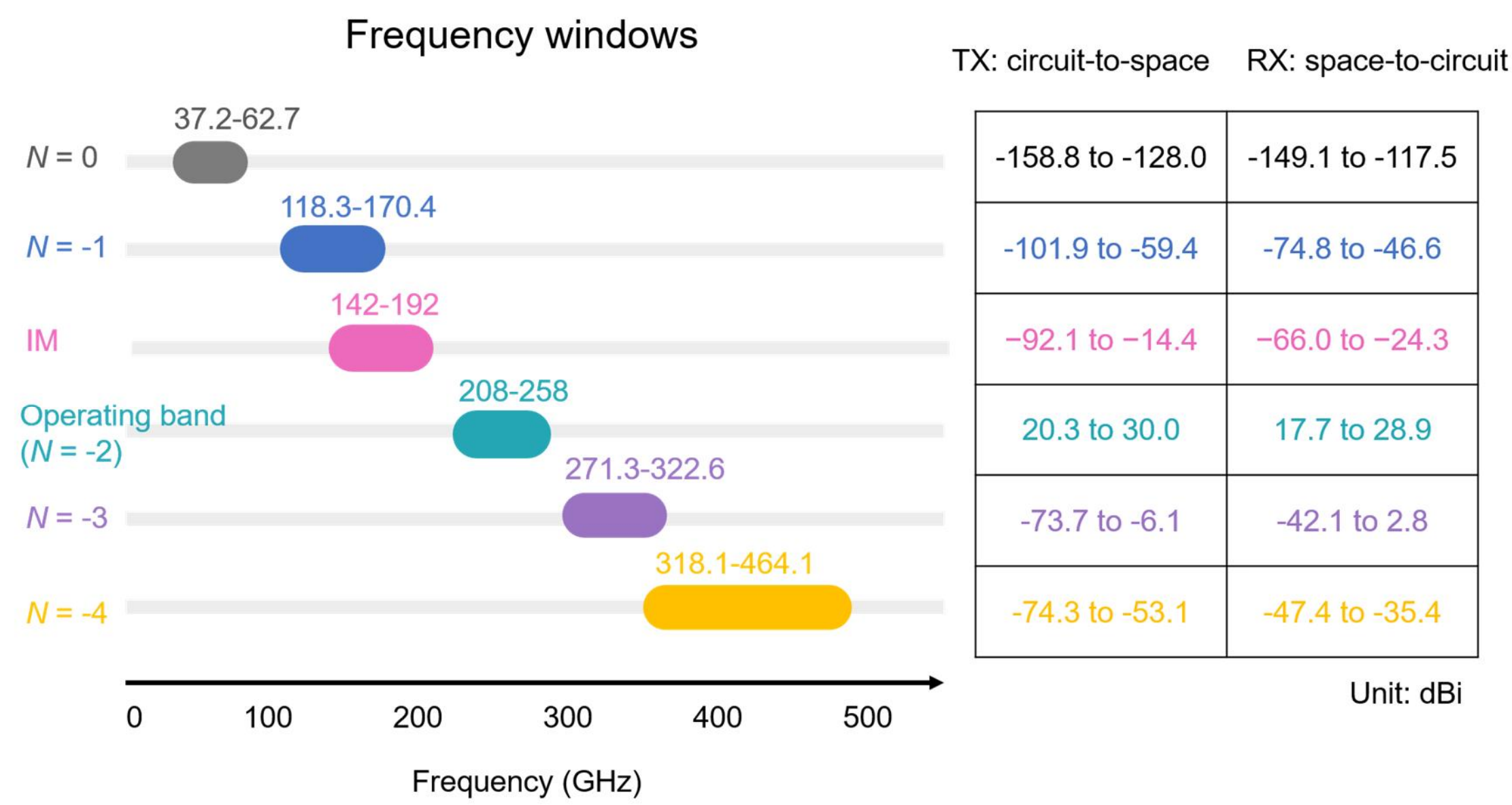

| | TX: circuit-to-space | RX: space-to-circuit |
|---|---|---|
| N = 0 | -158.8 to -128.0 | -149.1 to -117.5 |
| N = -1 | -101.9 to -59.4 | -74.8 to -46.6 |
| IM | −92.1 to −14.4 | −66.0 to −24.3 |
| Operating band (N = -2) | 20.3 to 30.0 | 17.7 to 28.9 |
| N = -3 | -73.7 to -6.1 | -42.1 to 2.8 |
| N = -4 | -74.3 to -53.1 | -47.4 to -35.4 |

**Supplementary Fig. S2-9 | Frequency windows and integrated conversion-gain ranges.**

Supplementary Fig. S2–9 summarises the desired operating band (208-258 GHz), the image-frequency band (142-192 GHz), and the other Floquet radiation windows, all calculated according to Eq. (6) in the main text for scan angles ranging from -50° to 50°, together with the corresponding ranges of the overall TX and RX conversion gains. The overall TX and RX conversion gains are estimated by combining the simulated HLM gain with the simulated PA and LNA gains, respectively.

# 3. SFDMA communication

## 3.1 One-to-one communication measurement setup and representative results

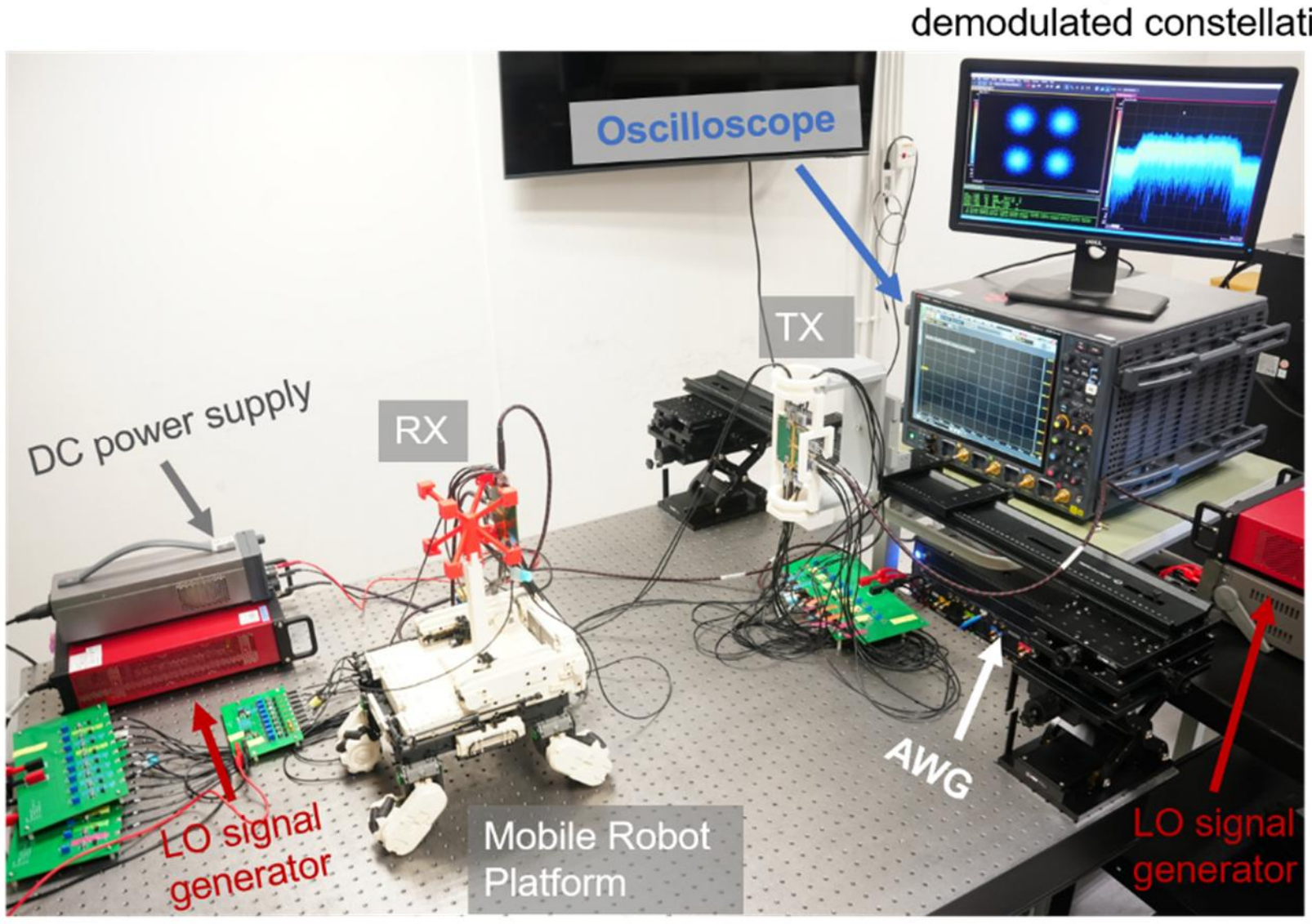

**Supplementary Fig. S3-1 | One-to-One THz communication measurement setup**.

Supplementary Fig. S3-1 shows the experimental setup for the THz communication link. The TRX chip operating as RX is mounted on a mobile robotic platform, enabling spatial movement during the measurement, whereas the TRX chip operating as TX remains fixed. An arbitrary waveform generator (AWG) generates the transmitted waveform, while an oscilloscope for subsequent demodulation captures the received signal. External LO sources and DC supplies provide coherent LO signals and bias voltages for both TRX chips. To establish the maximum experimentally demonstrated data rate for each frequency–spatial channel, short-range communication measurements using 32-QAM and 64-QAM were conducted at link distances of 8–32 cm, with the results summarized in Figs. 4b and c. At longer link distances, the reduced signal-to-noise ratio can be accommodated by employing a lower modulation order and/or a narrower signal bandwidth. Supplementary Figs. S3-2 and S3-3 present the demodulated constellation diagrams of two representative one-to-one THz links.

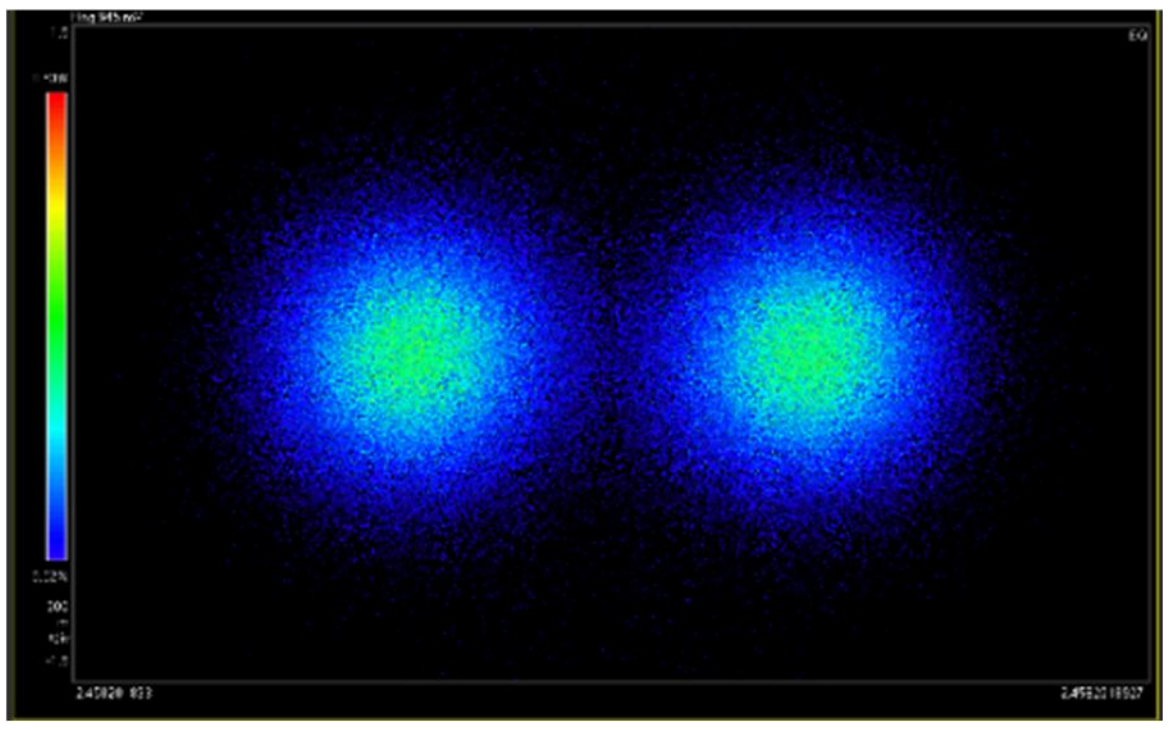

**Supplementary Fig. S3-2 | One-to-one THz communication performance over long distance.** Demodulated constellation centered at 234.3 GHz with a 1-Gbaud (CH20) BPSK signal, corresponding to a data rate of 1 Gbps. The measured error vector magnitude (EVM) is 42.93%, below the BPSK threshold of 55.21%, over an 82-cm wireless link.

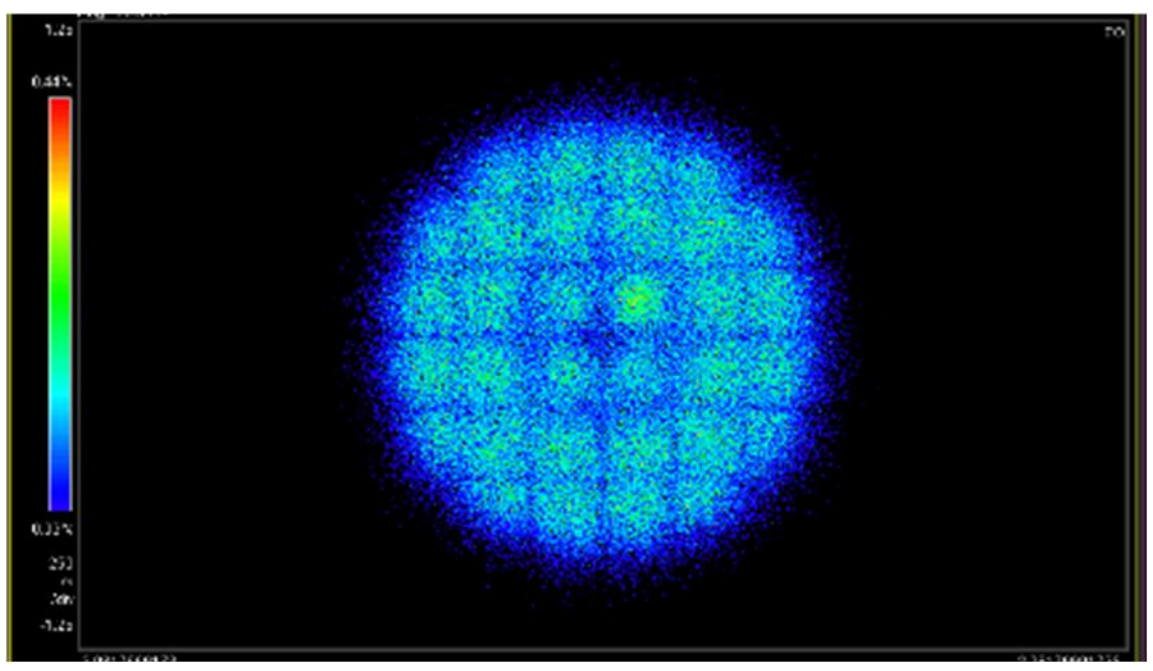

**Supplementary Fig. S3-3 | One-to-one THz communication performance at high data rate.** Demodulated constellation centered at 233 GHz with a 5-Gbaud 32-QAM signal, corresponding to a data rate of 25 Gbps. The measured EVM is 9.57%, below the 32-QAM threshold of 10.5%, over an 8-cm wireless link.

## 3.2 Link budget analysis

The IF-output noise power within the $i$-th channel is modeled as

$$P^{n}_{RX_IF,i} = k_B T_0 B_i F_{RX,i} G_{LNA,i} G_{d,i} \tag{S3-1}$$

Here, $k_B$ is Boltzmann's constant, $T_0$ is the reference noise temperature, $B_i$ is the receiver noise-equivalent bandwidth used for the corresponding channel, and $F_{RX}$ is the receiver noise factor referred to the RX input. Therefore, the complete measured IF power can be written as

$$P_{RX_IF,i} = P^{\mathrm{sig}}_{RX_IF,i} + P^{n}_{RX_IF,i} \tag{S3-2}$$

Combining Eq. (9) in the main text, the signal-to-noise ratio is expressed as:

$$SNR_i = \frac{P_{RX_IF,i}^{\mathrm{Sig}}}{P_{RX_IF,i}^{n}} = \frac{P_{TX_IF,i}^{\mathrm{comm}}\, G_{u,i}\, G_{PA,i}\, G_{TXap,i}(\theta)\left(\frac{c}{4\pi f_i^{\mathrm{comm}} R}\right)^2 G_{RXap,i}(\theta)}{k_B T_0 B_i F_{RX,i}} \tag{S3-3}$$

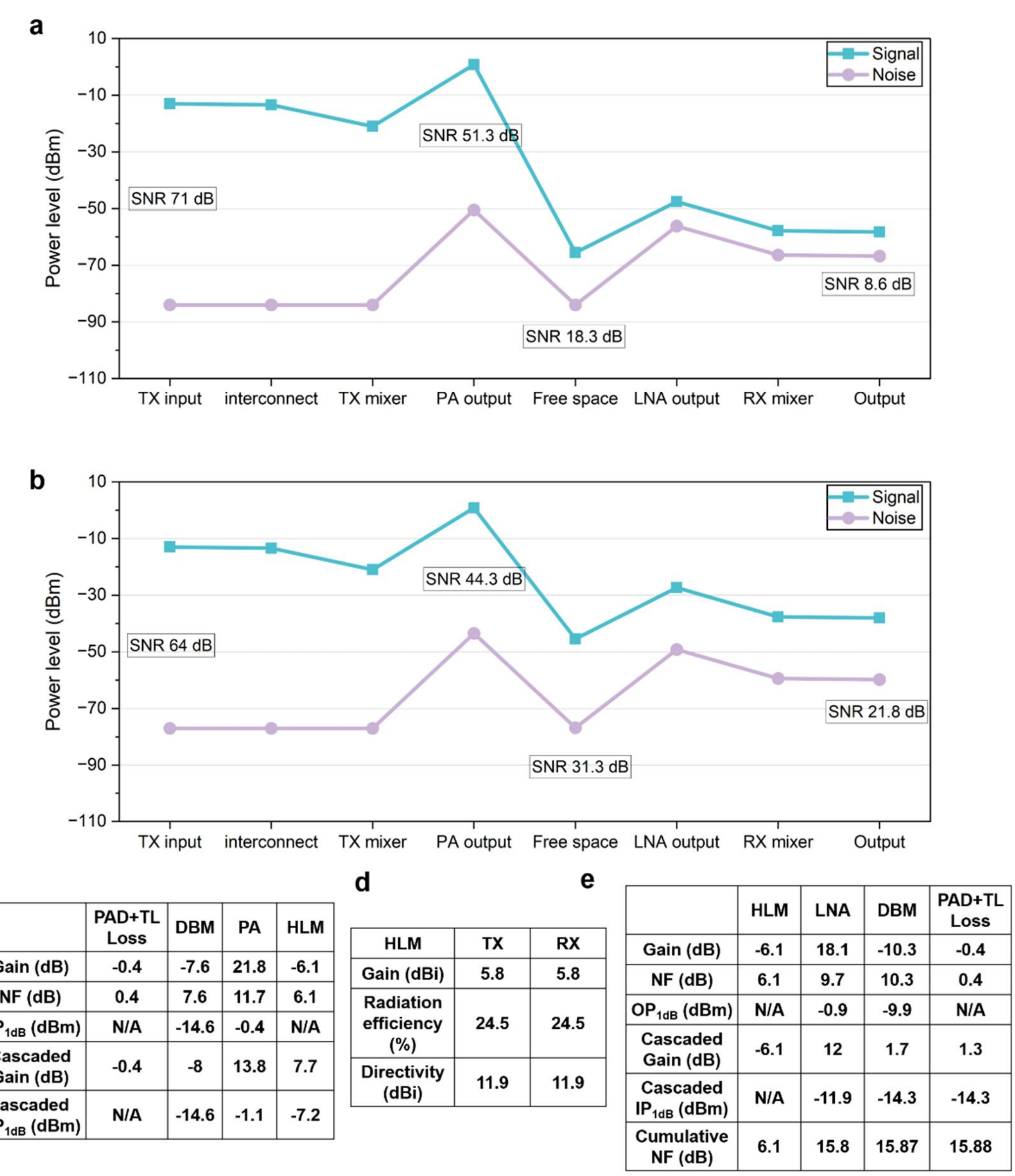

**c**

| | PAD+TL Loss | DBM | PA | HLM |
|---|---|---|---|---|
| Gain (dB) | -0.4 | -7.6 | 21.8 | -6.1 |
| NF (dB) | 0.4 | 7.6 | 11.7 | 6.1 |
| $OP_{1dB}$ (dBm) | N/A | -14.6 | -0.4 | N/A |
| Cascaded Gain (dB) | -0.4 | -8 | 13.8 | 7.7 |
| Cascaded $OP_{1dB}$ (dBm) | N/A | -14.6 | -1.1 | -7.2 |

**d**

| HLM | TX | RX |
|---|---|---|
| Gain (dBi) | 5.8 | 5.8 |
| Radiation efficiency (%) | 24.5 | 24.5 |
| Directivity (dBi) | 11.9 | 11.9 |

**e**

| | HLM | LNA | DBM | PAD+TL Loss |
|---|---|---|---|---|
| Gain (dB) | -6.1 | 18.1 | -10.3 | -0.4 |
| NF (dB) | 6.1 | 9.7 | 10.3 | 0.4 |
| $OP_{1dB}$ (dBm) | N/A | -0.9 | -9.9 | N/A |
| Cascaded Gain (dB) | -6.1 | 12 | 1.7 | 1.3 |
| Cascaded $IP_{1dB}$ (dBm) | N/A | -11.9 | -14.3 | -14.3 |
| Cumulative NF (dB) | 6.1 | 15.8 | 15.87 | 15.88 |

**Supplementary Fig. S3-4 | link-budget analysis. a,b**, Stage-by-stage signal level, noise level, and SNR for the 82-cm, 1-Gbaud binary phase-shift keying (BPSK) link and the 8-cm, 5-Gbaud 32-QAM link, respectively. **c**, TX-stage gain, noise-figure, and compression metrics. **d**, TX and RX aperture characteristics. **e**, RX-stage gain, noise-figure, and compression metrics.

In Eq. (S3-3), $P_{TX_IF,i}^{\mathrm{comm}}$ is the TX IF communication-signal power, $G_{u,i}$ and $G_{PA,i}$ are the upconversion and PA gains, and $G_{TXap,i}(\theta)$ and $G_{RXap,i}(\theta)$ are the directional TX and RX aperture gains.

A bit-error-rate (BER) target below $10^{-3}$ is assumed[S20]. For an M-QAM waveform, the required SNR is estimated using the AWGN approximation[S21]

$$SNR_{req} = \frac{log_2 M(M-1)}{3log_2\sqrt{M}}\left[erfc^{-1}\left(\frac{BERlog_2\sqrt{M}}{1-1/\sqrt{M}}\right)\right]^2 \qquad \text{(S3-4)}$$

where

$$erfc(x) = \frac{2}{\sqrt{\pi}}\int_x^{\infty} e^{-t^2}\,dt \qquad \text{(S3-5)}$$

Following the EVMrms-based criterion adopted in Supplementary Ref. S20, the SNR requirements used in this work are 19.58 dB for 32-QAM and 5.16 dB for BPSK, corresponding to EVMrms limits of 10.5% and 55.21%, respectively. All EVM values reported in this work refer to root-mean-square EVM.

Fig. S3-4 evaluates the two operating modes at $T_0$ = 290 K. The noise-equivalent bandwidth is 1 GHz for the 1-Gbaud BPSK mode and 5 GHz for the 5-Gbaud 32-QAM mode, giving input thermal-noise levels of approximately -84 dBm and -77 dBm, respectively. In both cases, a -13-dBm TX IF input experiences 0.4-dB pad and transmission-line loss, and 7.6-dB mixer conversion loss before the 21.8-dB PA gain raises the signal to 0.8 dBm.

The stage metrics in Supplementary Fig. S3-4c–e provide the electronic and aperture parameters used in the calculation. Including the -6.1-dB guided-to-space efficiency, the TX cascade provides 7.7-dB net gain and a cascaded OP1dB of -7.2 dBm. Each of the TX and RX apertures exhibits 11.9-dBi directivity, 24.5% efficiency, and 5.8-dBi realized gain. On the receiver side, the -6.1-dB space-to-guided efficiency, 18.1-dB LNA gain, -10.3-dB downconversion gain, and -0.4-dB output loss yield 1.3-dB net gain, 15.88-dB cumulative noise figure, and a -14.3-dBm cascaded input 1-dB compression point at the final IF node.

For the 8-cm link in Supplementary Fig. S3-4b, the net spatial-loss term is 46.25 dB after the TX and RX aperture gains are included. The final IF signal and noise levels are -38.05 dBm and -59.8 dBm, respectively, giving an output SNR of 21.8 dB. This exceeds the 19.58-dB 32-QAM requirement by 2.22 dB. At 5 Gbaud, 32-QAM carries five bits per symbol and therefore provides an uncoded raw data rate of 25 Gb/s.

For the 82-cm link in Supplementary Fig. S3-4a, the net spatial-loss term increases to 66.46 dB. The final IF signal and noise levels are -58.26 dBm and -66.8 dBm, yielding an output SNR of 8.6 dB. The resulting 3.44-dB margin above the 5.16-dB BPSK requirement supports a 1-Gb/s uncoded raw data rate at 1 Gbaud. The two budgets, therefore, establish complementary high-throughput short-range and robust extended-range operating modes at the nominal carrier frequency.

### 3.3 THz multimedia transmission measurement setup

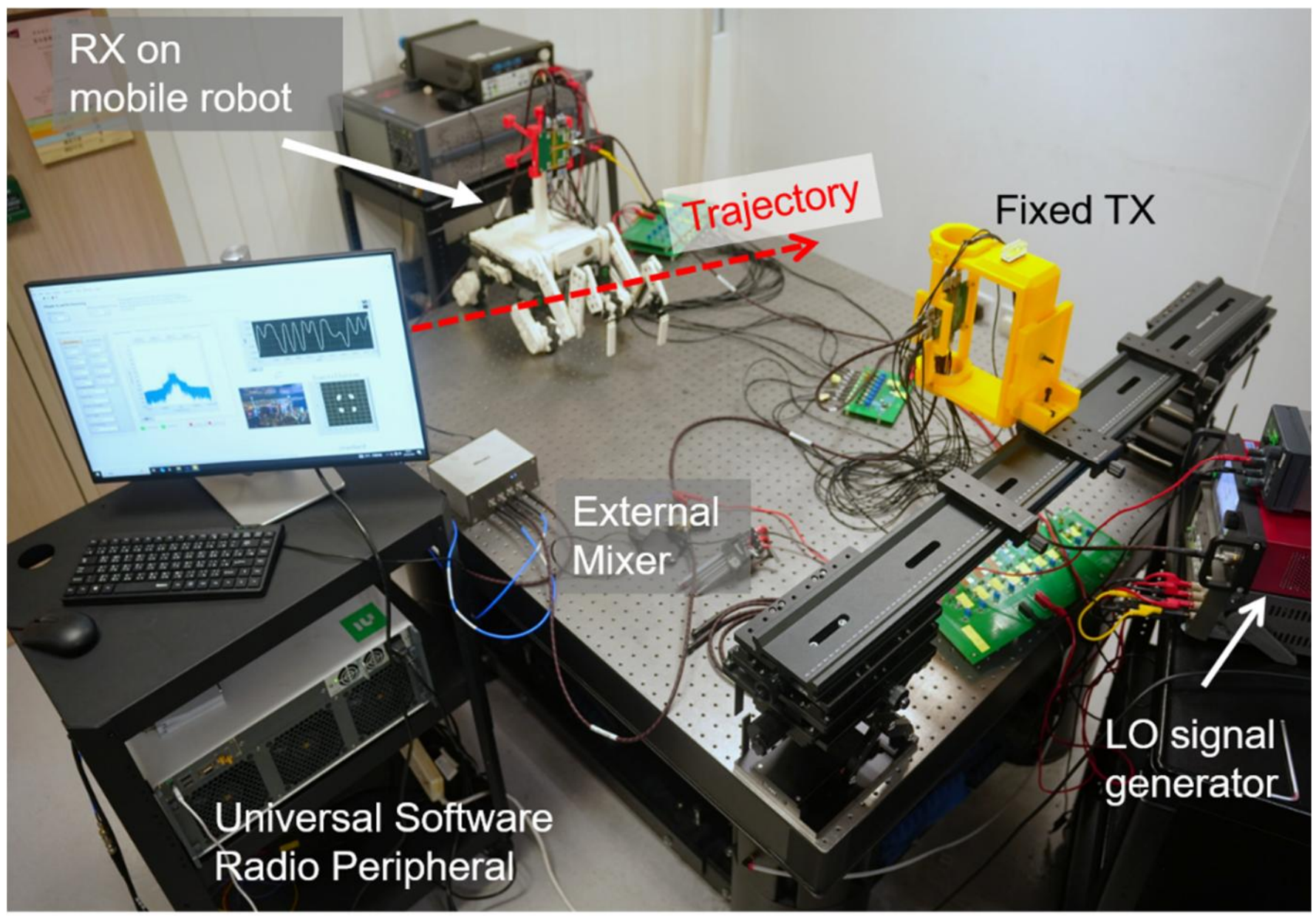

**Supplementary Fig. S3-5 | Experimental setup for THz multimedia transmission from a fixed TX to an RX mounted on a mobile robot.**

Supplementary Fig. S3-5 illustrates the experimental setup for THz multimedia transmission with a fixed TX and a mobile robot-mounted RX. A universal software radio peripheral (USRP)-2944 generates quadrature phase-shift keying (QPSK)-modulated image and video streams, with an external mixer translating the signals to the required IF range. Owing to test-equipment limitations, these experiments are restricted to a symbol rate of 100 kSymbol $s^{-1}$ and one-to-one communication. By exploiting the frequency–space mapping of the proposed HLM aperture, different RX positions along the robot trajectory are served by distinct SFDMA channels.

# 4. Neural-network-assisted AoA estimation and 2D trajectory reconstruction

## 4.1 AoA sample acquisition measurement and setup

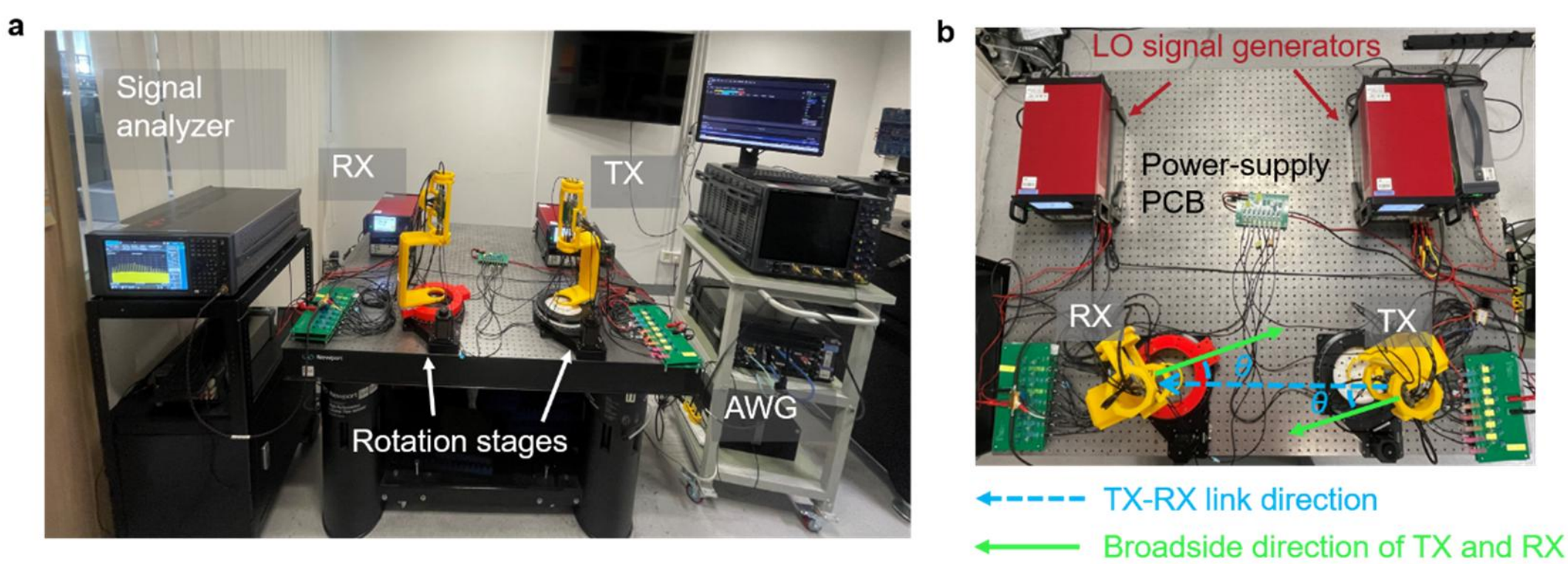

**Supplementary Fig. S4-1 | Photograph of the AoA sample acquisition setup with TX and RX mounted on two rotation stages. a**, overview and **b**, top view.

Supplementary Fig. S4-1 shows the angle-of-arrival (AoA) sample-acquisition setup of two chips. An AWG supplies 50 equal-amplitude low-frequency tones to the TX IF input. The TX performs on-chip upconversion and amplification, and then the transmit HLM aperture radiates the resulting RF tones from 207.8 to 256.8 GHz with 1-GHz spacing. Following over-the-air propagation, the THz tones are captured by the receive HLM aperture, amplified, and downconverted into low-frequency IF signals, which are recorded by a spectrum analyzer. Two signal generators provide quarter-rate LO signals at the same frequency to the TX and RX. During sample acquisition, the TX and RX module boards are mounted on two independent rotation stages, whose orientations are adjusted simultaneously to keep the two chips parallel. The angle $\theta$ is defined as the angle between the TX-RX link direction and the broadside direction of either module board.

Due to the limited bandwidth of the signal analyzer, the 50-tone signal with 1-GHz spacing was measured in two separate frequency spans. In the first span, the signal generator was set to 49.95 GHz to provide the quarter-rate LO signal, while the AWG generated equal-amplitude IF tones from 8 to 32 GHz with 1-GHz spacing. In the second span, the signal generator was set to 51.95 GHz, while the AWG generated equal-amplitude IF tones from 25 to 49 GHz with 1-GHz spacing. Combining the two spans gives 50 discrete RF tones from 207.8 to 256.8 GHz with 1-GHz spacing.

In the first span, the signal-generator frequency was set to 49.95 GHz, intentionally offset from 50 GHz. If a 50-GHz source were used, a 25-GHz spur internally generated by the signal generator could enter the transmitter LO path and mix with the 200-GHz LO, producing an undesired 225-GHz component. This component would overlap with the desired RF tone upconverted from the 25-GHz IF signal and interfere with the measurement. The same consideration applies to the second measurement.

This tone-based probing strategy enhances the per-tone SNR by simultaneously increasing the useful signal power and reducing the integrated measurement noise. For the same total transmitted power, concentrating the energy into discrete tones produces stronger received spectral peaks than a continuous wideband waveform. Meanwhile, each tone is measured through the signal analyzer's narrow resolution-bandwidth filter, so the associated noise power is integrated only over the analyzer bandwidth rather than over the entire sensing bandwidth. Consequently, the received spectral signatures become more robust, enabling a longer effective sensing range.

### 4.2 Distance-invariant feature extraction for AoA estimation

In the sensing experiments, the frequency of the $j$-th RF tone for sensing is given by

$$f_j^{\mathrm{sen}} = 207.8+(j-1)\ \mathrm{GHz},\ j = 1, 2, \ldots, 50 \tag{S4-1}$$

Following the communication link model Eq. (9), the measured RX IF power of the $j$-th sensing tone can be written as

$$P_{RX_IF,j}^{sen} = P_{TX_IF,j}\, G_{u,j}\, G_{PA,j}\, G_{TXap,j}(\theta)\left(\frac{c}{4\pi f_j^{\mathrm{sen}} R}\right)^2 G_{RXap,j}(\theta) G_{LNA,j}\, G_{d,j} \tag{S4-2}$$

Here, the subscript $j$ denotes the sensing-tone index. The noise term is neglected because the per-tone SNR is sufficiently high under the adopted measurement conditions. This expression shows that the angle information is encoded in the frequency-dependent received spectrum rather than only in the strongest tone; therefore, we develop a neural-network-assisted AoA estimator to learn these angle-dependent spectral signatures of the received THz signals.

Since identical leaky-metasurface apertures are used for TX and RX and the two modules are kept parallel during the experiments, the TX and RX aperture responses are approximated as

$$G_{TXap,j}(\theta) G_{RXap,j}(\theta) \approx G_{ap,j}(\theta)^2 \tag{S4-3}$$

Substituting Eq. (S4-3) into Eq. (S4-2) gives

$$P_{RX_IF,j}^{sen} = P_{TX_IF,j}\, G_{u,j}\, G_{PA,j}\, G_{ap,j}(\theta)^2\, G_{LNA,j}\, G_{d,j}\, (\frac{c}{4\pi\, f_j^{\text{sen}}\, R})^2 \qquad \text{(S4-4)}$$

The expression (S4-4) can be decomposed into a distance-dependent scaling term and a frequency-angle-dependent spectral term as

$$P_{RX_IF,j}^{sen} = \frac{1}{R^2} Q_j(\theta) \qquad \text{(S4-5)}$$

where

$$Q_j(\theta) = P_{TX_IF,j}\, G_{u,j}\, G_{PA,j}\, G_{ap,j}(\theta)^2\, G_{LNA,j}\, G_{d,j}\, (\frac{c}{4\pi\, f_j^{\text{sen}}})^2 \qquad \text{(S4-6)}$$

Here, $Q_j(\theta)$ collects all frequency-dependent front-end responses, the frequency-angle-dependent aperture response, and the frequency-dependent part of the free-space propagation term. This notation highlights that, within a measured spectrum, the propagation distance $R$ mainly contributes a common multiplicative scaling factor, $1/R^2$, across all frequency tones, whereas the useful AoA information is encoded in the relative frequency-dependent distribution $Q_j(\theta)$, which varies with the TX-RX angle $\theta$.

To mathematically cancel the distance-dependent $1/R^2$ factor, a probability density function (PDF) feature is obtained by normalizing the measured linear-power spectrum by its total power:

$$p_j = \frac{P_{RX_IF,j}^{sen}}{\sum_{n=1}^{N} P_{RX_IF,n}^{sen}} \qquad \text{(S4-7)}$$

Substituting the separated link model gives

$$p_j = \frac{\frac{1}{R^2} Q_j(\theta)}{\sum_{n=1}^{N} \frac{1}{R^2} Q_n(\theta)} = \frac{Q_j(\theta)}{\sum_{n=1}^{N} Q_n(\theta)} \qquad \text{(S4-8)}$$

Therefore, the distance-dependent factor $1/R^2$ is cancelled by the PDF normalization, while the relative spectral distribution associated with $\theta$ is retained.

To additionally capture how the spectral energy accumulates along the frequency axis, the cumulative distribution function (CDF) is derived from the PDF and appended to the feature representation:

$$c_j = \sum_{n=1}^{j} p_n \qquad \text{(S4-9)}$$

Finally, the PDF and CDF are concatenated as the model input feature:

$$x = [p_1, \dots, p_{50}, c_1, \dots, c_{50}]^T \in \mathbb{R}^{100} \qquad \text{(S4-10)}$$

This PDF-CDF representation preserves the repeatable angle-dependent spectral signature while reducing sensitivity to absolute power fluctuations, making it suitable for robust AoA estimation over the operating bandwidth.

As an example, Supplementary Fig. S4-2 shows the measured spectrum at a TX-RX relative angle of $\theta = 16°$. The signal analyzer records the RX IF output spectrum; for clarity, the frequency axis in the figure is converted to the corresponding RF frequency by adding the LO frequency. Supplementary Fig. S4-3a and b show the calculated PDF and the CDF of the sample.

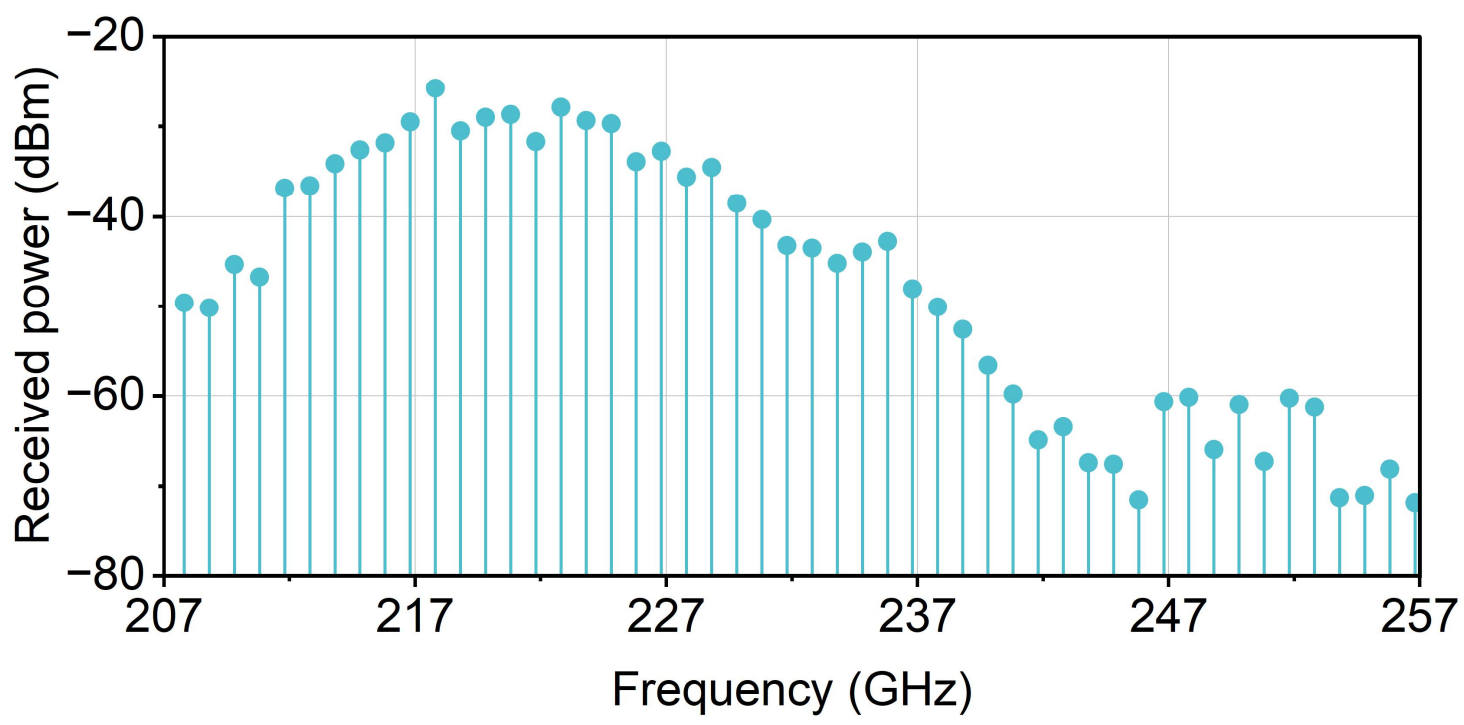

**Supplementary Fig. S4-2 | Example of a measured 50-tone spectrum from 207.8 to 256.8 GHz with 1-GHz spacing at $\theta = 16°$.**

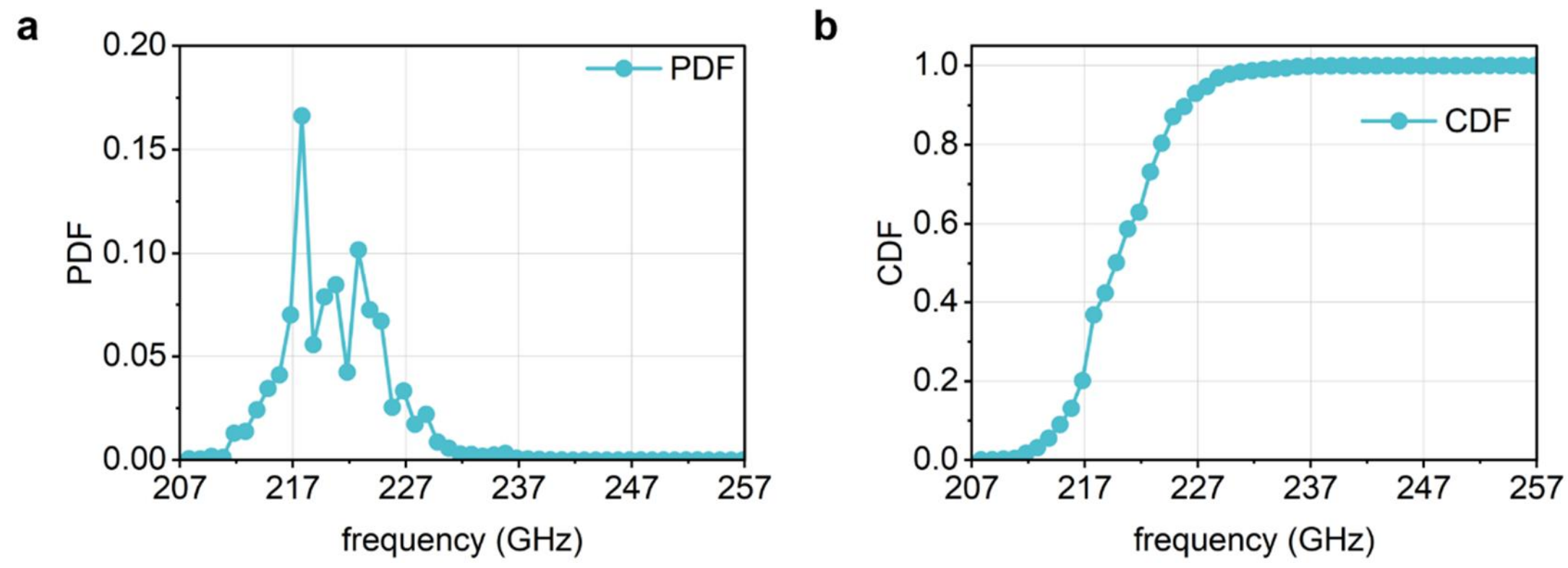

**Supplementary Fig. S4-3 | Feature of the sample of $\theta = 16°$. a,** PDF. **b,** CDF.

### 4.3 Prototype-similarity regression

The model contains a positive learnable feature-weight vector $\boldsymbol{w}$ and a positive similarity-scale parameter $T$. For an unknown feature vector $\boldsymbol{x}$ and the $k$th prototype feature vector $x_k$, the weighted representations are

$$\boldsymbol{z} = \boldsymbol{w} \odot \boldsymbol{x},\ \boldsymbol{z}_k = \boldsymbol{w} \odot \boldsymbol{x}_k \tag{S4-11}$$

where $\odot$ denotes element-wise multiplication. Their cosine-similarity score is

$$s_k = \cos(\boldsymbol{z}, \boldsymbol{z}_k) = \frac{\boldsymbol{z}^T \boldsymbol{z}_k}{\|\boldsymbol{z}\|_2 \|\boldsymbol{z}_k\|_2} \tag{S4-12}$$

The similarity scores are converted into normalized prototype weights using a softmax operation:

$$\alpha_k = \frac{\exp(T s_k)}{\sum_{m=1}^{N_p} \exp(T s_m)} \tag{S4-13}$$

where $N_p$ is the number of prototypes. The resulting weights are non-negative and sum to one. The estimated AoA is then obtained using the similarity-weighted interpolation defined in Eq. (10) of the main text.

### 4.4 Model training, cross-validation, and feature-ablation analysis of the AoA estimator

The target-range dataset contains 85 measured spectra acquired between -40° and +40°. Specifically, the target-range set comprises 81 spectra measured at 1° intervals from -40° to +40°, together with four additional off-grid measurements acquired at 3.5°, 14.5°, 19.5°, and 32.5°. These half-degree samples introduce sub-degree angular diversity and help verify that the estimator learns a continuous spectral-to-angle mapping rather than relying on an integer-angle lookup grid. Ten additional boundary-support spectra, acquired from -45° to -41° and from +41° to +45°, improve prototype support near the edges of the target range and were excluded from validation and included in the training set of every fold.

For five-fold cross-validation, the 85 target-range samples are divided into five fixed, non-overlapping folds of 17 samples each. In each fold, 68 target-range samples and all 10 boundary-support samples form a 78-sample training set, while the remaining 17 target-range samples constitute the held-out validation set. The validation and boundary-support angles are listed in Supplementary Table S4-1.

At each training epoch, leave-one-out prediction is performed within the 78-sample training set: each training sample is used once as the query, while the remaining 77 samples serve as prototypes. The mean loss over the 78 leave-one-out predictions is backpropagated to optimize $\boldsymbol{w}$ and $T$.

Each held-out validation spectrum is subsequently treated as an unseen input and compared with the 78 prototypes from the corresponding training fold. This procedure ensures that every target-range spectrum is evaluated exactly once without being included among its validation prototypes. After cross-validation, the final estimator is retrained using all 95 measured spectra, which subsequently serve as prototypes during inference.

**Supplementary Table S4-1 | Fixed held-out validation angles and boundary-support training angles for five-fold cross-validation.**

| Fold | Held-out validation angles (°) | Boundary-support angles included in training (°) |
|---|---|---|
| **Fold 1** | -40, -35, -30, -25, -20, -15, -10, -5, 0, 4, 9, 14, 18, 22, 27, 32, 36 | -45, -44, -43, -42, -41, 41, 42, 43, 44, 45 |
| **Fold 2** | -39, -34, -29, -24, -19, -14, -9, -4, 1, 5, 10, 14.5, 19, 23, 28, 32.5, 37 | |
| **Fold 3** | -38, -33, -28, -23, -18, -13, -8, -3, 2, 6, 11, 15, 19.5, 24, 29, 33, 38 | |
| **Fold 4** | -37, -32, -27, -22, -17, -12, -7, -2, 3, 7, 12, 16, 20, 25, 30, 34, 39 | |
| **Fold 5** | -36, -31, -26, -21, -16, -11, -6, -1, 3.5, 8, 13, 17, 21, 26, 31, 35, 40 | |

The proposed AoA estimator is trained using a Smooth-L1 regression loss with a weak regularization term on the feature weights. To evaluate its generalization capability and avoid reporting accuracy caused by accidental fitting to a specific training split, five-fold cross-validation is adopted. In each fold, the validation samples are excluded from model training and are estimated only by the model trained with the remaining samples. Therefore, the K-fold prediction error represents out-of-fold generalization performance rather than training-set fitting accuracy. As shown in Supplementary Fig. S4-4, the validation mean absolute error (MAE) remains below 0.75° across all five folds, indicating stable prediction performance over different train-validation splits.

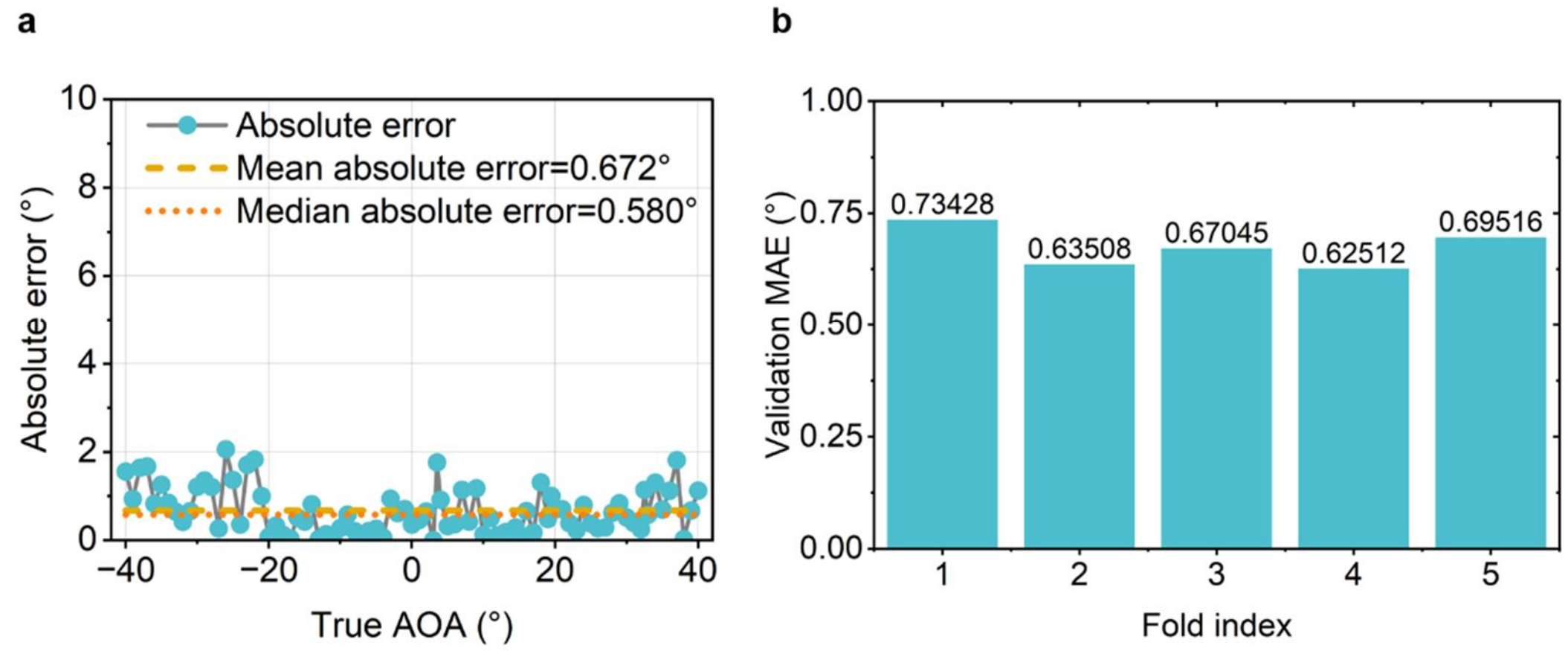

**Supplementary Fig. S4-4 | Five-fold cross-validation performance of the neural-network-assisted AoA estimator. a,** Absolute errors of the 85 out-of-fold AoA estimates, with the dashed and dotted lines denoting the overall mean and median absolute errors, respectively. **b,** Validation MAE for each fold, with 17 held-out target-range samples evaluated in every fold. 10 boundary-support samples are used only for training and are excluded from the validation statistics.

To further evaluate the contribution of the proposed PDF-CDF feature representation, we perform a feature-ablation study using the same prototype-similarity neural regressor and the same five-fold validation protocol. Two additional models are trained with reduced feature inputs: the PDF-only model uses only the 50-dimensional PDF feature, while the CDF-only model uses only the 50-dimensional CDF feature. Their performance is compared with that of the full 100-dimensional PDF-CDF feature. As shown in Supplementary Fig. S4-5, the combined PDF–CDF representation therefore provides the most accurate and robust performance. This result confirms that the PDF captures the local spectral-energy distribution, whereas the CDF provides complementary cumulative frequency-order information, and their combination forms a more discriminative feature representation for AoA estimation.

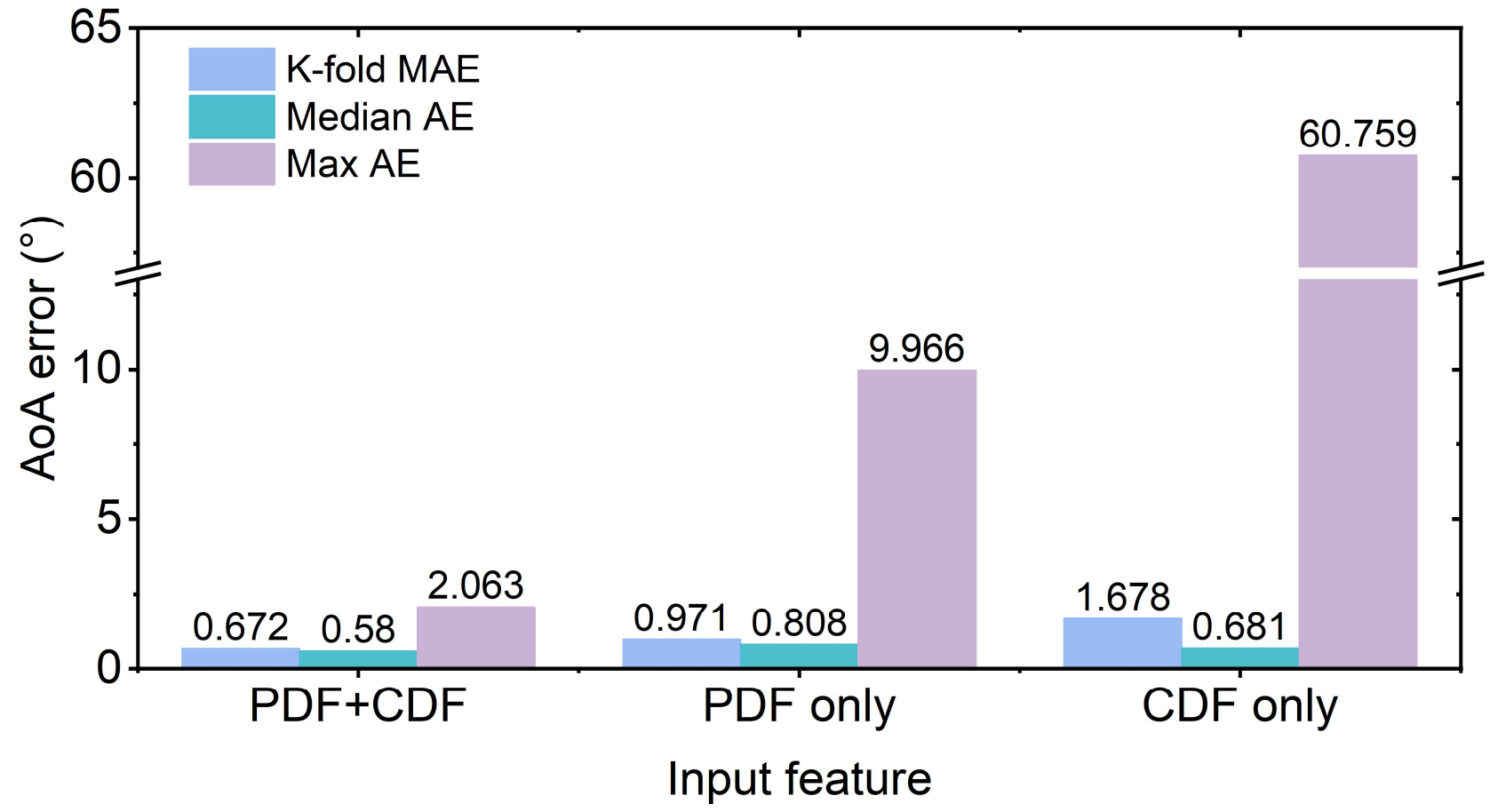

**Supplementary Fig. S4-5 | Feature-ablation comparison of AoA estimation accuracy.**

## 4.5 2D localization based on AoA triangulation

In this section, we perform a 2D localization experiment using two fixed transmitters and one moving receiver. The two transmitters are placed on the $x$-axis with a spacing of 56 cm, as shown in Fig. 5g. Their coordinates are defined as

$$r_1 = \left(x_{1,0}\right) = (-28\ \text{cm}, 0),\ \ r_2 = \left(x_{2,0}\right) = (28\ \text{cm}, 0) \tag{S4-14}$$

The AoA is defined with respect to the positive $y$-axis. For the $i$-th fixed transmitter located at ($x_i$, 0), the AoA estimated from the received spectral signature defines a line-of-bearing constraint for the moving receiver position ($x$, $y$):

$$x = x_i - y\tan\left(\hat{\theta}_i\right), i = 1, 2 \tag{S4-15}$$

The estimated 2D receiver position ($\hat{x}$, $\hat{y}$) is obtained from the intersection of the two bearing lines.

$$\hat{y} = \frac{x_2 - x_1}{\tan(\hat{\theta}_2) - \tan(\hat{\theta}_1)} \tag{S4-16}$$

$$\hat{x} = x_1 - \hat{y}\tan\left(\hat{\theta}_1\right) \tag{S4-17}$$

The ten measured points form an S-shaped trajectory, covering different receiver positions and bearing-angle combinations. Supplementary Fig. S4-6 presents the AoA and localization errors along the ten-point trajectory. The mean AoA errors associated with TX1 and TX2 are 0.77° and 0.72°, respectively. The mean and maximum 2D localization errors are 1.49 cm and 1.81 cm, respectively. All spectra used in the 2D trajectory experiment were acquired independently and were excluded from the training, cross-validation, and prototype datasets, serving only as unseen inputs during inference.

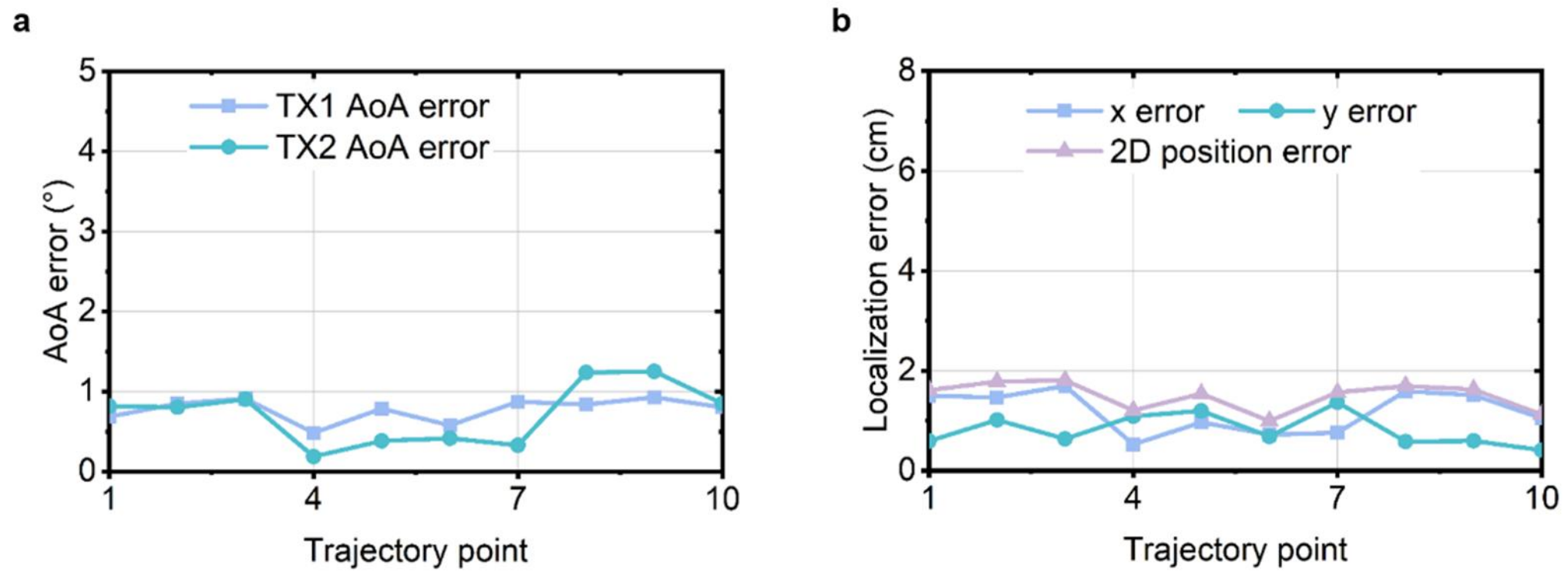

**Supplementary Fig. S4-6 | 2D localization errors along an S-shaped mobile-receiver trajectory. a,** AoA prediction errors for the two bearing links from TX1 and TX2. **b,** Absolute *x*- and *y*-coordinate errors and the corresponding 2D position error.